\documentclass[%
  aps,
  prd,
  nofootinbib,
  preprintnumbers,
  superscriptaddress,
  floatfix,
]{revtex4-2}

\usepackage{graphicx}
\usepackage{dcolumn}
\usepackage{bm}
\usepackage{amsmath}
\usepackage{amssymb}
\usepackage{amsthm}
\usepackage{xcolor}
\usepackage{physics}
\usepackage{tensor}
\usepackage{cases}
\usepackage{multirow}
\usepackage{appendix}

\usepackage[%
  bookmarksnumbered,
  pdfpagelabels=true,
  plainpages=false,
  colorlinks=true,
  linkcolor=blue,
  citecolor=blue,
  urlcolor=blue,
]{hyperref}

\newcommand{\be}{\begin{equation}}
\newcommand{\ee}{\end{equation}}

\newcommand\m[1]{\begin{pmatrix}#1\end{pmatrix}}

\begin{document}


\title{Solving BDNK diffusion using physics-informed neural networks}

\author{Vicente Chomal\'i-Castro}
 \email{faridc2@illinois.edu}
 \affiliation{Illinois Center for Advanced Studies of the Universe and Department of Physics, \\ University of Illinois Urbana-Champaign, Urbana, IL 61801, USA}
\author{Nick Clarisse}
 \email{nclari2@illinois.edu}
 \affiliation{Illinois Center for Advanced Studies of the Universe and Department of Physics, \\ University of Illinois Urbana-Champaign, Urbana, IL 61801, USA}
\author{Nicki Mullins}
 \email{nmmulli2@ncsu.edu}
 \affiliation{Department of Physics, North Carolina State University, Raleigh, NC 27695, USA}
\author{Jorge Noronha}
 \email{jn0508@illinois.edu}
 \affiliation{Illinois Center for Advanced Studies of the Universe and Department of Physics, \\ University of Illinois Urbana-Champaign, Urbana, IL 61801, USA}

\date{\today}

\begin{abstract}
In this work, we reformulate the relativistic BDNK (Bemfica-Disconzi-Noronha-Kovtun) diffusion equation in flux-conservative form, and solve the resulting equations in $(1+1)$D using both a second-order Kurganov-Tadmor finite volume scheme and physics-informed neural networks (PINNs). In particular, we introduce the SA-PINN-ACTO framework, which combines the self-adaptive PINN technique with an exact enforcement of initial and periodic boundary conditions through an algebraic transform of the network's raw output, allowing the network to focus solely on minimizing the PDE residual. We test both approaches on smooth and discontinuous initial data, for both trivial and dynamically evolving velocity and temperature BDNK backgrounds, and for two characteristic speeds. The SA-PINN-ACTO method matches the converged Kurganov-Tadmor solutions for smooth profiles, while for discontinuous profiles the errors increase, reflecting an expected limitation of PINNs near sharp gradients.
\end{abstract}


\maketitle

\section{Introduction}\label{sec:introduction}

Hydrodynamics describes the long-wavelength, long-timescale behavior of systems governed by the dynamics of conserved quantities~\cite{landau1987fluid}. Its power and universality arise from the fact that the hydrodynamic equations of motion follow directly from conservation laws which, in the nonrelativistic setting, encode the local conservation of mass, energy, and momentum. For ideal fluids, the system closes once an equation of state is specified, and one can show that local solutions exist and are unique for appropriate initial data~\cite{AnileBook}. Beyond equilibrium, additional input is required to determine the behavior of dissipative fluxes and solve the corresponding system of partial differential equations. Hydrodynamics provides a reliable effective description whenever a clear scale separation exists, enabling a systematic gradient expansion of the constitutive relations~\cite{landau1987fluid}. This approximation applies when the characteristic gradient length scale $\ell$ of the macroscopic variables is much larger than the microscopic interaction scale $l_{\text{micro}}$ (e.g.\ the mean free path in a dilute gas). This corresponds to the small Knudsen-number regime, $\mathrm{Kn} = l_{\text{micro}}/\ell \ll 1$, in which hydrodynamics is expected to hold~\cite{landau1987fluid,Rocha:2023ilf}. Truncating the gradient expansion at first order yields the well-known Navier--Stokes equations~\cite{landau1987fluid}, which successfully describe a wide range of phenomena from laminar flows to turbulent motion~\cite{Frisch_1995}.

The same reasoning may be extended to incorporate relativity and determine how it modifies the equations of motion of fluid dynamics~\cite{Rezzolla_Zanotti_book}. At zero chemical potential, relativistic hydrodynamics is built around the covariant conservation of the energy-momentum tensor,
\begin{equation}
\partial_\mu T^{\mu\nu} = 0.
\end{equation}
For an ideal relativistic fluid, the energy-momentum tensor takes the form
\begin{equation}
T^{\mu\nu}_0 = \varepsilon u^\mu u^\nu + P \Delta^{\mu\nu},
\label{eq:ideal_hydro_Tmunu}
\end{equation}
where $u^\mu$ is the fluid four-velocity (a future-directed timelike vector normalized as $u_\mu u^\mu = -1$), $\Delta_{\mu\nu} = \eta_{\mu\nu} + u_\mu u_\nu$ projects orthogonally to $u^\mu$, $\eta_{\mu\nu}$ is the (Minkowski) spacetime metric, $\varepsilon$ is the energy density, and $P = P(\varepsilon)$ is the equilibrium pressure determined by the equation of state. Energy-momentum conservation then gives four equations, matching the four independent dynamical variables of ideal hydrodynamics, yielding a closed system. Dissipative effects can be incorporated via corrections encoded in a tensor $\Pi^{\mu\nu}$, leading to
\[
T^{\mu\nu} = T^{\mu\nu}_0 + \Pi^{\mu\nu}.
\]
These viscous corrections depend on transport coefficients such as the shear viscosity $\eta$ and the bulk viscosity $\zeta$, which characterize the response to spacetime gradients. Modern formulations of relativistic dissipative hydrodynamics fall broadly into two classes: derivative-expansion formulations and second-order (Israel--Stewart-type) theories. The earliest derivative-expansion theories—those of Eckart~\cite{Eckart:1940te} and Landau--Lifshitz~\cite{landau1987fluid}—define $\Pi^{\mu\nu}$ using first-order derivatives of the ideal hydrodynamic variables, with the two formulations corresponding to different hydrodynamic frames~\cite{Kovtun:2012rj}. While these equations reduce to Navier--Stokes theory in the nonrelativistic limit, they face acausality and display unphysical instabilities around equilibrium~\cite{Hiscock:1985zz} that prevent their use in most applications.

The problems of the relativistic Navier--Stokes equations motivated early on a shift to frameworks where the dissipative fluxes are promoted to new dynamical variables. The prototypical example is Israel--Stewart (IS) theory~\cite{Israel:1979wp}, in which relaxation-type evolution equations for $\Pi^{\mu\nu}$ are derived either by expanding a non-equilibrium entropy current to second order in deviations from equilibrium or via a truncation of the relativistic Boltzmann equation~\cite{Israel:1979wp,PhysRevD.85.114047}. For a review, see~\cite{Rocha:2023ilf}. Generalizations of the Israel--Stewart framework~\cite{Baier:2007ix,PhysRevD.85.114047} are widely used in modeling the quark-gluon plasma formed in ultrarelativistic heavy-ion collisions, see~\cite{Romatschke:2017ejr}. Israel--Stewart-type theories also naturally arise in astrophysics, prominently in studies of non-conserved currents and bulk-viscous effects in neutron star mergers~\cite{Gavassino_2021,Alford:2017rxf,Most:2021zvc,Celora:2022nbp,Camelio:2022fds,Camelio:2022ljs,Most_2024,Chabanov:2023blf,Gavassino:2023xkt,Ripley:2023qxo,Yang:2023ogo,Yang:2025yoo}.

Second-order theories enjoy significantly better physical and mathematical properties than the first-order Eckart or Landau--Lifshitz formulations. In the linear regime, causality and stability can be achieved with appropriate parameter choices~\cite{Hiscock_Lindblom_stability_1983}, and for theories derived from an information-current principle~\cite{Gavassino:2021cli,Gavassino:2021kjm}, strong hyperbolicity also holds near equilibrium~\cite{Gavassino:2023odx}. Only recently have general nonlinear results become available. Indeed, general statements concerning causality and strong hyperbolicity have been rigorously established in the purely bulk-viscous case~\cite{Bemfica:2019cop}, in resistive relativistic magnetohydrodynamics around black holes~\cite{Cordeiro:2023ljz}, and for various second-order theories involving diffusion~\cite{Cordeiro:2025mtg} and shear/bulk viscosity~\cite{Bemfica:2020xym} (see also earlier work in~\cite{Floerchinger:2017cii}). Nevertheless, the full set of conditions ensuring causality, strong hyperbolicity, and local well-posedness—including all dissipative channels—remains unknown, with complexity further compounded by the large number of second-order contributions in general theories~\cite{PhysRevD.85.114047}. Global well-posedness statements have been obtained in the bulk-viscous case in Ref.~\cite{disconzi2023breakdown} and, in~\cite{Bemfica:2025gws}, finite-time gradient blow-up and shock formation in Israel--Stewart theory with (1+1)-dimensional plane symmetry with bulk, shear, and diffusion were investigated.  

The Bemfica-Disconzi-Noronha-Kovtun (BDNK) theory~\cite{Bemfica:2017wps,Kovtun:2019hdm,Bemfica:2019knx,Hoult:2020eho,Bemfica:2020zjp} offers a simpler and more tractable approach to relativistic viscous hydrodynamics, at least near equilibrium. BDNK is the most general causal, stable, and strongly hyperbolic first-order formulation of relativistic hydrodynamics that includes shear, bulk, and diffusion effects while retaining only first-order derivatives of the hydrodynamic variables. Local well-posedness was proven in~\cite{Bemfica:2019hok,Bemfica:2020gcl,Bemfica:2020zjp,Disconzi-Shao-2024}, and global well-posedness for small initial data in Minkowski space was established in~\cite{Sroczinski,Freistuhler-Sroczinski-2025}. These results show that BDNK behaves precisely as a relativistic viscous hydrodynamic theory should, where small perturbations evolve nonlinearly toward equilibrium in a manner consistent with relativity. In that context, BDNK is unique among relativistic viscous frameworks because of this level of mathematical control.

A distinguishing feature of BDNK is the presence of terms that involve first-order time derivatives even in the fluid rest frame, illustrating the flexibility in choosing the hydrodynamic frame. The additional transport parameters of BDNK, which supplement the usual $\eta$, $\zeta$, and conductivity $\sigma$, encode this frame choice~\cite{Bemfica:2017wps,Kovtun:2019hdm}. A broad set of parameter choices can be used to maintain causality and stability for this theory, see~\cite{Bemfica:2020zjp,Abboud:2023hos}.

In BDNK hydrodynamics, hydrodynamic frame parameters can be chosen so that causality holds automatically, a striking contrast to Israel--Stewart-type theories, where causality must be checked through evolving nonlinear inequalities~\cite{Bemfica:2020xym,Plumberg:2021bme}. This structural simplicity makes BDNK not only mathematically appealing but also, at least in principle, highly convenient for numerical simulation. As a result, it has been applied in a variety of high-energy and astrophysical contexts—including linearized studies~\cite{Redondo-Yuste:2024vdb,Caballero:2025omv,Redondo-Yuste:2025ktt}—and has been used in numerous recent numerical works~\cite{Pandya_2022a,Pandya_2022b,Bantilan:2022ech,Bea:2023rru,Bea:2025eov,Bhambure:2024axa,Fantini:2025gnm,Clarisse:2025lli,Keeble:2025bkc,Shum:2025jnl}.

Diffusion plays a central role in relativistic hydrodynamics as it governs the relaxation of conserved-charge imbalances and determines how matter and charge are transported in fluids~\cite{landau1987fluid}.
In this work, we adapt a flux-conservative formulation of BDNK relativistic viscous hydrodynamics introduced in~\cite{Clarisse:2025lli}, specifically to the case where only diffusion is taken into account.
The flux-conservative equations are valid for a general $U(1)$ conserved current, up to first order in derivatives. This flux-conservative formulation for relativistic diffusion is the first new result in this paper; the second novelty is of numerical origin, as here we show how to solve BDNK diffusion using neural networks.

Many problems in theoretical and computational physics are governed by systems of partial differential equations (PDEs) that encode conservation laws, transport phenomena, or field dynamics. Traditional numerical methods, such as finite difference, finite volume, and spectral methods, are highly effective, but their implementation can be extremely challenging when handling contrived geometries and boundary conditions. Moreover, their accuracy and stability depend strongly on the discretization of the domain, whose refinement in space or time causes a rapid increase in computational cost, especially in multiphysics setups or in problems of high dimensionality.
Despite these challenges, finite volume methods remain the standard tool for simulating systems governed by conservation laws, mainly because of their robustness and well-understood numerical properties. An example of such a method is the Kurganov-Tadmor (KT) central scheme~\cite{KURGANOV2000241, Bazow:2016yra}, which achieves second-order accuracy and remains numerically stable and robust under shock formation. 

In recent years, physics‐informed neural networks (PINNs)~\cite{RAISSI2019686} have emerged as a powerful alternative paradigm for tackling both forward and inverse problems governed by partial differential equations. The technique, which consists of embedding physical laws directly into the learning process of a neural network, is particularly useful because it constitutes a universal framework capable of learning solutions to a wide range of PDEs, such as nonlinear, stochastic, backward-in-time, or even fractional ones~\cite{RAISSI2019686,https://doi.org/10.1029/2020WR029479,doi:10.1137/19M1260141,doi:10.1137/18M1229845}.
Further, experimental or incomplete data can be incorporated into the training process.

The way physical laws are embedded into the training process of a PINN is by constructing loss functions that penalize violations of the governing equations (such as the Navier--Stokes equations, Einstein’s field equations, or, in this work, BDNK diffusion equation), as well as of any given initial and boundary conditions. Namely, one must define a residual---that is, a measure of how far the PINN's predicted solution is from satisfying some constraints---and train the network to minimize this residual. In other words, one trains a neural network to generate data that approximately respects auxiliary conditions and the fundamental laws that govern some physical system.
Unlike traditional numerical solvers, PINNs work with a mesh-free formulation, allowing one to easily handle complex geometries without requiring structured grids. Moreover, the continuous functional representation makes the approach well suited for problems involving many spatial or temporal variables~\cite{Karniadakis:2021nrp}.

A central motivation for using PINNs is that they incorporate the physical structure of the problem directly into the training process, unlike applications of neural networks in physics that are trained purely on historical data (as in weather-prediction models~\cite{gmd-12-2797-2019, 10.1098/rsta.2020.0097, pathak2022fourcastnetglobaldatadrivenhighresolution, bi2023accurate}), which behave as a ``black-box’’~\cite{wetzel2025interpretablemachinelearningphysics, roxlo2018opening}. Such purely data-driven neural-network models have been widely applied both in relativistic hydrodynamics~\cite{Huang:2018fzn, Hirvonen:2023lqy, Pang:2016vdc} and in heavy-ion collision studies~\cite{Mallick:2023vgi, Sun:2024lgo}.
In black-box settings, the network may learn correlations that reproduce the correct output but offer no insight into the underlying physical processes, and it is often unclear what features the model has actually captured. For physics applications where one is also interested in dynamical evolution rather than end-point predictions, such as in heavy-ion collisions~\cite{Shen:2020mgh, Busza:2018rrf}, this opacity may hide the underlying physics contained in the solution of the hydrodynamic PDEs. PINNs avoid this problem because embedding the governing equations in the loss function forces the network to generate solutions consistent with the physical model. In this sense, PINNs do not operate as black boxes that capture features of empirical data, but as models that output solutions that approximately satisfy the governing laws of some physical system.

Despite these advantages, PINNs come with significant drawbacks. Their training can be computationally expensive, especially as the governing PDEs become more complex or involve multiple coupled fields, and their precision is generally lower than that of well-resolved finite volume simulations. Moreover, PINNs struggle to accurately approximate solutions containing sharp gradients or discontinuities~\cite{liu2024discontinuity,ABBASI2025131440,doi:10.1137/22M1522504}. In this sense, PINNs offer universality and flexibility, while finite volume methods such as KT provide speed and accuracy for equations that can be cast in conservative form.

In the context of physics, PINNs have been successfully applied to a wide range of systems including non-relativistic viscous fluid dynamics~\cite{RAISSI2019686, 10.1063/5.0244094}, plasma physics~\cite{Lu:2021deep}, and relativistic ideal hydrodynamics~\cite{FERRERSANCHEZ2024116906, Urb_n_2025}. These methods not only often provide numerical accuracy comparable to classical solvers but also allow differentiability and smoothness properties advantageous for optimization and uncertainty quantification. Overall, PINNs bridge the gap between traditional numerical analysis and modern machine learning, enabling a new class of highly flexible solvers that respect fundamental physical principles.

In this paper we employ PINNs for the first time to solve relativistic viscous hydrodynamic equations. We compare finite volume methods to PINNs by considering the case of BDNK diffusion~\cite{Bemfica:2017wps, Kovtun:2019hdm, Bemfica:2019knx, Hoult:2020eho, Bemfica:2020zjp}. In particular, we focus on relativistic BDNK diffusion in ($1+1$) dimensions. Several examples are considered: a simple Gaussian initial condition, a shock profile, and a Gaussian with non-trivial BDNK background constructed using the code presented\footnote{We do not solve the combined BDNK equations for energy-momentum and baryon number conservation. In this first work, we use the BDNK background of~\cite{Clarisse:2025lli} in some of our simulations of diffusion, \emph{neglecting} the backreaction to the energy-momentum sector.} in~\cite{Clarisse:2025lli}. Different scales of chemical potential (relative to temperature) are considered, which allows us to assess the hydrodynamic frame robustness of the simulations. The finite volume benchmark simulations are performed using standard techniques with the KT algorithm, while the neural network approach applies a new self-adaptive PINN introduced in this work, called SA-PINN-ACTO, which has exact enforcement of initial data and periodic boundary conditions. This handling of initial and boundary conditions via post-processing allows the neural network to focus solely on minimizing the residual of the partial differential equation, increasing the efficiency and accuracy of our simulations.

This paper is organized as follows. In Sec.~\ref{sec:BDNK-theory}, we discuss the theory of BDNK diffusion. In particular, in Sec.~\ref{ssec:diffusion-of-conserved-charge} we present the equations that describe the diffusion of a conserved charge in BDNK hydrodynamics, in Sec.~\ref{ssec:flux_conservative_formulation} we present the flux-conservative formulation of the equations, and in Sec.~\ref{ssec:choice-of-eos} we state our choice of equation of state and transport parameters. Next, in Sec.~\ref{sec:numerical-methods}, we describe the numerical methods we employ: in Sec.~\ref{ssec:KT} we discuss our implementation of the Kurganov-Tadmor scheme, and in Sec.~\ref{ssec:PINNs} we present a detailed discussion of PINNs, first explaining what the vanilla PINN technique consists of, then introducing what we call the ACTO modification (implemented through two consecutive output transforms), and finally reviewing the SA-PINN technique~\cite{MCCLENNY2023111722}. Together, ACTO and SA-PINN define our new SA-PINN-ACTO framework. Then, in Sec.~\ref{sec:test_cases_and_results} we present our numerical results for both the KT and SA-PINN-ACTO simulations. Finally, in Sec.~\ref{sec:conclusions} we state our conclusions and outlook. Further details about our simulations are presented in the appendices.

\subsection*{Notation}
We use a mostly plus metric signature, $\eta_{\mu\nu} = \mathrm{diag} (-1,1,1,1)$, in $(3+1)$D Minkowski spacetime with Cartesian coordinates, and natural units $\hbar=k_B=c=1$. 

\section{BDNK theory for diffusion}\label{sec:BDNK-theory}

Diffusion describes the motion of conserved quantities within a fluid from regions of high concentration to low concentration. Such a system is typically described by a single charge density $n$, which is governed by a dynamical equation defined by the conservation of charge. For a relativistic system, we consider a single conserved current $J^{\mu}$ which obeys 
\begin{equation}
    \partial_{\mu} J^{\mu} = 0. 
\end{equation}
In equilibrium, the only quantities are the density $n$, as well as the chemical potential $\mu$, temperature $T$, and the background fluid velocity $u^{\mu}$. The density is related to the chemical potential and temperature by an equation of state $n=n(\mu,T)$. The conserved current in equilibrium takes the form~\cite{landau1987fluid}
\begin{equation}
    J_{\mathrm{eq}}^{\mu} = n u^{\mu}. 
\end{equation}
In this work, we consider a ``probe approximation'' where the temperature and fluid velocity are taken to be background fields, and backreactions between the diffusion and energy-momentum sector are neglected for simplicity.

Dissipative terms can be included through a gradient expansion in terms of derivatives of the chemical potential divided by temperature. In BDNK theory, all possible terms are included up to first order in derivatives~\cite{Bemfica:2017wps, Kovtun:2019hdm, Bemfica:2019knx, Bemfica:2020zjp}. Neglecting the variation of $T$ and $u^\mu$ in the probe limit results in the conserved current~\cite{Mullins:2023ott}
\begin{equation} \label{eq:Jmu_BDNK}
    J^{\mu} = \left( n + \lambda T u^{\nu} \partial_{\nu} \alpha \right) u^{\mu} - \sigma T \Delta^{\mu\nu} \partial_{\nu} \alpha , 
\end{equation}
where $\alpha = \mu / T$.
In this construction, we have introduced two new parameters, $\lambda$ and $\sigma$, which can both be functions of the temperature and density. By comparison to the nonrelativistic diffusion equation~\cite{Kovtun:2012rj}, we can identify $\sigma$ as the charge conductivity. The parameter $\lambda$ is a hydrodynamic frame parameter coefficient that appears in the relativistic case, which in a first-order theory can be removed by making a hydrodynamic frame transformation\footnote{For an out-of-equilibrium system, the definition of the thermodynamic variables such as the temperature and density are not unique~\cite{Kovtun:2022vas}. This means that such quantities can be freely redefined as long as such a redefinition recovers the same value in equilibrium. A definition of the relevant thermodynamic quantities is known as a hydrodynamic frame. More on hydrodynamic frames can be found in~\cite{Kovtun:2019hdm}.} of the form 
\begin{equation}
    n \rightarrow n - \frac{\lambda T}{\chi} u^{\mu} \partial_{\mu} n.
\end{equation}
where $\chi = \partial n / \partial \mu$ is the charge susceptibility.
The fact that $\lambda$ can be removed by such a transformation, at the level of precision of the first-order theory, does not mean that it is immaterial.
In fact, it can be shown that the corresponding equation of motion is causal, hyperbolic, and stable if and only if~\cite{Hoult:2020eho,Mullins:2023ott}
\begin{subequations}
\begin{align}
    \lambda, \sigma & > 0, \\
    1 > \frac{\sigma}{\lambda} & > 0.
\end{align}
\end{subequations}
When these inequalities are satisfied, the first-order theory defined by the constitutive relation \eqref{eq:Jmu_BDNK}  provides a theory for diffusion that preserves causality and is stable around equilibrium for all Lorentz observers~\cite{Gavassino:2021owo}. The charge conductivity $\sigma$ is a physical parameter in this first-order formulation that can be extracted from microscopic theories~\cite{Rocha:2023ilf}, while $\lambda$ plays the role of a UV regulator that encodes our ignorance about high-order derivative effects. The regime of validity of the theory lies in the so-called frame robust regime, where solutions of the nonlinear equations of motion do not significantly depend on the choice of $\lambda$~\cite{Bea:2023rru,Clarisse:2025lli}.

\subsection{Relativistic BDNK diffusion equation}\label{ssec:diffusion-of-conserved-charge}

We now consider the equation of motion for BDNK diffusion. Computing $\partial_{\mu} J^{\mu}=0$ in a Minkowski background, we find that
\begin{equation}\label{eq:2nd-order-PDE}
    \partial_{\mu} (n u^{\mu}) + \partial_{\mu} \left( \lambda T u^{\mu} u^{\nu} \partial_{\nu} \alpha \right) - \partial_{\mu} \left( \sigma T \Delta^{\mu\nu} \partial_{\nu} \alpha \right)  = 0.
\end{equation}
For a general observer in Minkowski spacetime, the fluid velocity can be decomposed in the form $u^{\mu} = \gamma (1,\vec{v})$, where $\gamma = (1 - v^2)^{-1/2}$ is the Lorentz factor. 
In a general Lorentz frame with arbitrary background, one finds
\begin{multline}
    \partial_t (\gamma n) + \partial_i (\gamma n v^i)  + \partial_t [\gamma^2 \lambda T (\partial_t \alpha + v^i \partial_i \alpha) ] + \partial_j [\gamma^2 v^j \lambda T (\partial_t \alpha + v^i \partial_i \alpha) ] \\- \partial_t \{ \sigma T [-\partial_t \alpha + \gamma^2 (\partial_t \alpha + v^i \partial_i \alpha)] \} - \partial_j \{ \sigma T [\partial^j \alpha + \gamma^2 v^j (\partial_t \alpha + v^i \partial_i \alpha)] \} = 0.
\end{multline}
Note that $T, \vec{v}$ are not determined by the equations of motion, rather they are background fields that must be specified in our probe approximation.

\subsection{Flux-conservative formulation}\label{ssec:flux_conservative_formulation}

For numerical applications, it is useful to write the equations of motion in flux-conservative form, meaning that they take the form $\partial_t \mathcal{N} + \partial_i \mathcal{J}^i = \Phi$, for some set of densities $\mathcal{N}$, fluxes $\mathcal{J}^i$, and sources $\Phi$. The fluxes and sources should be uniquely determined by the densities to truly have a flux-conservative structure. It might seem that the equation of motion as written already takes this form since it comes from a conservation law
\begin{equation}
    \partial_t J^0 + \partial_i J^i = 0.
\end{equation}
However, both the source $J^i$ and the density $J^0$ in \eqref{eq:Jmu_BDNK} contain spatial and time derivatives. It is therefore convenient to perform an order-reduction by introducing the new field
\begin{equation} \label{eq:def_N}
    N_{\mu} = - \partial_{\mu} \alpha,
\end{equation}
so that
\begin{equation}
    \partial_t \alpha = - N_0,
\end{equation}
is in flux-conservative form. This equation along with the conservation law are still not closed because the flux $J^i$ depends on $N_i$, so we also need an equation for $N_i$. This can be obtained by taking a time derivative of $N_i$ and using Eq.~\eqref{eq:def_N} to write 
\begin{equation}
    \partial_t N_i + \partial_i \partial_t \alpha = 0 \quad \rightarrow \quad \partial_t N_i - \partial_i N_0 = 0,  
\end{equation}
which is again in conservative form. 
We have thus converted the single BDNK diffusion equation to a first-order system in conservative form with densities $J^0, \alpha, N_i$. 

To proceed, we would like to write both the fluxes and sources as functions of these densities.
This can be done by writing the BDNK conserved current of Eq.~\eqref{eq:Jmu_BDNK} in terms of $N_{\mu}$ as 
\begin{equation}\label{eq:Jmu}
    J^{\mu} = n u^{\mu} + \left( -\lambda T u^{\mu} u^{\nu} + \sigma T \Delta^{\mu\nu} \right) N_{\nu}. 
\end{equation}
Inverting this expression to find $N_{\mu}$ as a function of $J^{\mu}$ yields
\begin{equation}
    N_{\nu} = \left( -\frac{1}{\lambda T} u_{\mu} u_{\nu} + \frac{1}{\sigma T} \Delta_{\mu\nu} \right) \left( J^{\mu} - n u^{\mu} \right). 
\end{equation}
Using these relations and the equations derived above, we can write the equations of motion in flux-conservative form 
\begin{equation}
    \partial_t \begin{pmatrix}
        J^0 \\
        \alpha \\
        N_i
    \end{pmatrix} + \partial_j \begin{pmatrix}
        J^j \\
        0 \\
        -\delta^j_i N_0
    \end{pmatrix} = \begin{pmatrix}
        0 \\
        -N_0 \\
        0
    \end{pmatrix},
\end{equation}
where the fluxes are determined by\footnote{Note that, in our metric signature, the spatial components satisfy $N^i=N_i$.}
\begin{equation}
    N_0 = \frac{-J^0 + n \gamma +(\sigma - \lambda) T \gamma^2 v^i N_i}{\sigma T + (\lambda - \sigma)T \gamma^2},
\end{equation}
\begin{equation}
    J^i = \gamma n v^i + \sigma T N^i + \gamma^2 T (-\lambda + \sigma) v^i v^j N_j + \gamma^2 T (-\lambda + \sigma) v^i N_0.
\end{equation}
Substituting the form of $N_0$, one may also write
\begin{equation}
    J^i  = \gamma n v^i + \sigma T N^i + \gamma^2 T (-\lambda + \sigma) v^i v^j N_j + \gamma^2 T (-\lambda + \sigma) v^i \left(\frac{-J^0 + n \gamma +(\sigma - \lambda) T \gamma^2 v^j N_j}{\sigma T + (\lambda - \sigma)T \gamma^2}\right) 
\end{equation}
so that it is clear that $J^i = J^i(J^0,\alpha,N^i)$ and $N_0 = N_0(J^0,\alpha,N^i)$ in the general case when $v^i \neq 0$. 
This fully specifies the equations of motion of BDNK diffusion in a form that can be simulated using standard approaches for flux-conservative equations. These equations are valid for arbitrary transport coefficients and background fields $T, \vec{v}$, even if these background fields are functions of spacetime.  

\subsubsection{$(1+1)$D dynamics}

In the simple case in $(1+1)$D, where the background velocity vanishes, i.e., the local rest frame (LRF), and the temperature $T=T_0$ is constant, one finds the simple equations
\begin{equation}\label{eq:flux_conservative_formulation}
    \partial_t \begin{pmatrix}
        J^0 \\
        \alpha \\
        N^x
    \end{pmatrix} + \partial_x \begin{pmatrix}
        J^x \\
        0 \\
        - N_0
    \end{pmatrix} = \begin{pmatrix}
        0 \\
        -N_0 \\
        0
    \end{pmatrix},
\end{equation}
where the fluxes are determined by 
\begin{subequations}
\begin{align}
    N_0 & = \frac{-J^0 + n}{\lambda T_0}, \\
    J^x & = \sigma T_0 N^x.
\end{align}
\end{subequations}
We note that when introducing extra dynamical fields defined in terms of spatial derivatives, such as $\vec{N} = -\nabla \alpha$, constraints such as $\nabla \times \vec{N}=0$ do not appear in the flux-conservative dynamical equations. This demands greater care to ensure these constraints are propagated in the numerical evolution, see~\cite{Clarisse:2025lli}. In this work, since we limit ourselves to (1+1)D simulations, this is not an issue. 
 
\subsection{Choice of equation of state and transport parameters}\label{ssec:choice-of-eos}
For simplicity, in this work we use as the equation of state (EOS) the case of a massless gas of $N_f$ quarks and gluons with $N_c$ colors where the pressure $P$ is given by~\cite{ratti2021deconfinement}
\begin{equation}
    P = \left[2(N_c^2 - 1)+\frac{7}{2}N_c N_f\right]\frac{\pi^2 T^4}{90} + N_c N_f \frac{\mu^2 T^2}{54} + N_c N_f\frac{\mu^4}{972 \pi^2}.
\end{equation}
The energy density $\varepsilon = 3P$ and the baryon density is given by 
\begin{equation}
    n = \left(\frac{\partial P}{\partial \mu}\right)_T = N_c N_f T^3 \left(\frac{\alpha}{27} + \frac{\alpha^3}{243 \pi^2}\right)
\end{equation}
where we used that $\alpha = \mu/T$. Note that we can write $P$ and $\varepsilon$ as functions of $T$ and $\alpha$. In this work, we use $N_c=3$ and $N_f=3$. 

For the conductivity, we use the expression employed in numerical studies of baryon diffusion in the quark-gluon plasma, see~\cite{PhysRevC.98.034916}, where 
\begin{equation}
    \sigma(T, \alpha) = \frac{C_B n}{T^2} \left[ \frac{1}{3} \coth \left( \alpha \right) - \frac{n T}{\varepsilon + P} \right].
\end{equation}
This expression is derived from the relativistic Boltzmann equation in the relaxation-time approximation, and $C_B$ is a parameter that depends on the form of the relaxation time. We note that this definition is valid even for large values of the fugacity $\alpha$. However, this expression can be further simplified and the form of $C_B$ can be determined for small $\alpha$.
In that limit, one finds that 
\begin{equation}
\sigma = \frac{5 C_B}{3 T}\frac{n}{\mu},
\end{equation}
with
\begin{equation}
C_B =\frac{\eta T}{\varepsilon+P},
\end{equation}
for shear viscosity $\eta$. For this small $\alpha$ result, the parameter $C_B$ is approximately equal to the specific shear viscosity $\eta / s$, so we will set $C_B = 1 / 4\pi$ for the simulation in the small fugacity limit (\ref{ssec:third_setup}), in accordance with~\cite{Kovtun:2004de}; for the higher-fugacity simulations (\ref{ssec:first_setup},~\ref{ssec:second_setup}), we use $C_B=0.4$.

In our BDNK simulation, we also need to define $\lambda$, which parameterizes the hydrodynamic frame. For simplicity, we may use here that $c_\mathrm{ch}^2 = \sigma/\lambda$ is a constant (a parameter of our simulation) with $c_\mathrm{ch} \in (0,1)$ for causality and stability. 
Thus, in our setup, our simulation has as free parameters $C_B$ and $c_\mathrm{ch}$, besides the background temperature and flow velocities.

\section{Numerical methods}\label{sec:numerical-methods}
For the sake of completeness, we briefly review the numerical tools used in this work. Neither the Kurganov-Tadmor scheme nor physics-informed neural networks are new methods in themselves; our contributions are limited to problem-specific choices and minor variations.\footnote{All our simulation codes are publicly available at \url{https://github.com/vchomalicastro/1-1D-BDNK-diffusion-simulations/}.}

\subsection{Kurganov-Tadmor scheme}\label{ssec:KT}
\subsubsection{Flux formulation and reconstruction}
To solve our system of conservation equations~\eqref{eq:flux_conservative_formulation} numerically, we use a central KT scheme that is second order in space (piecewise-linear reconstruction with a Total Variation Diminishing, TVD, limiter) and second order in time (SSP-RK2)~\cite{KURGANOV2000241}. The KT algorithm can be thought of as a particular choice of numerical flux and, unlike traditional upwind methods that rely on approximate Riemann solvers and characteristic decompositions, this scheme uses local speed estimates to compute numerical fluxes. This allows for high-resolution shock-capturing while maintaining a simpler, eigenstructure-free formulation.

The semi-discrete form of the $(1+1)$D KT scheme is
\begin{equation}
\frac{d}{dt}\, \mathbf q_{i} = - \frac{ \mathbf H_{i+1/2} - \mathbf H_{i-1/2}}{\Delta x} + \mathbf S[\mathbf q_{i}],
\end{equation}
where $\mathbf q$ is the conserved variable vector and is composed of cell-averaged solutions, $\mathbf S$ the source term, and $\mathbf H$ are the numerical fluxes incorporating local wave speeds. Second-order accuracy is achieved by using linear reconstructions and non-oscillatory limiters (e.g., minmod) to suppress spurious oscillations near discontinuities.

Each flux is computed as
\begin{equation}\label{eq:KTflux}
    \begin{aligned}
        \mathbf H_{i\pm 1/2} &= \frac{ a^+_{i\pm 1/2}\,\mathbf F(\mathbf q^-_{i\pm 1/2}) - a^-_{i\pm 1/2}\,\mathbf F(\mathbf q^+_{i\pm 1/2}) }{ a^+_{i\pm 1/2} - a^-_{i\pm 1/2} }
        + \frac{ a^+_{i\pm 1/2} a^-_{i\pm 1/2} }{ a^+_{i\pm 1/2} - a^-_{i\pm 1/2} } \bigl( \mathbf q^+_{i\pm 1/2} - \mathbf q^-_{i\pm 1/2} \bigr),
    \end{aligned}
\end{equation}
where $\mathbf q^\pm_{i\pm1/2}$ are the left/right reconstructed states at the interface.
In our implementation, we use symmetric one-sided speeds
\begin{gather}
    a^+_{i\pm1/2} = c_{i\pm1/2},\quad a^-_{i\pm1/2} = -\,c_{i\pm1/2},
    \\ \text{with}\quad c_{i\pm1/2} = \sqrt{\frac{\sigma_{i\pm1/2}}{\lambda_{i\pm1/2}}},
\end{gather}
which reduces the flux to the standard ``single-$a$'' form commonly encountered in the literature~\cite{Bazow:2016yra}:
\begin{equation}
    \label{eq:KT_single_a_flux}
    \begin{aligned}
        \mathbf H_{i\pm 1/2}
        &=
        \tfrac{1}{2}
        \bigl[\mathbf F(\mathbf q^{-}_{i\pm 1/2})
              + \mathbf F(\mathbf q^{+}_{i\pm 1/2})\bigr]
        - \tfrac{1}{2}\,a_{i\pm1/2}
          \bigl(\mathbf q^{+}_{i\pm 1/2}
                - \mathbf q^{-}_{i\pm 1/2}\bigr).
    \end{aligned}
\end{equation}
In $(1+1)$D, the left and right states at cell interfaces are reconstructed using a slope-limited linear interpolation:
\begin{align*}
    \mathbf q^{+}_{i+1/2} &= \mathbf q_{i+1} - \frac{\Delta x}{2} (\mathbf q_x)_{i+1}, \\
    \mathbf q^{-}_{i+1/2} & = \mathbf q_i + \frac{\Delta x}{2} ( \mathbf q_x)_i, \\
    \mathbf q^{+}_{i-1/2} & = \mathbf q_i - \frac{\Delta x}{2} (\mathbf q_x)_i,\\ 
    \mathbf q^{-}_{i-1/2} & = \mathbf q_{i-1} + \frac{\Delta x}{2} (\mathbf q_x)_{i-1}.
\end{align*}
At each interface we evaluate the quantities $T$, $v$, $\sigma$, and $\lambda$ by averaging their corresponding neighboring cell-centered values, and use these to compute $\mathbf F$, $N_0$, and $c=\sqrt{\sigma/\lambda}$ at interfaces.
Periodic boundary conditions are imposed by a simple index wrap-around ($ \mathbf q_{-1}\equiv \mathbf q_{N_x-1}$ and $\mathbf q_{N_x}\equiv \mathbf q_{0}$).

The slope $(\mathbf q_x)_i$ is computed with the TVD minmod limiter applied to forward/backward differences:
\begin{gather}
    (\mathbf q_x)_i = \operatorname{minmod}\!\left(\frac{\mathbf q_i - \mathbf q_{i-1}}{\Delta x},\, \frac{\mathbf q_{i+1} - \mathbf q_i}{\Delta x}\right),
    \\ \operatorname{minmod}(a,b) = \tfrac12\bigl(\operatorname{sign}(a)+\operatorname{sign}(b)\bigr)\min(|a|,|b|).
\end{gather}

All our KT simulations were executed on an Apple M1 Pro (ARM64, 8-core CPU, 16 GB memory) using a NumPy/Numba CPU implementation in double precision, with no GPU acceleration.

\subsubsection{Grid and periodic boundary conditions}
We discretize $x\in[-L,L]$ into $N$ finite volume cells of width $\Delta x = 2L/N$, so we store $N+1$ edge points, with $x_0=-L$ and $x_N=L$. However, because we use periodic boundary conditions, $x_0$ and $x_N$ represent the same physical location, so there are only $N$, and not $N+1$, distinct interfaces, even though computationally we must create arrays that store $N+1$ of them. Throughout the paper, all simulations are performed in grids with $N$, $2N$ and $4N$ spatial cells, with $N=1000$ for simulations with continuous initial data and $N=2000$ in the case of discontinuous initial data.

\subsubsection{Adaptive time step}
We use adaptive time step sizes in our KT implementation. At each time step, we compute the maximum local characteristic speed
\begin{equation}
    c_{\rm max}(t) = \max_{x}\sqrt{\frac{\sigma(t)}{\lambda(t)}}
\end{equation}
and set
\begin{equation}
    \Delta t
    =
    \frac{\Delta x}{8 c_{\rm max}(t)}.
\end{equation}
This time step size is held fixed throughout the two SSP-RK2 stages and recomputed for the next time step after updating $T$, $v$, $\alpha$, and $\sigma$. In practice, however, since BDNK theory fixes some $c_{\rm ch}^2=\sigma/\lambda={\rm const}\in(0,1)$ parameter, we end up with a constant time step:
\begin{equation}
    \Delta t = \frac{\Delta x}{8c_{\rm ch}}.
\end{equation}

\subsubsection{Convergence tests}\label{sssec:convergence_tests}
We verify the convergence of our KT simulations using the Richardson extrapolation method~\cite{1911RSPTA.210..307R,1927RSPTA.226..299R} applied to grid refinements. Convergence testing provides a systematic procedure for verifying that numerical solutions approach the continuum limit as the grid is refined. The central idea is that a consistent and stable discretization of order $p$ should yield an error that decreases as a power of the grid spacing, $|u_N - u| \sim (\Delta x)^p$, where $u_N$ denotes the numerical solution obtained on a mesh with $N$ spatial cells (and $\Delta x = 2L/N$) and $u$ represents the continuum solution. In practice, since exact solutions are rarely available for nonlinear systems such as relativistic hydrodynamics, convergence is verified by comparing simulations performed at different resolutions and by measuring how the differences between them scale with grid refinement. A numerical method that exhibits the expected scaling is regarded as convergent within the tested regime. So, we compute
\begin{equation}
Q_n(t)
=
\log_2{
\frac{\|n_{N}(t)-n_{2N}(t)\|_1}{\|n_{2N}(t)-n_{4N}(t)\|_1}
},
\end{equation}
where $\|\cdot\|_1$ is an $L_1$ norm, and the sub-indices $N$, $2N$ and $4N$ denote the runs with those numbers of spatial cells. For smooth solutions we expect $Q_n(t) \to 2$, whereas for solutions with discontinuities we expect $Q_n(t) \to 1$.

The choice of the $L_1$ norm is motivated by its robustness: it provides a global measure of the integrated error across the computational domain, it is less sensitive than the $L_\infty$ norm to isolated pointwise deviations, and it offers a clearer interpretation in the presence of shocks or sharp gradients than the $L_2$ norm. The subtractions are performed on the base (shared, coarse) resolution corresponding to the run with $N$ spatial cells: the higher-resolution solutions ($2N$, $4N$) are first restricted to the $N$-cell grid by cell-averaging before forming the differences. Snapshot times are the same across all runs, so no temporal interpolation is required. The $L_1$ norm is computed as $|f|_1=\sum_j |f_j|\,\Delta x$ with $\Delta x=2L/N$ on the $N$-cell grid. All the results for convergence tests can be found in Appendix~\ref{app:convergence_tests_results}.

\subsection{Self-adaptive physics-informed neural network with algebraic enforcement of auxiliary conditions through transforms to the output (SA-PINN-ACTO)}\label{ssec:PINNs}
\subsubsection{Vanilla PINN}\label{sssec:vanilla_PINN}
In this work, we implement a physics-informed neural network to tackle the forward $(1+1)$D BDNK diffusion problem. For clarity, we first discuss the standard (``vanilla'') formulation of PINNs.

As discussed in Sec.~\ref{sec:introduction}, PINNs can learn an approximate, mesh-free (i.e. continuous) solution to a given partial differential equation by attempting to enforce in its predictions the differential equations that govern the problem, as well as the chosen initial and boundary conditions. Let us consider a PDE in its most general form,
\begin{equation}
  \mathcal{F}\!\left(t,\mathbf{x},\mathbf{u},
  \partial_t\mathbf{u},\ldots,\partial_t^{k}\mathbf{u},
  \nabla_{\mathbf{x}}\mathbf{u},\ldots\right)=0,
  \; \mathbf{x}\in\Omega\subset\mathbb{R}^d,\;t\in[0,T],
\end{equation}
with initial and boundary conditions
\begin{align}
    \partial_t^{\,j}\mathbf{u}(t=0,\mathbf{x})
    &=\mathbf{u}^{(j)}_{\rm IC}(\mathbf{x}),\;\mathbf{x}\in\Omega,\; j=0,\dots,k-1, \\
    {\mathcal{B}}[\mathbf{u}]&=0,\;\mathbf{x}\in\partial\Omega, \;t\in[0,T].
\end{align}
Here, $\mathbf{u}(t,\mathbf{x})$ is the solution to the differential equation, $\mathcal{F}[\cdot]$ is a differential operator, and ${\mathcal{B}}[\cdot]$ is a boundary operator.

The PINN solution, called the prediction, here symbolized by the sub-index $\theta$, is generated by training the neural network on minimizing a loss or residual $\mathcal{L}$ given by
\begin{equation}\label{eq:loss}
    \mathcal{L}
    \equiv
    \mathcal{L}_{\rm PDE}
    + \lambda_{\rm IC} \mathcal{L}_{\rm IC}
    + \lambda_{\rm BC} \mathcal{L}_{\rm BC},
\end{equation}
where
\begin{gather}
    \mathcal{L}_{\rm PDE}
    \equiv \frac{1}{|N_{\rm PDE}|}\sum_{i\in N_{\rm PDE}}
       \left|\mathcal{F}\!\left(t_i,x_i,\mathbf{u}_\theta,
       \partial_t\mathbf{u}_\theta,\dots,\partial_t^{k}\mathbf{u}_\theta,
       \nabla_{\!x}\mathbf{u}_\theta,\dots\right)\right|^2,
    \\
    \mathcal{L}_{\rm IC}
    \equiv \frac{1}{|N_{\rm IC}|}\sum_{i\in N_{\rm IC}}
       \sum_{j=0}^{k-1}
       \left|\partial_t^{\,j}\mathbf{u}_\theta(0,x_i)
       -\mathbf{u}^{(j)}_{\rm IC}(x_i)\right|^2,
    \label{eq:loss_ic}
\end{gather}
and, for example, for periodic boundary conditions,
\begin{equation}
    \mathcal{L}_{\rm BC}
    \equiv \frac{1}{|N_{\rm BC}|}\sum_{i\in N_{\rm BC}}
      \left|\mathbf{u}_\theta(t_i,L) - \mathbf{u}_\theta(t_i,-L)\right|^2.
    \label{eq:loss_bc}
\end{equation}
Here, $\mathbf{u}_\theta(t,x)$ is the network's output (i.e., the prediction), and $(t,x)\in[0,T]\times[-L,L]$ is the network's input. Numerically, we compute the derivatives of $\mathbf{u}_\theta(t,x)$ via automatic differentiation, which, unlike finite difference schemes, delivers exact gradients up to machine precision. $\lambda_{\rm IC}$ and $\lambda_{\rm BC}$ are simply weighting coefficients to adjust the magnitude of the different loss terms according to their importance, numerically speaking. $N_{\rm PDE}$, $N_{\rm IC}$, and $N_{\rm BC}$ are the sets of collocation points for each term, and $|\cdot|$ denotes their cardinalities. The collocation points are the points in spacetime on which the PINN is trained. The IC collocation points are placed at $t=0$ at random $x_i$, the BC collocation points are placed at $x=\pm L$ at random $t_i$, and the PDE collocation points are placed at random $x_i$ and $t_i$, all throughout the domain. To ensure that these collocation points are a representative random sample of their respective domains, we use Latin Hypercube Sampling (LHS)~\cite{fbe9475b-1452-3229-a386-04540442a9c3}, a random sampling method that distributes samples evenly over the sampling space. We choose different $|N_{\rm PDE}|$ for the different simulations.

From Eq.~\eqref{eq:flux_conservative_formulation}, and from our choice of periodic boundary conditions and arbitrary initial conditions for $\alpha$ and $J^0$, it follows that, for the problem we are solving, we can define
\begin{gather}
    R_{{\rm PDE},{i}}^2 \equiv
    \left|
    \partial_t J^0_{\theta}+\partial_x J^x_{\theta}
    \right|_i^2
    +
    \left|
    \partial_t \alpha_{\theta}+N_{0,\theta}
    \right|_i^2
    ,
    \\
    R_{{\rm BC},{i}}^2
    \equiv
    \left|
        \mathbf{u}_\theta(t_i,L)-
        \mathbf{u}_\theta(t_i,-L)
    \right|^2,
    \\
    R_{{\rm IC},{i}}^2
    \equiv
    \left|
        \mathbf{u}_\theta(0,x_i)-
        \mathbf{u}_{\rm IC}(x_i)
    \right|^2,
\end{gather}
where
\begin{equation}
    \mathbf{u}=\begin{pmatrix}
        J^0 \\ \alpha
    \end{pmatrix},
\end{equation}
to then construct our PDE, BC, and IC losses, respectively, as
\begin{gather}\label{eq:loss_pde}
    \mathcal{L}_{\rm PDE}
    \equiv \frac{1}{|N_{\rm PDE}|}
    \sum_{i\in N_{\rm PDE}}
    {\!\!\!R_{{\rm PDE},i}^2},
    \\
    \mathcal{L}_{\rm BC}
    \equiv \frac{1}{|N_{\rm BC}|}
    \sum_{i\in N_{\rm BC}}
    {\!R_{{\rm BC},i}^2},
    \\
    \mathcal{L}_{\rm IC}
    \equiv \frac{1}{|N_{\rm IC}|}
    \sum_{i\in N_{\rm IC}}
    {\!R_{{\rm IC},i}^2}.
\end{gather}
Note that for our $(1+1)$D BDNK diffusion problem, $d=1$, so $\mathbf{x}\equiv x\in[-L,L]$. 

\begin{figure*}
    \centering
    \includegraphics[width=1.0\linewidth]{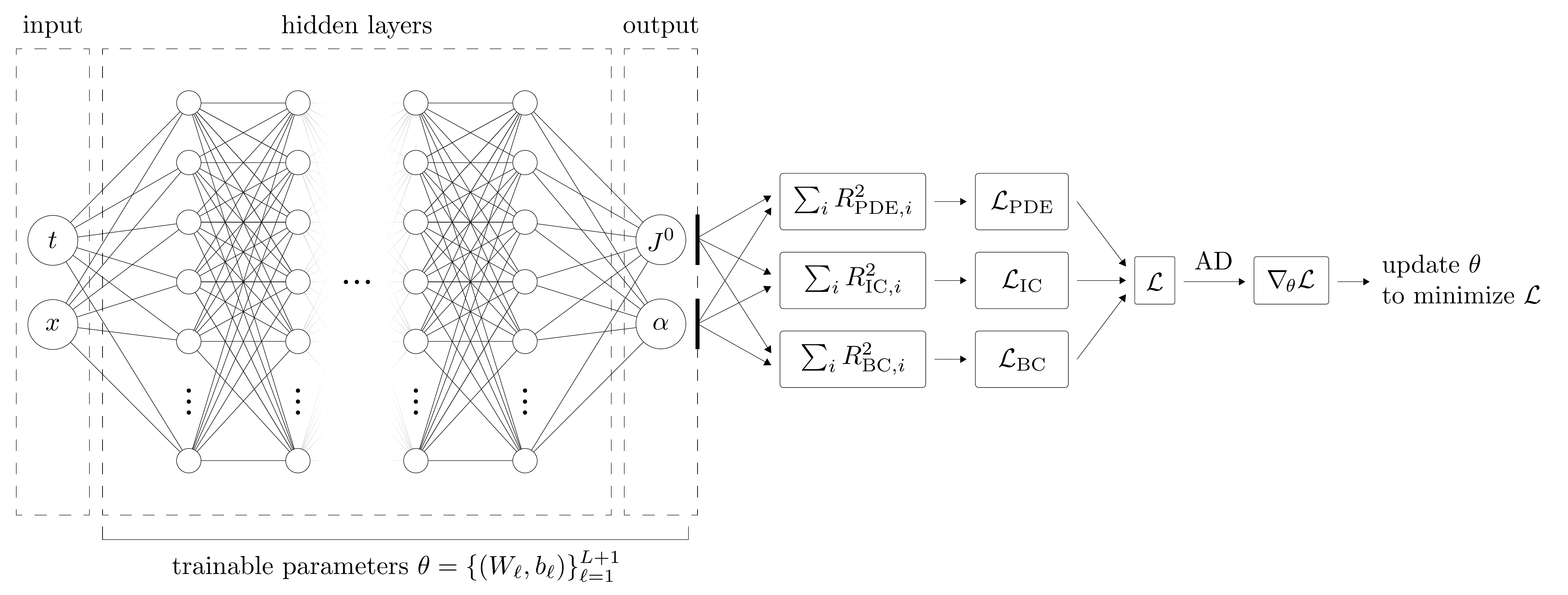}
    \caption{Vanilla PINN architecture and loss construction for the $(1+1)$D BDNK diffusion problem. 
    The network takes spacetime inputs $(t,x)$ and predicts $(J^0,\alpha)$. It is fundamental that the quantities $J^x$ and $N_0$ are obtained from the constitutive relations from Sec.~\ref{ssec:flux_conservative_formulation} and not from the identities $N_0=-\partial_t\alpha$ and $\partial_x J^x = - \partial_t J^0$, since that would decouple the output variables $\alpha$ and $J^0$ and incorrectly put them on independent grounds. Afterwards, automatic differentiation (AD) yields $\partial_t J^0$, $\partial_t\alpha$, and $\partial_x J^x$. The residuals $\mathcal{L}_{\rm IC}$, $\mathcal{L}_{\rm BC}$ and $\mathcal{L}_{\rm PDE}$ are calculated as shown in Eqs.~\eqref{eq:loss_ic}-\eqref{eq:loss_pde}, respectively, and the total loss $\mathcal{L}\equiv\mathcal{L}_{\rm PDE}+\lambda_{\rm IC}\mathcal{L}_{\rm IC}+\lambda_{\rm BC}\mathcal{L}_{\rm BC}$ is minimized by computing $\nabla_\theta \mathcal{L}$ and updating $\theta$ using gradient descent (Adam), followed by quasi-Newton refinement (L-BFGS).}
    \label{fig:bdnk-vanilla-pinn}
\end{figure*}

In practice, to implement our PINN, we utilize the open-source machine learning framework PyTorch~\cite{NEURIPS2019_bdbca288}. For the architecture of our PINN, we use a feed-forward multilayer perceptron (see Appendix~\ref{app:mlp}). The input layer of our MLP has $2$ neurons: one neuron outputs $x$ and the other outputs $t$. Then, we use $L=10$ hidden layers with $N=70$ neurons each, with a $\sigma(\cdot)=\tanh(\cdot)$ activation. At the end, we have an output layer with $2$ neurons with a $\sigma(\cdot)={\rm id}(\cdot)$ (i.e., linear) activation; one neuron outputs $J^0$ and the other outputs $\alpha$. This is illustrated in Fig.~\ref{fig:bdnk-vanilla-pinn}.

To train the PINN, on the other hand, we adopt a common two-stage optimization strategy: first, we perform Adam pre-training~\cite{kingma2017adammethodstochasticoptimization}, which rapidly drives down the loss and places the network parameters in a good region for fine-tuning. For this stage of training, we initialize the Adam learning rate at $3\times 10^{-3}$ and schedule it to reduce to $40\%$ of its previous value whenever the loss plateaus for $700$ Adam epochs (though we limit this decrease to $1/100$ of the initial learning rate). We use a different number of Adam pre-training epochs for each of the three simulations. Afterwards, once the loss has plateaued during Adam, we switch to the quasi-Newton limited-memory Broyden-Fletcher-Goldfarb-Shanno (L-BFGS) algorithm~\cite{Liu:1989esw} which, using full-batch second-order information, fine-tunes all parameters to ensure a higher accuracy satisfaction of all three PDE, initial, and boundary condition residuals. For all our numerical simulations, we use up to $1,000$ L-BFGS iterations, exiting either when the optimizer plateaus or when non-finite losses appear, which halts the L-BFGS procedure. Once training is finalized, we restore the best model. All our PINN simulations are run on an NVIDIA H200-SXM5-141GB GPU.

\subsubsection{PINN with algebraic enforcement of initial conditions}\label{sssec:heic}
A problem one might encounter when training a PINN, however, is that the initial condition might not be learned to the desired level of precision. This has been commonly reported in the literature~\cite{hao2025stabilitytrainingpinnsstiff,CAO2024117222}, and it raises a concern: even if the PDE and BC residuals are low, the solution might not be of interest if it is associated with predicted initial conditions that do not exactly match the ones the network was trained to replicate. This problem can be solved, nevertheless, by doing a hard or algebraic enforcement of the initial conditions~\cite{hao2025stabilitytrainingpinnsstiff, ren2024improvingpinnsalgebraicinclusion}.

This is accomplished by applying a transformation to the raw output of the PINN (which so far we had treated as the final answer): letting $\hat{\mathbf{u}}_\theta$ be the raw PINN output (which, in Sec.~\ref{sssec:vanilla_PINN}, we had called simply $\mathbf{u}_\theta$) and $\mathbf{u}_{\rm IC}$ the desired initial condition, we perform the transformation
\begin{equation}\label{eq:algebraic_enforcement_of_ICs}
    \tilde{\mathbf{u}}_\theta(t,x) 
    =
    \mathbf{u}_{\rm IC}\cdot
    e^{-\beta t}
    +
    \hat{\mathbf{u}}_{\theta}(t,x)\cdot
    (1-e^{-\beta t}).
\end{equation}
where, together, $e^{-\beta t}$ and $1-e^{-\beta t}$ represent an arbitrary partition of unity, meaning that $\tilde{\mathbf{u}}_\theta$ is a nice blend of the IC and the network raw output. 
This aligns closely with the technique introduced by Ren \textit{et al.}~\cite{ren2024improvingpinnsalgebraicinclusion}, as they also use two asymptotic functions of time that partition unity, though different ones. For our choice of functions, $\beta>0$ is simply an adjustable parameter to control the partition dynamics. We use $\beta=1$.

Obviously, by construction, this transformation makes the transformed output of the network, $\tilde{\mathbf{u}}_\theta(t,x)$, exactly equal to $\mathbf{u}_{\rm IC}$ at $t=0$ and, therefore,
\begin{equation}
    \mathcal{L}_{\rm IC}=0,
\end{equation}
from which it follows that the loss function is now simply
\begin{equation}
    \mathcal{L}
    =
    \mathcal{L}_{\rm PDE}
    + \lambda_{\rm BC} \mathcal{L}_{\rm BC}.
\end{equation}

\subsubsection{PINN with algebraic enforcement of boundary conditions}\label{sssec:hebc}
We employ one further technique to also algebraically enforce periodic boundary conditions, instead of imposing them as a soft, learned constraint. Here, we follow the idea of Hao \textit{et al.}~\cite{hao2025stabilitytrainingpinnsstiff}, though with substantially different methodology; instead of augmenting the input features to the network with a Fourier-like embedding, we simply apply yet another transformation to the result from Eq.~\eqref{eq:algebraic_enforcement_of_ICs}:
\begin{equation}\label{eq:algebraic_enforcement_of_BCs}
    \mathbf{u}_\theta(t,x)
    =
    \tilde{\mathbf{u}}_\theta(t,x)
    -
    \frac{x+L}{2L}
    \left[
        \tilde{\mathbf{u}}_\theta(t,L)
        -
        \tilde{\mathbf{u}}_\theta(t,-L)
    \right].
\end{equation}
This transformation guarantees that $\mathbf{u}_\theta(t,-L)=\mathbf{u}_\theta(t,L)$ for all $t$, without having to modify the network inputs or architecture. Therefore,
\begin{equation}
    \mathcal{L}_{\rm BC}=0
\end{equation}
by construction, meaning that now,
\begin{equation}
    \mathcal{L} = \mathcal{L}_{\rm PDE}.
\end{equation}

That is, the PINN now only has to learn the solution to the PDE. Both auxiliary (initial and boundary) conditions are now exactly enforced via a transform to the output: we introduce this combination of techniques as the PINN-ACTO.

Note that algebraic (also known as hard) enforcement of auxiliary conditions is not new: prior work enforces initial conditions using output transforms~\cite{hao2025stabilitytrainingpinnsstiff, CAO2024117222, ren2024improvingpinnsalgebraicinclusion}, and algebraically enforces either periodic boundary conditions by augmenting the inputs with Fourier features~\cite{hao2025stabilitytrainingpinnsstiff}, or Dirichlet boundary conditions through an output transform~\cite{ren2024improvingpinnsalgebraicinclusion}. In contrast, what our ACTO variant constitutes is an explicit output transform that enforces periodic boundary conditions, thus requiring no input embeddings or architectural changes, and, specifically, the combination of this periodic BC-enforcing output transform with the standard IC-enforcing output transform.

\subsubsection{Self-adaptive PINN and network-internal normalization}\label{sssec:SA_PINN}
So far, we have introduced the vanilla PINN in Sec.~\ref{sssec:vanilla_PINN} and two additional techniques in Secs.~\ref{sssec:heic} and~\ref{sssec:hebc} that, together, give us the PINN-ACTO. Here, we take two further steps: we adopt the self-adaptive PINN (SA-PINN) technique proposed by McClenny \textit{et al.}~\cite{MCCLENNY2023111722}, and implement network-internal normalization of the magnitudes of the relevant fields ($\alpha$ and $J^0$), similar to the hierarchically normalized PINN (hnPINN) framework~\cite{LEDUC2024108400}; the former modifies the loss function of the PINN by introducing trainable and pointwise weights that make the PINN focus on points of spacetime where the solution is resulting to be harder to learn, while the latter is a simple though crucial numerical rescaling step that ensures that the network operates on well-conditioned numerical ranges, improving gradient conditioning and training stability.

For the $(1+1)$D BDNK diffusion problem, we do this by redefining the first-order PDE residual at the $i$th collocation point, $(t_i,x_i)$, as
\begin{equation}\label{eq:sa_pinn_residuals}
    R_{{\rm PDE},i}^2
    \equiv 
    \frac{\left|\partial_t J^0_{\theta}+\partial_x J^x_{\theta}\right|_i^2}{s_{J^0}^2}
    +
    \frac{
    \left|\partial_t \alpha_{\theta}+N_{0,\theta}\right|^2_i}{s_{\alpha}^2},
\end{equation}
where $s_{J^0}$ and $s_{\alpha}$ (for network-internal normalization) are the magnitudes of the $J^0$ and $\alpha$ fields, respectively, determined by the maximum magnitude of their initial conditions ($s_\alpha = \max_x |\alpha(t=0,x)|$ and $s_{J^0} = \max_x |J^0(t=0,x)|$), to then redefine the PDE term of the loss function, $\mathcal{L}_{\rm PDE}$, as
\begin{equation}\label{eq:loss_pde_sa}
    \mathcal{L}_{\rm PDE}
    \equiv \frac{1}{|N_{\rm PDE}|}
    \sum_{i\in N_{\rm PDE}}
    {\!\!\!\lambda_{i}^2 R_{{\rm PDE},i}^2},
\end{equation}
where each $\lambda_{i}$ is the weighting factor for the PDE residual at the $i$th collocation point. 

This differs from the original (vanilla) PINN technique in two senses: in the first place, in the vanilla PINN, the $\lambda$ weights are simply scalars that multiply the different loss terms (see Eq.~\eqref{eq:loss}) to assign different numerical importance to each one of them; secondly, and more importantly, in the SA-PINN framework, the pointwise weights $\lambda_{i}$ are not manually assigned; rather, they are learned by the PINN jointly with the set of network parameters $\theta$. To keep the weights $\lambda_i$ non-negative and numerically stable, we represent them by unconstrained parameters (logits) that get passed through a smooth, positive mapping. In this work we choose the $\rm softplus$ function~\cite{NIPS2000_44968aec},
\begin{equation}
  \lambda_{i} = \operatorname{softplus}(z_i)
  = \ln\!\left(1+e^{z_i}\right),
\end{equation}
where $z_i$ is the logit associated with the $i$th collocation point. During training, the logits $\{z_i\}_{i\in N_{\rm PDE}}$ are optimized jointly with the network parameters $\theta$, using the same gradient‐based procedure. 

The effect of this construction is that collocation points with larger residuals naturally receive larger adaptive weights, leading the neural network to give more attention to regions of spacetime where the solution is harder to learn. This self-adaptive mechanism has been shown to accelerate convergence and improve the accuracy of PINNs.

\begin{figure*}
    \centering
    \includegraphics[width=1\linewidth]{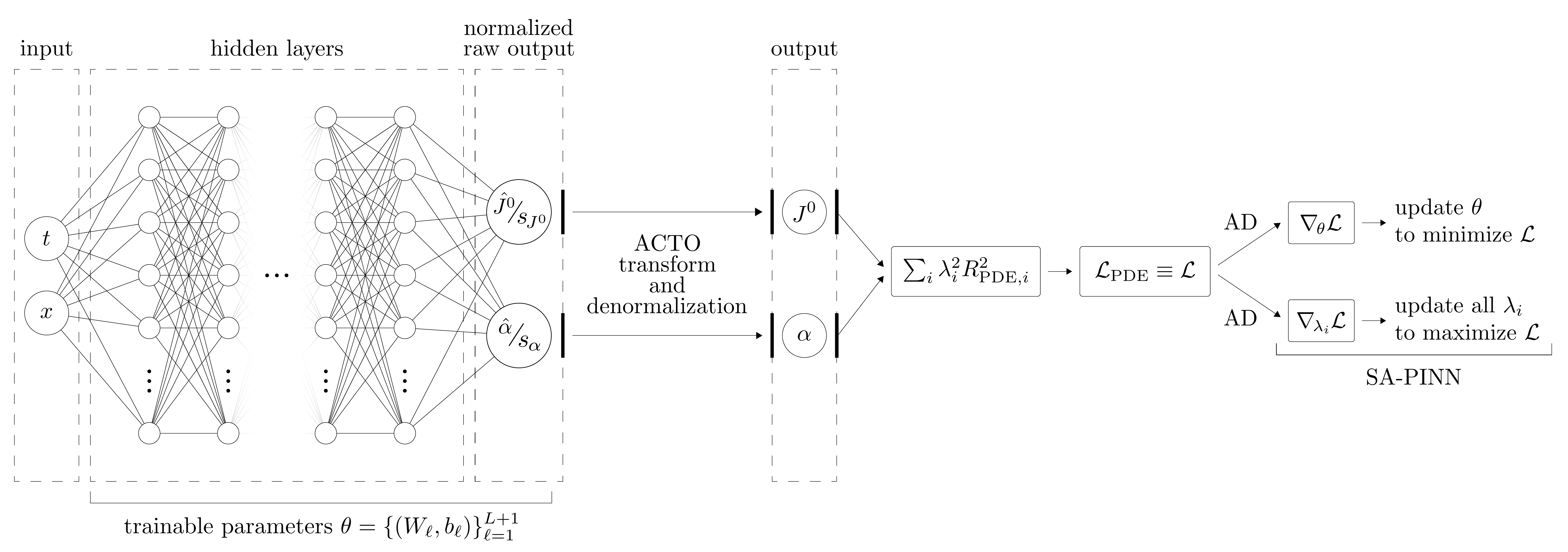}
    \caption{SA-PINN-ACTO architecture and loss construction for the $(1+1)$D BDNK diffusion problem. As in the vanilla PINN (see Fig.~\ref{fig:bdnk-vanilla-pinn}), the network takes spacetime inputs $(t,x)$, but unlike it, the SA-PINN-ACTO predicts normalized raw outputs $\hat{J}^0/s_{J^0}$ and $\hat{\alpha}/s_{\alpha}$. These are first denormalized, then passed through the IC-enforcing transform in Eq.~\eqref{eq:algebraic_enforcement_of_ICs}, and then through the BC-enforcing transform in Eq.~\eqref{eq:algebraic_enforcement_of_BCs} (together, the ACTO transform), yielding physical outputs $(J^0,\alpha)$ that exactly satisfy the prescribed initial data and periodic boundary conditions. Automatic differentiation (AD) is then used to compute the normalized PDE residuals $R_{{\rm PDE},i}$ defined in Eq.~\eqref{eq:sa_pinn_residuals}; these are weighted pointwise by the self-adaptive factors $\lambda_i^2$ to build the PDE loss $\mathcal{L}_{\rm PDE}\equiv\mathcal{L}$ in Eq.~\eqref{eq:loss_pde_sa}. The logits underlying $\lambda_i=\mathrm{softplus}(z_i)$ are optimized jointly with the network parameters $\theta$: AD provides $\nabla_{\theta}\mathcal{L}$ to update $\theta$ so as to minimize $\mathcal{L}$ and $\nabla_{\lambda_i}\mathcal{L}$ to adapt the weights, focusing learning on collocation points with larger residuals.}
    \label{fig:bdnk-sa-pinn-acto}
\end{figure*}

As for network-internal normalization, we also multiply the network raw outputs for $J^0$ and $\alpha$ by $s_{J^0}$ and $s_\alpha$, respectively, so that, in some sense, we are asking the network to predict $J^0$ and $\alpha$ not in their physical units, but in a dimensionless, normalized form, of order $\sim 1$. These normalized predictions are then rescaled back to their physical units before the computation of spatial and temporal derivatives, and thus before evaluating the PDE residuals. Therefore, we say that the network raw outputs are $\hat{J}^0/s_{J^0}$ and $\hat{\alpha}/s_\alpha$, and the processed outputs (denormalized and ACTO-transformed) are $J^0$ and $\alpha$. This ensures that the network operates on well-conditioned numerical ranges while all physical laws remain expressed in ${\rm GeV}$ units. Although this normalization is conceptually independent of our ACTO technique, we find it practically indispensable for achieving robust and stable training across test cases involving widely separated physical scales.
All our numerical simulations are thus run on our modified SA-PINN, which we call the SA-PINN-ACTO. We illustrate it in Fig.~\ref{fig:bdnk-sa-pinn-acto}.

Finally, we augment the set of pseudo-randomly collocated PDE points, $N_{\rm PDE}$, by appending $500$ evenly spaced points along the slice $t=0$ ($x\in[-L,L]$). We do this because our PINN ultimately solves a second-order PDE for $\alpha$ and, thus, beyond exactly reproducing the initial values of $\alpha$ and $J^0$ through our ACTO technique, we also place special emphasis on correctly capturing the second-order initial conditions of $\alpha$ by accurately learning the solution to the coupled ODE system for $\alpha$ and $J^0$ at $t=0$.

\section{Numerical results for BDNK diffusion}\label{sec:test_cases_and_results}

In this section, we numerically solve three setups (\ref{ssec:first_setup},~\ref{ssec:second_setup},~\ref{ssec:third_setup}) using the SA-PINN-ACTO and benchmark the results against a high-resolution KT solver validated through convergence tests (see Appendix~\ref{app:convergence_tests_results}). All runs use periodic boundaries on $[-L,L]$ with $L=50\,{\rm GeV^{-1}}$ and evolve up to a setup-specific $t_{\rm end}$. For each of the three setups, we vary the maximum characteristic speed, $c_{\rm ch}\in\{0.5,0.9\}$, which sets the propagation rate of structures; there are thus six tests in total. We track total charge conservation $\int_{-L}^{L}{J^0\,dx}$, verify grid-refinement convergence for KT, and, for the SA-PINN-ACTO, monitor PDE residual decay during training.

Moreover, as a validation metric, we report for each one of the six tests the relative $L^2$ error between the KT and SA-PINN-ACTO solutions over the full spacetime domain $[0,t_{\rm end}]\times[-L,L]$,
\begin{equation}\label{eq:l2_validation}
    E_{\rm rel}[\phi]=
    \frac{\left\|\phi_{\rm PINN}-\phi_{\rm KT}\right\|_2}
    {\left\|\phi_{\rm KT}\right\|_2},
\end{equation}
where $\phi\in\{n,J^0\}$ and $\|\cdot\|_2$ denotes the spacetime $L^2$ norm
\begin{equation}
\|\phi\|_2^2 = \int_0^{t_{\rm end}} \!\!\int_{-L}^{L} |\phi(t,x)|^2\, dx\, dt.
\end{equation}
These errors are reported in Table~\ref{tab:relL2}.

\subsection{First setup: Gaussian initial condition}\label{ssec:first_setup}
First, we simulate a Gaussian initial condition for both $n$ and $J^0$, described by
\begin{equation}
    \begin{cases}
        n(t=0,x)
        =
        \left(0.2e^{-\left(\frac{7x}{L}\right)^2}+1\right)\,{\rm GeV}^{3},
        \\
        J^0(t=0,x)
        =\left(
        0.05e^{-\left(\frac{10x}{L}\right)^2}+1.05\right)\,{\rm GeV}^{3}.
    \end{cases}
\end{equation}
That is, $n(t=0,x)$ is a Gaussian of amplitude $0.2\,{\rm GeV}^{3}$ and standard deviation $L/(7\sqrt{2})$ standing on a pedestal of $1\,{\rm GeV}^{3}$, and $J^0(t=0,x)$ is a Gaussian of amplitude $0.05\,{\rm GeV}^{3}$ and standard deviation $L/(10\sqrt{2})$ standing on a pedestal of $1.05\,{\rm GeV}^{3}$.
We perform this simulation first for $c_{\rm ch}=0.5$, and then for $c_{\rm ch}=0.9$. In this simulation, both the field temperature $T$ and field velocity $\vec{v}$ are fixed: $T=0.3\,{\rm GeV}$ and $\vec{v}=0$. The simulations were run up to $t=t_{\rm end}=20\,{\rm GeV}^{-1}$. Here, $C_B=0.4$.

\begin{figure*}[!t]
\centering

\begin{minipage}[t]{0.49\textwidth}
  \centering
  \includegraphics[width=\linewidth]{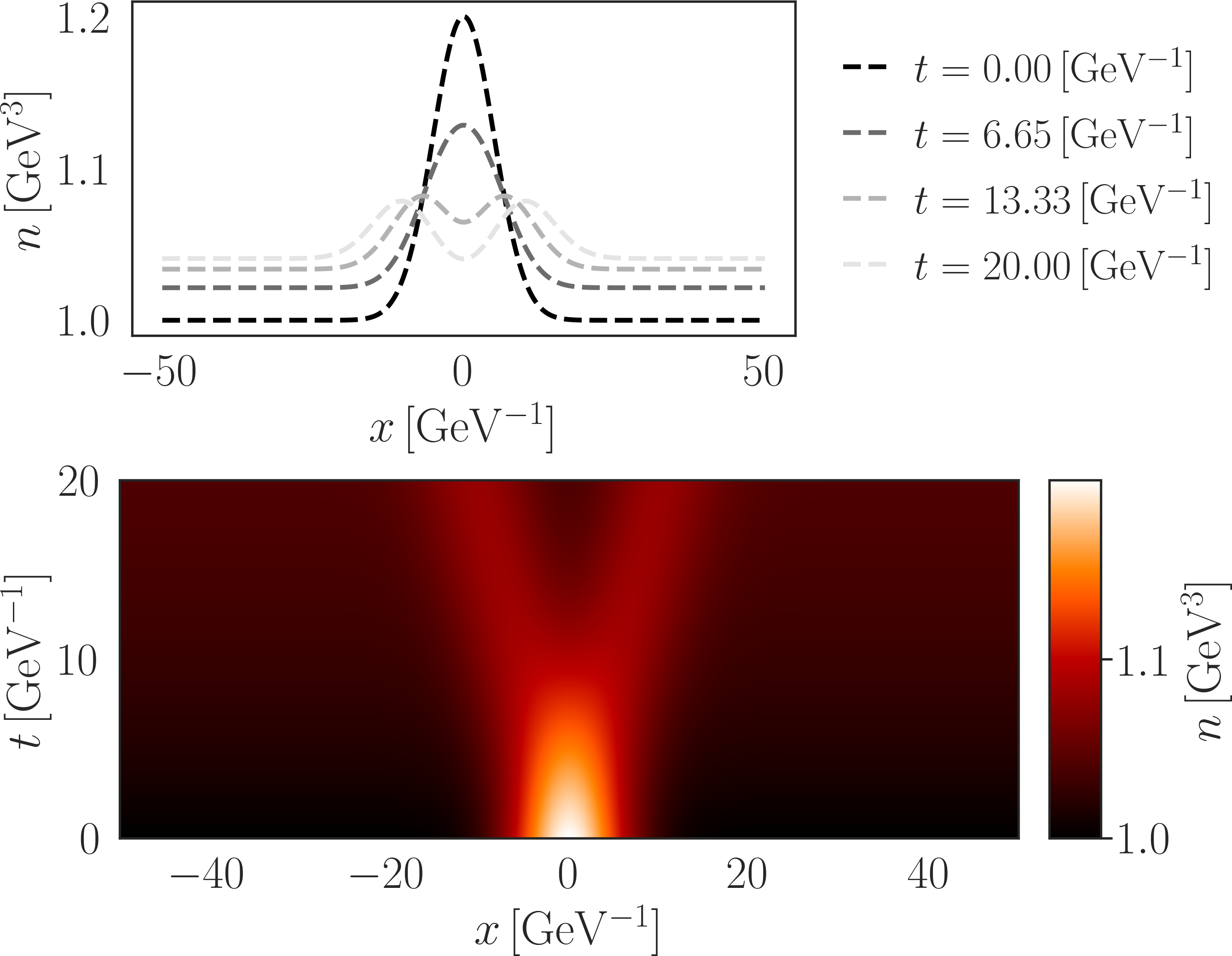}
  \makebox[0pt][l]{\small (a)}
\end{minipage}\hfill
\begin{minipage}[t]{0.49\textwidth}
  \centering
  \includegraphics[width=\linewidth]{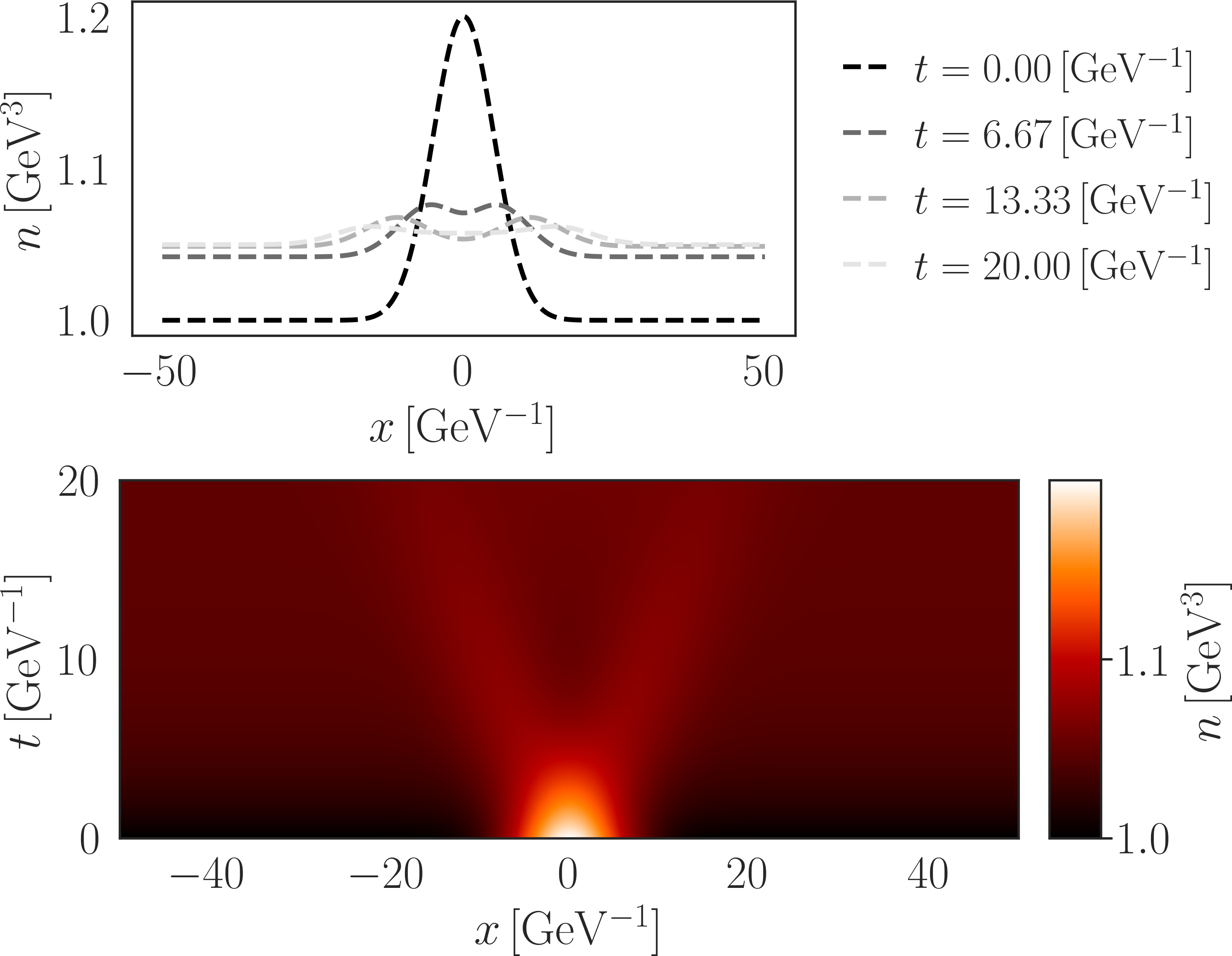}
  \makebox[0pt][l]{\small (b)}
\end{minipage}

\vspace{0.75em}

\begin{minipage}[t]{0.49\textwidth}
  \centering
  \includegraphics[width=\linewidth]{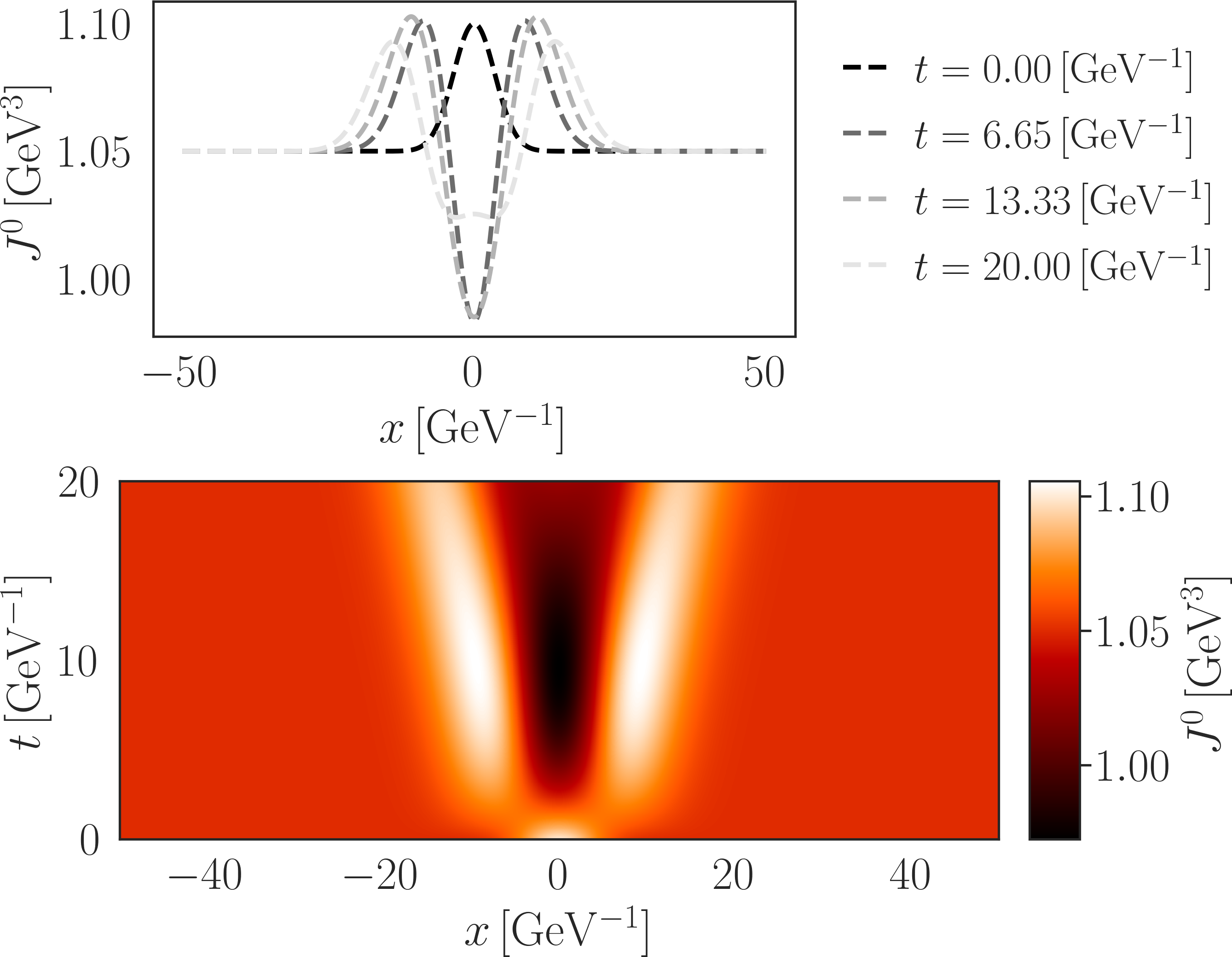}
  \makebox[0pt][l]{\small (c)}
\end{minipage}\hfill
\begin{minipage}[t]{0.49\textwidth}
  \centering
  \includegraphics[width=\linewidth]{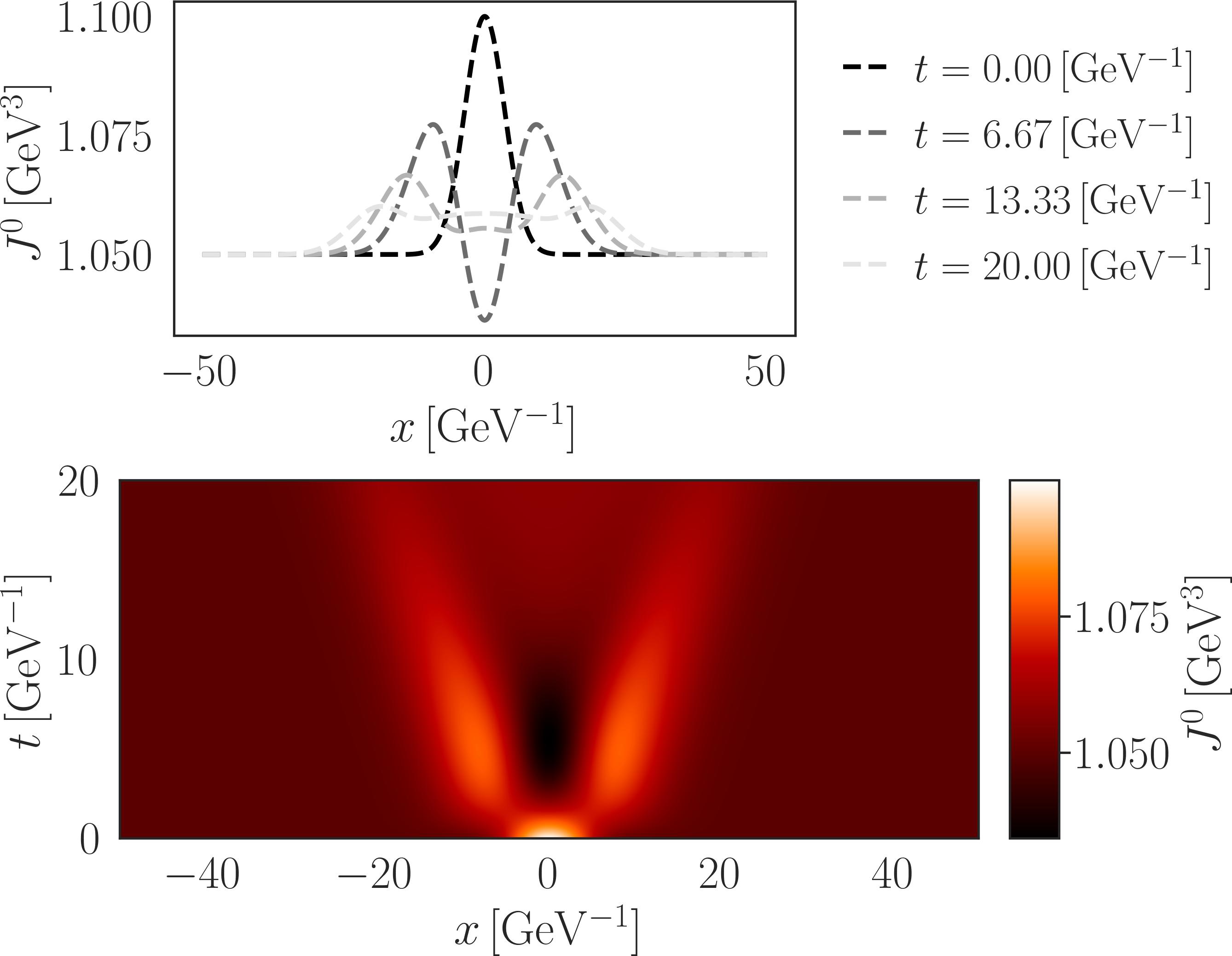}
  \makebox[0pt][l]{\small (d)}
\end{minipage}

\caption{Results from KT for the first setup.
(a) Evolution of $n$ for $c_{\rm ch}=0.5$.
(b) Evolution of $n$ for $c_{\rm ch}=0.9$.
(c) Evolution of $J^0$ for $c_{\rm ch}=0.5$. Total charge $\int_{-L}^{L}{J^0\,dx}$ conserved at all times up to a fraction of $4.4\times 10^{-15}$ of the initial charge.
(d) Evolution of $J^0$ for $c_{\rm ch}=0.9$. Total charge $\int_{-L}^{L}{J^0\,dx}$ conserved at all times up to a fraction of $4.9\times 10^{-15}$ of the initial charge.}
\label{fig:setup1-KT}
\end{figure*}

\begin{figure*}[!t] 
\centering

\begin{minipage}[t]{0.49\textwidth}
  \centering
  \includegraphics[width=\linewidth]{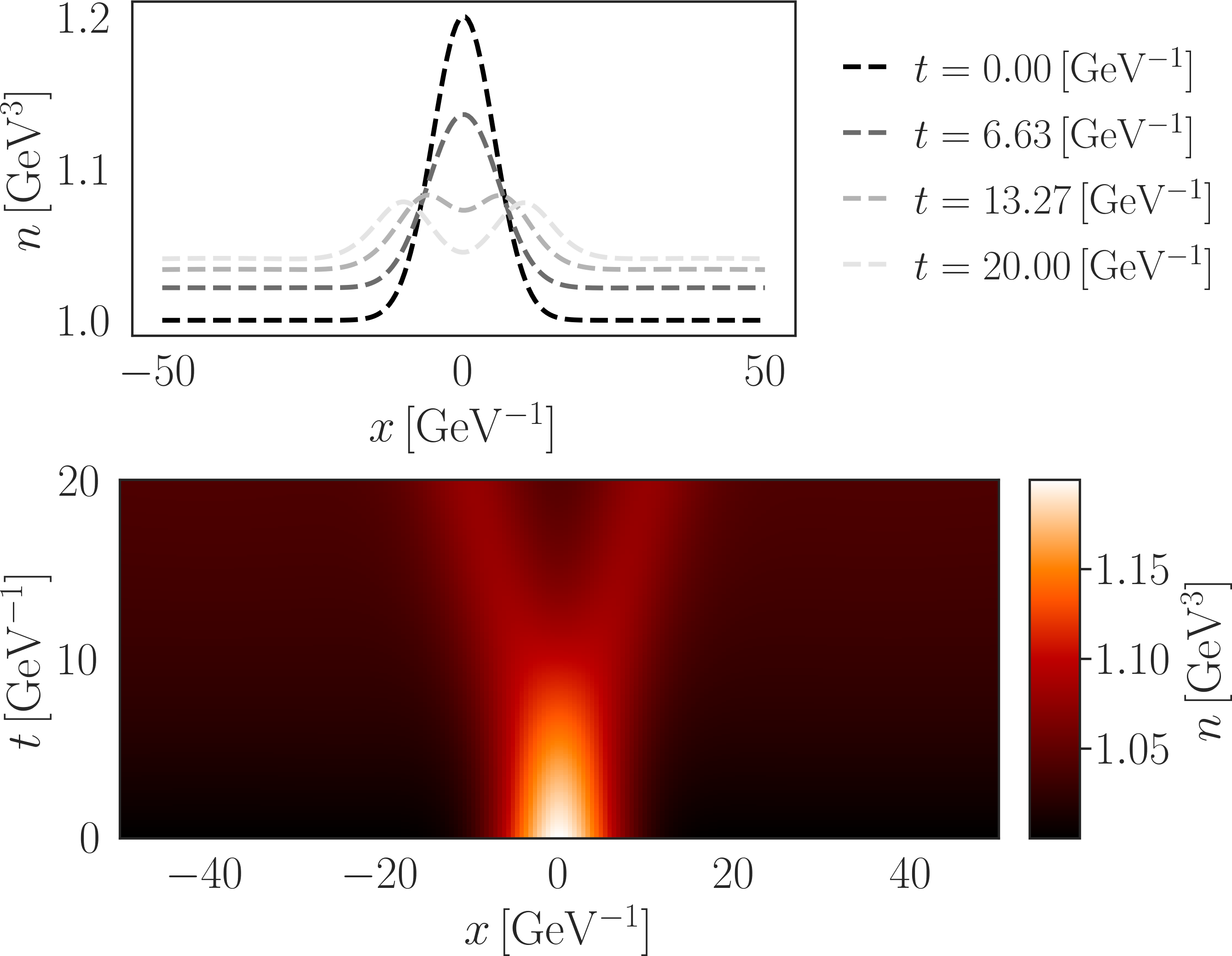}
  \makebox[0pt][l]{\small (a)}
\end{minipage}\hfill
\begin{minipage}[t]{0.49\textwidth}
  \centering
  \includegraphics[width=\linewidth]{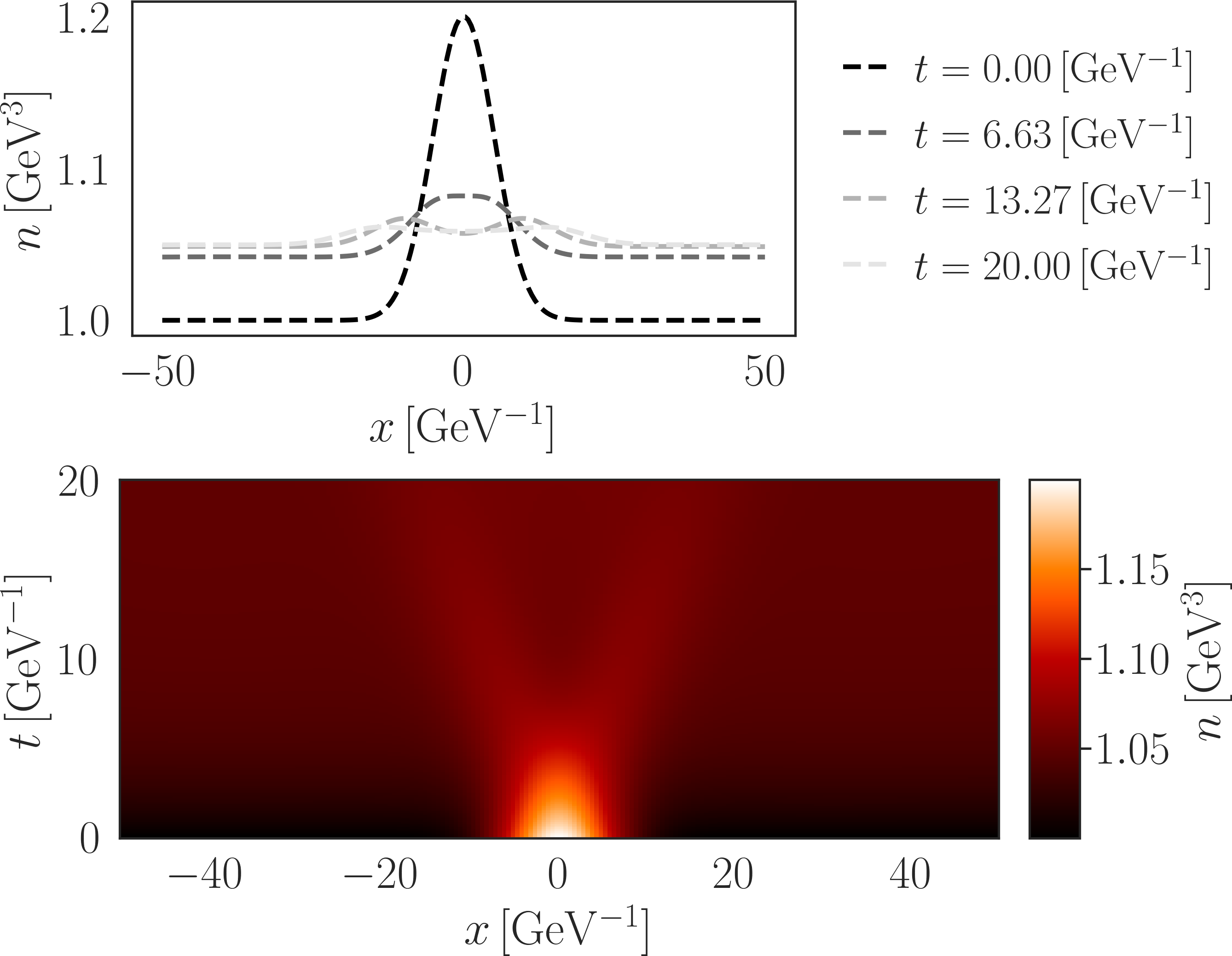}
  \makebox[0pt][l]{\small (b)}
\end{minipage}

\vspace{0.75em}

\begin{minipage}[t]{0.49\textwidth}
  \centering
  \includegraphics[width=\linewidth]{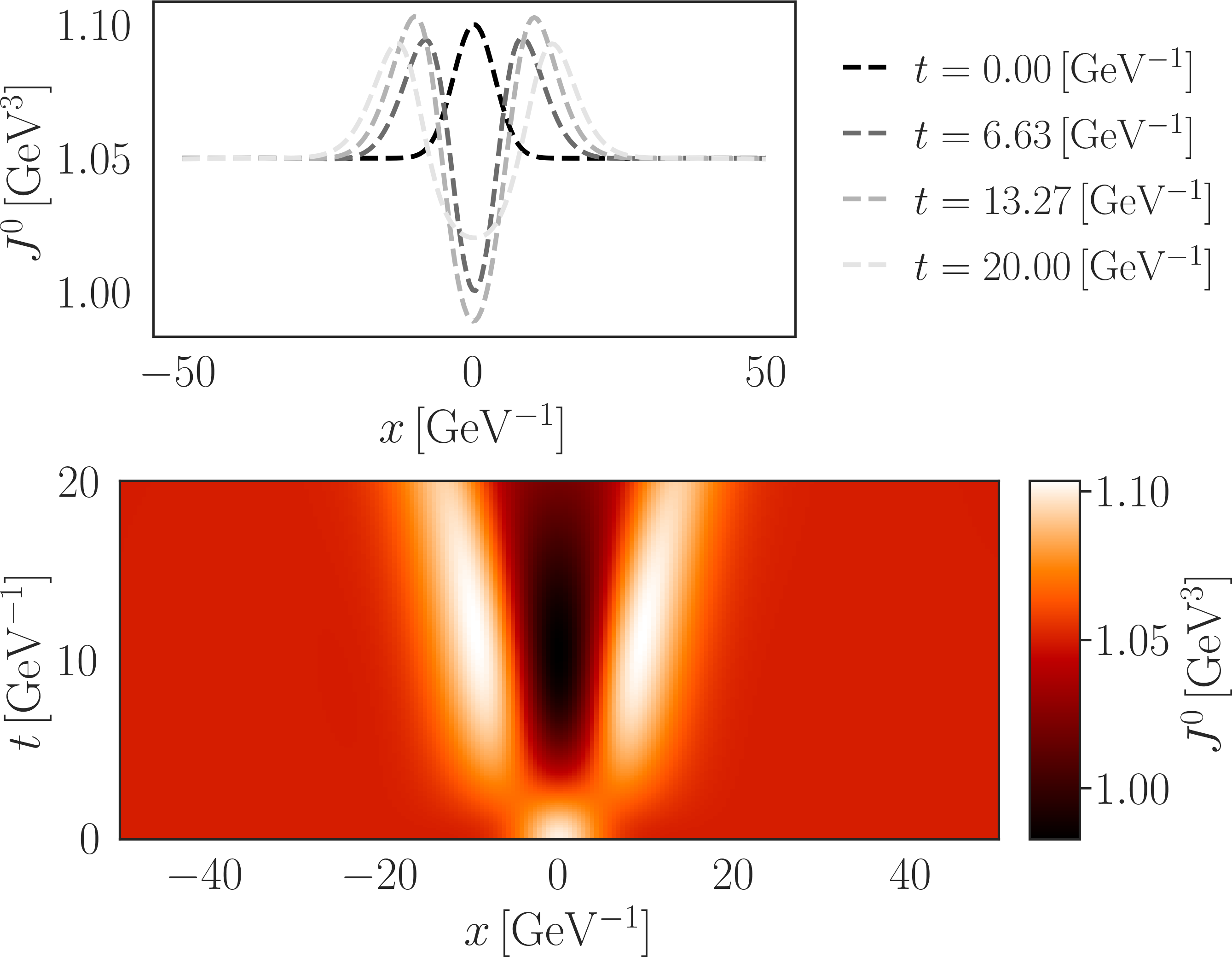}
  \makebox[0pt][l]{\small (c)}
\end{minipage}\hfill
\begin{minipage}[t]{0.49\textwidth}
  \centering
  \includegraphics[width=\linewidth]{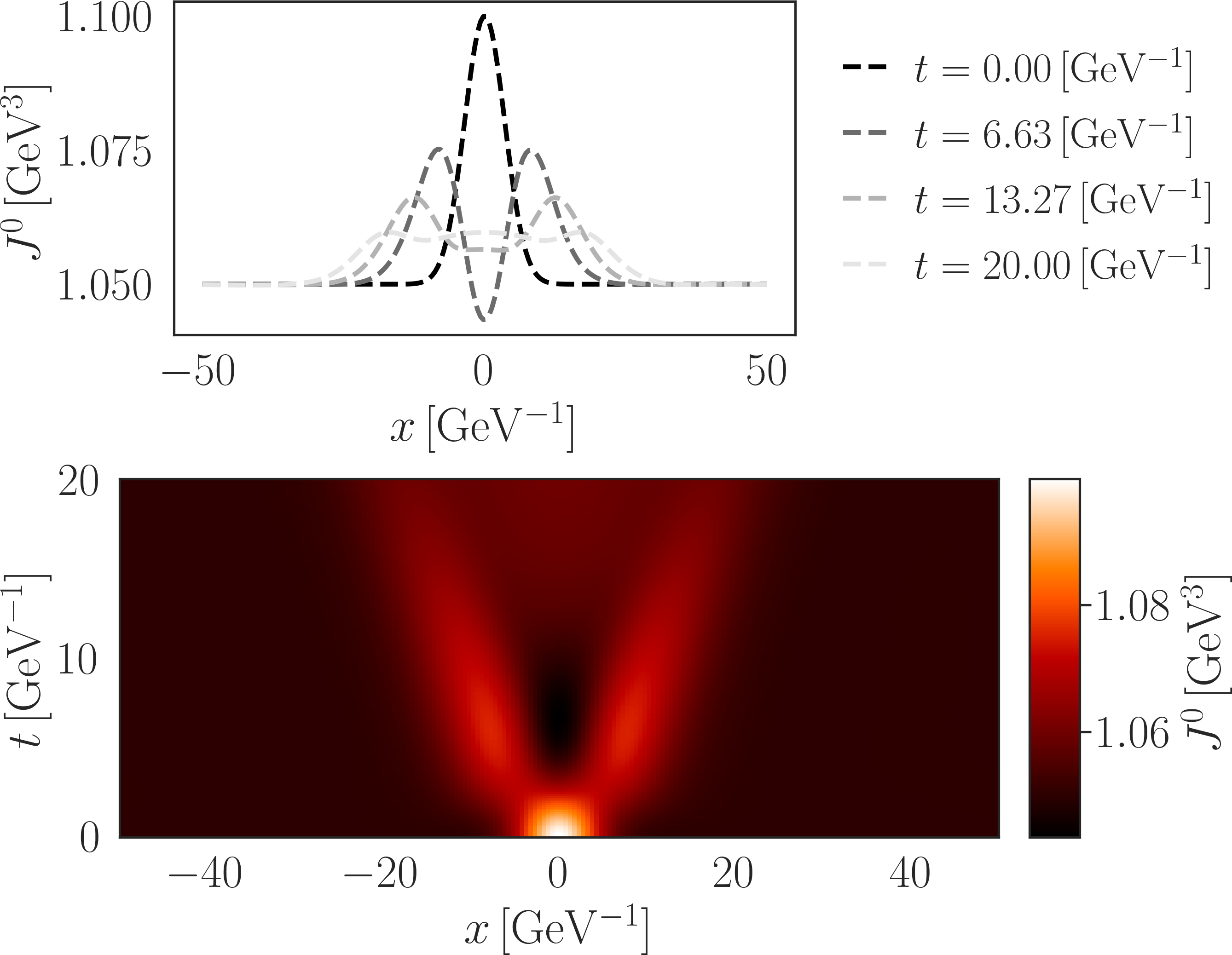}
  \makebox[0pt][l]{\small (d)}
\end{minipage}

\vspace{0.75em}

\begin{minipage}[t]{0.49\textwidth}
  \centering
  \includegraphics[width=\linewidth]{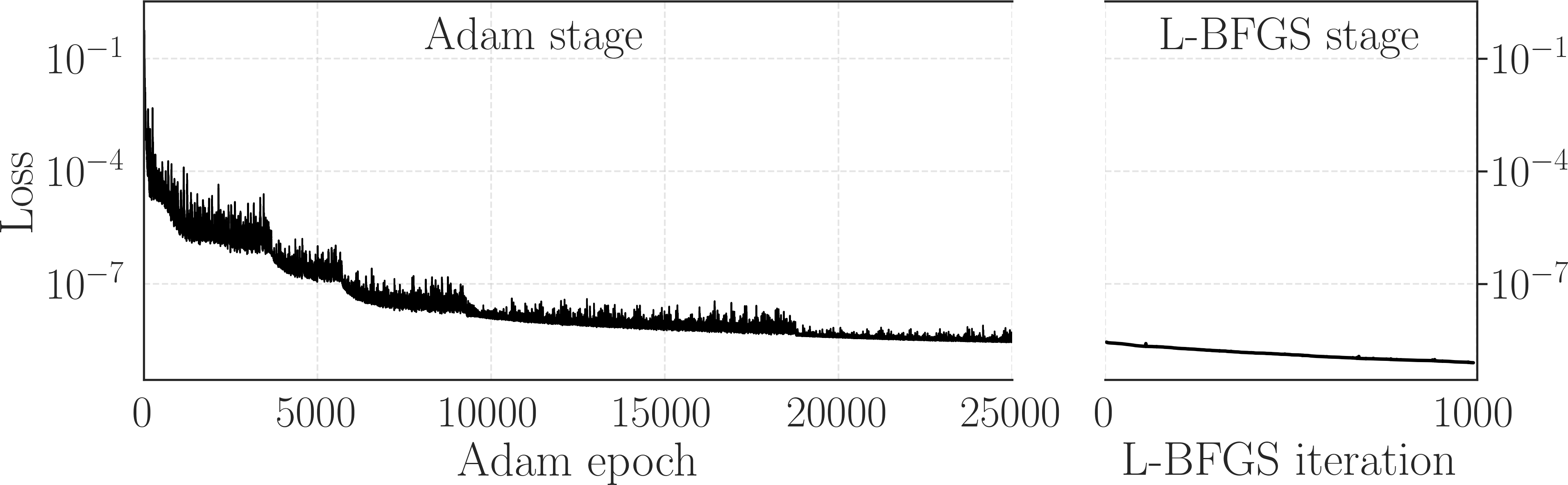}
  \makebox[0pt][l]{\small (e)}
\end{minipage}\hfill
\begin{minipage}[t]{0.49\textwidth}
  \centering
  \includegraphics[width=\linewidth]{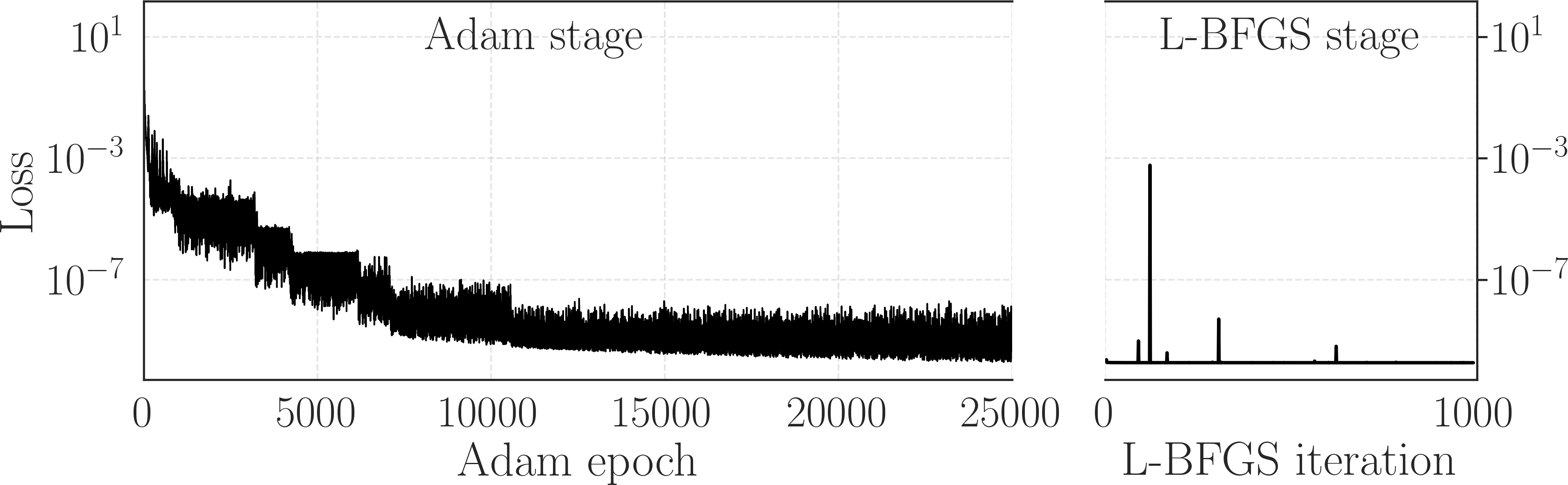}
  \makebox[0pt][l]{\small (f)}
\end{minipage}

\caption{Results from the SA-PINN-ACTO for the first setup. For these simulations, we use $|N_{\rm PDE}|=20,000+500$, and $25,000$ Adam epochs, followed by L-BFGS fine-tuning.
(a) Evolution of $n$ for $c_{\rm ch}=0.5$.
(b) Evolution of $n$ for $c_{\rm ch}=0.9$.
(c) Evolution of $J^0$ for $c_{\rm ch}=0.5$. Total charge $\int_{-L}^{L}{J^0\,dx}$ conserved at all times up to a fraction of $1.1\times 10^{-4}$ of the initial charge.
(d) Evolution of $J^0$ for $c_{\rm ch}=0.9$. Total charge $\int_{-L}^{L}{J^0\,dx}$ conserved at all times up to a fraction of $8.6\times 10^{-5}$ of the initial charge.
(e) Total loss history for $c_{\rm ch}=0.5$. Best loss was $7.752\times 10^{-10}$. Total training time: $434.64$ seconds (Adam: $411.78$ seconds; L-BFGS: $22.86$ seconds).
(f) Total loss history for $c_{\rm ch}=0.9$. Best loss was $1.862\times 10^{-10}$. Total training time: $429.12$ seconds (Adam: $406.95$ seconds; L-BFGS: $22.17$ seconds).}
\label{fig:setup1-PINN}
\end{figure*}

In Fig.~\ref{fig:setup1-KT} we present the evolution of $n$ and $J^0$ for this setup obtained with the KT finite volume method for two characteristic speeds, $c_{\text{ch}} = 0.5$ and $0.9$. The evolution of $n$, shown in subplots (a) and (b), is symmetric and retains its Gaussian character, with the minimum density rising in time and the extrema shrinking as the wave structure dissipates. This behavior is evident from both the time snapshots and the heatmap, where the color contrast decreases with time. In contrast, the evolution of $J^0$, shown in subplots (c) and (d), exhibits a rapid restructuring of the initial Gaussian profile into a configuration with a central minimum near $x=0$ and two symmetric local maxima. This new structure subsequently dissipates in time. Simulations with the larger characteristic speed yield smaller gradients in the evolution of both $n$ and $J^0$, as well as faster dissipation of the wave structure, as evidenced by the broader cone in the heatmaps for subplots (b) and (d). 

This difference is expected for large values of $\alpha$ ($\alpha \gtrsim 1$) because at those scales, gradients are large and we are likely outside the regime of applicability of the hydrodynamic theory. Looking forward to Fig.~\ref{fig:setup3-KT}, we can see that in the small $\alpha$ regime the value of $c_{\text{ch}}$ does not affect the numerical results, which indicates that in that case we are in the frame robust regime where changes between causal hydrodynamic frames do not significantly change the numerical simulations \cite{Bea:2023rru,Bea:2025eov,Clarisse:2025lli}.

In Fig.~\ref{fig:setup1-PINN} we investigate the same setup, but instead of using the KT finite volume algorithm, we employed the SA-PINN-ACTO proposed in Sec.~\ref{sec:numerical-methods}. Regarding the evolution described previously, the PINN-based approach is very similar and shares many of the qualitative conclusions we have drawn before. Both $n$ and $J^0$ dissipate with time and the evolution for $c_{\text{ch}} = 0.9$ has smaller gradients throughout. The evolution of both $n$ and $J^0$ is symmetric as before, where we have the left- and right-moving waves damp and spread. The $c_{\text{ch}} = 0.9$ case shows smaller extrema in the evolution of $n$ and $J^0$, consistent with the behavior observed in the KT simulations. This difference is harder to see in the heatmap, but it is much clearer in the snapshot evolution. The dynamics with the larger characteristic speed are more damped and we posit that this difference arises from hydrodynamic frame dependence dominating at larger $\alpha$. The field gradients propagate outward faster and dampen more rapidly in the $c_{\text{ch}} = 0.9$ subplots, see Fig.~\ref{fig:setup1-PINN} (b) and (d). It is therefore reasonable to conclude that the presence of larger characteristics can more rapidly damp the evolution, possibly even preventing discontinuous shock formation in a relativistic viscous theory. See Sec.~\ref{ssec:second_setup} for our investigation into near-discontinuous initial data. 

Overall, the PINN demonstrates strong numerical performance, with residuals of order $10^{-10}$. The loss functions reach their minimum in the L-BFGS stage, with the $c_{\mathrm{ch}}=0.9$ simulation achieving a slightly better (smaller) loss than the $c_{\mathrm{ch}}=0.5$ one. Moreover, as shown in Table~\ref{tab:relL2}, the relative $L^2$ error between the SA-PINN-ACTO and KT solutions is of order $10^{-3}$ for both $n$ and $J^0$.

\subsection{Second setup: Smooth shock profile}\label{ssec:second_setup}
We also simulate a smooth shock-like initial condition described by
\begin{equation}
    \begin{cases}
        n(t=0,x)
        =
        \left[
            1.1-0.1\tanh\!\left(s\left[\left(\frac{4x}{L}\right)^2-1\right]\right)
        \right]
        {\rm GeV}^{3},
        \\
        J^0(t=0,x)
        =
        1.05\;{\rm GeV}^{3}.
    \end{cases}
\end{equation}
This choice yields $n(t{=}0,x)\in C^{\infty}(\mathbb{R})$ and, as $s$ increases, $n$ approaches a piecewise constant profile with $n = 1.2\,{\rm GeV^{3}}$ for $|x|<L/4$ and $n = 1.0\,{\rm GeV^{3}}$ elsewhere. Above, the initial data for $J^0(t=0,x)$ is a pedestal of $1.05\,{\rm GeV}^{3}$. Here, we use $s=60$, since with this value we simultaneously achieve both numerical convergence and sufficient detail in the simulations.
We perform our simulations first for $c_{\rm ch}=0.5$, and then for $c_{\rm ch}=0.9$. For both cases, the field temperature $T$ and field velocity $\vec{v}$ are fixed: $T=0.3\,{\rm GeV}$ and $\vec{v}=0$. The simulations run up to $t=t_{\rm end}=15\,{\rm GeV}^{-1}$. Here, $C_B=0.4$.

\begin{figure*}[!t]
\centering

\begin{minipage}[t]{0.49\textwidth}
  \centering
  \includegraphics[width=\linewidth]{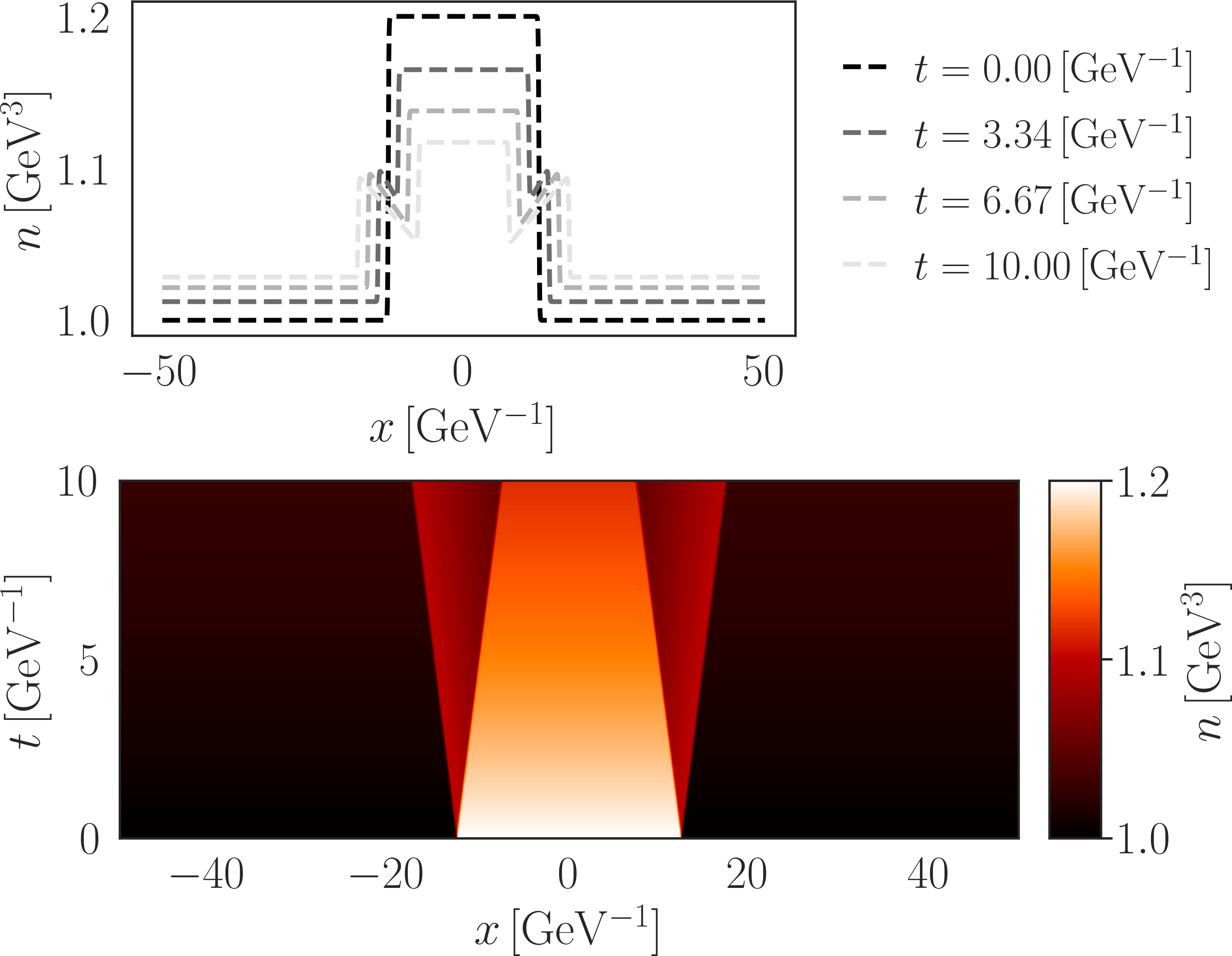}
  \makebox[0pt][l]{\small (a)}
\end{minipage}\hfill
\begin{minipage}[t]{0.49\textwidth}
  \centering
  \includegraphics[width=\linewidth]{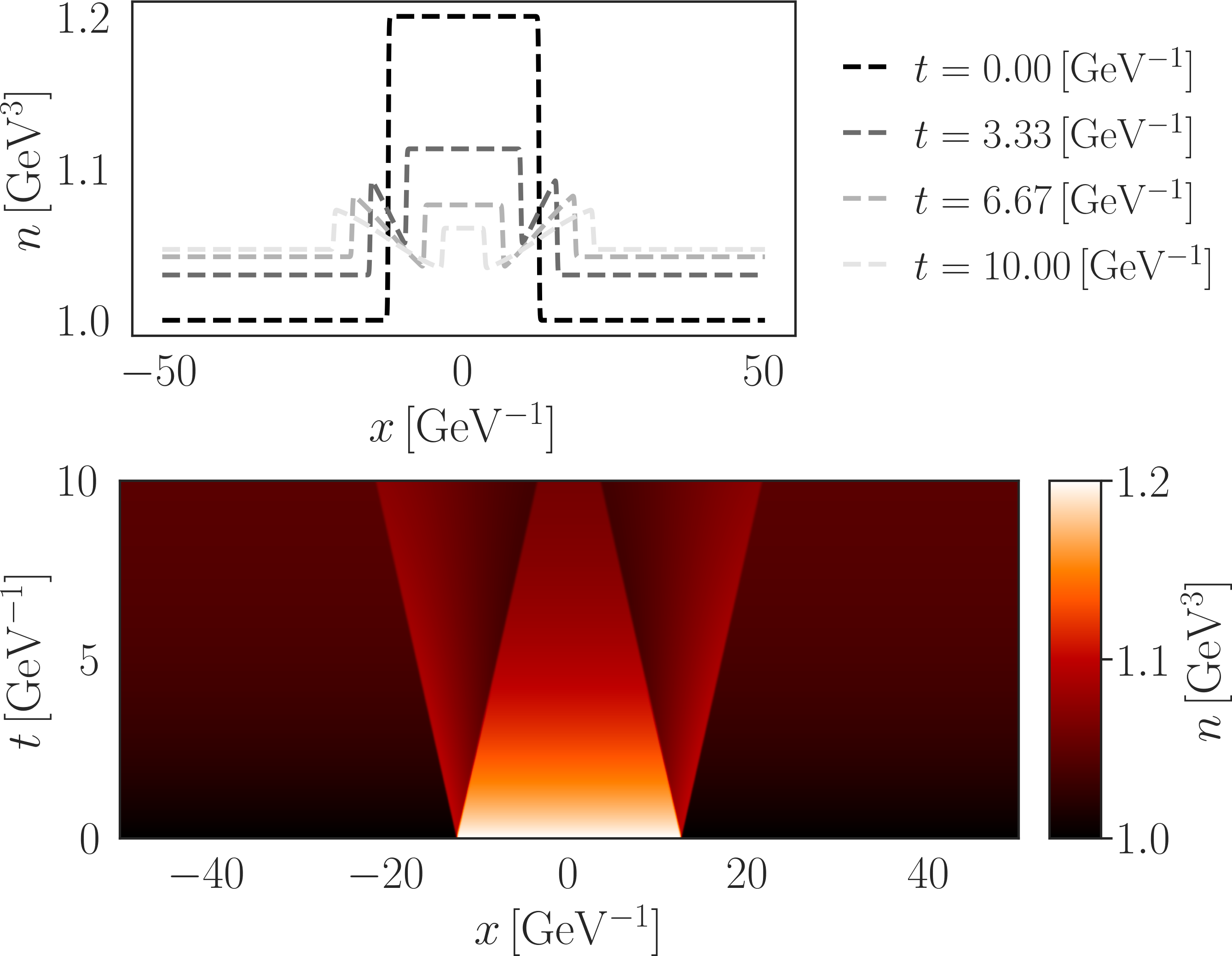}
  \makebox[0pt][l]{\small (b)}
\end{minipage}

\vspace{0.75em}

\begin{minipage}[t]{0.49\textwidth}
  \centering
  \includegraphics[width=\linewidth]{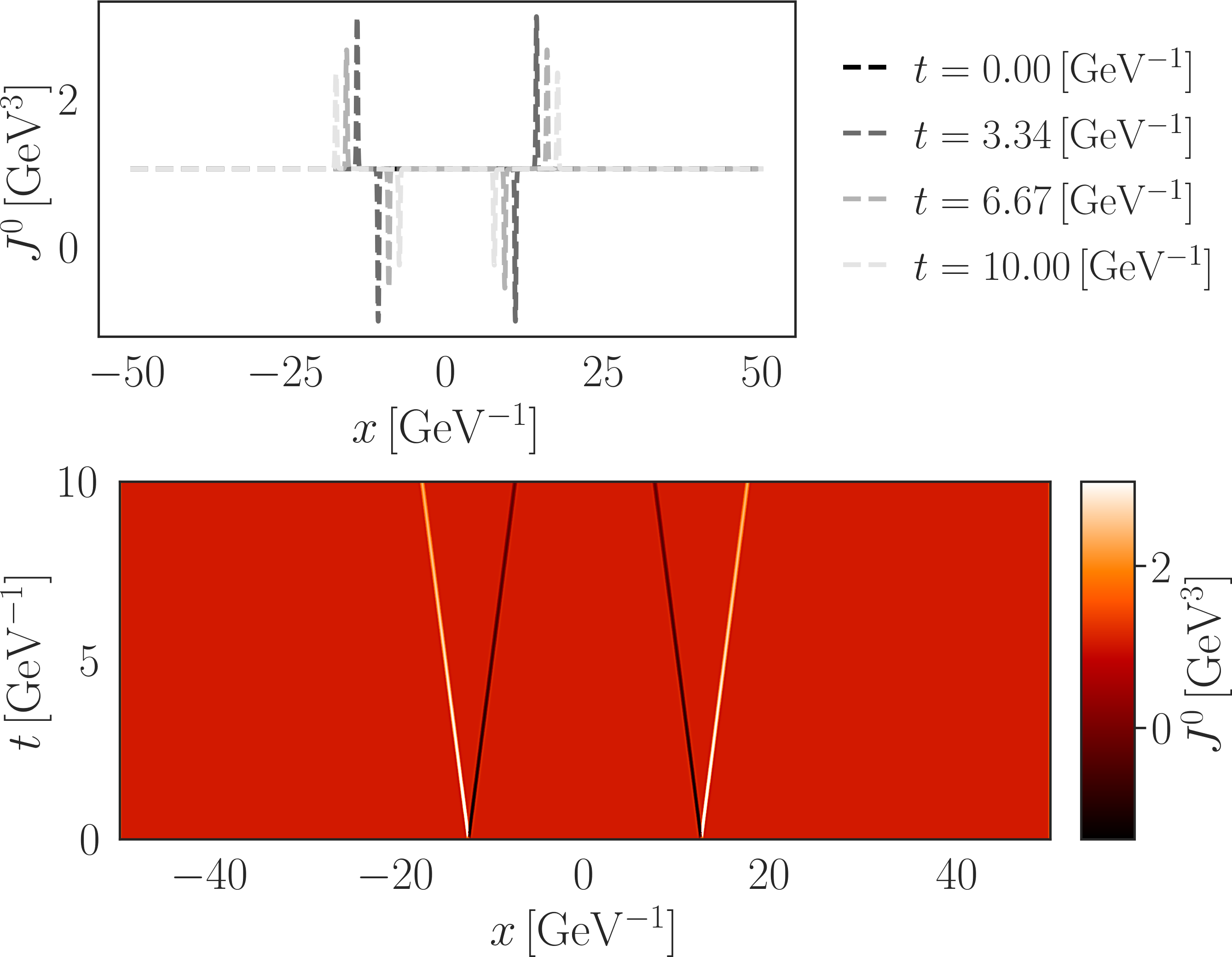}
  \makebox[0pt][l]{\small (c)}
\end{minipage}\hfill
\begin{minipage}[t]{0.49\textwidth}
  \centering
  \includegraphics[width=\linewidth]{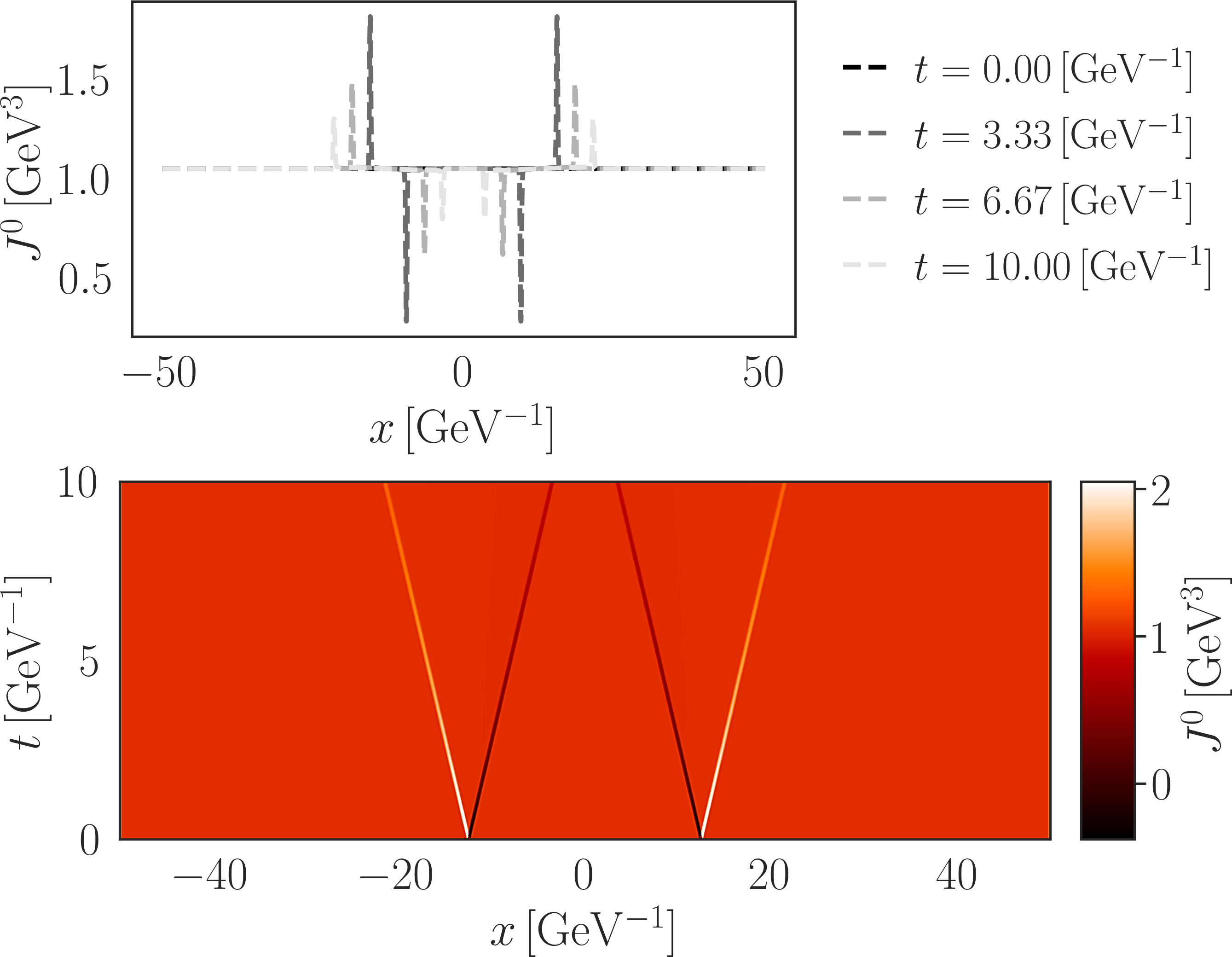}
  \makebox[0pt][l]{\small (d)}
\end{minipage}

\vspace{0.75em}

\caption{Results from KT for the second setup.
(a) Evolution of $n$ for $c_{\rm ch}=0.5$.
(b) Evolution of $n$ for $c_{\rm ch}=0.9$.
(c) Evolution of $J^0$ for $c_{\rm ch}=0.5$. Total charge $\int_{-L}^{L}{J^0\,dx}$ conserved at all times up to a fraction of $4.1\times 10^{-16}$ of the initial charge.
(d) Evolution of $J^0$ for $c_{\rm ch}=0.9$. Total charge $\int_{-L}^{L}{J^0\,dx}$ conserved at all times up to a fraction of $4.1\times 10^{-16}$ of the initial charge.
}
\label{fig:setup2-KT}
\end{figure*}

\begin{figure*}[!t]
\centering

\begin{minipage}[t]{0.49\textwidth}
  \centering
  \includegraphics[width=\linewidth]{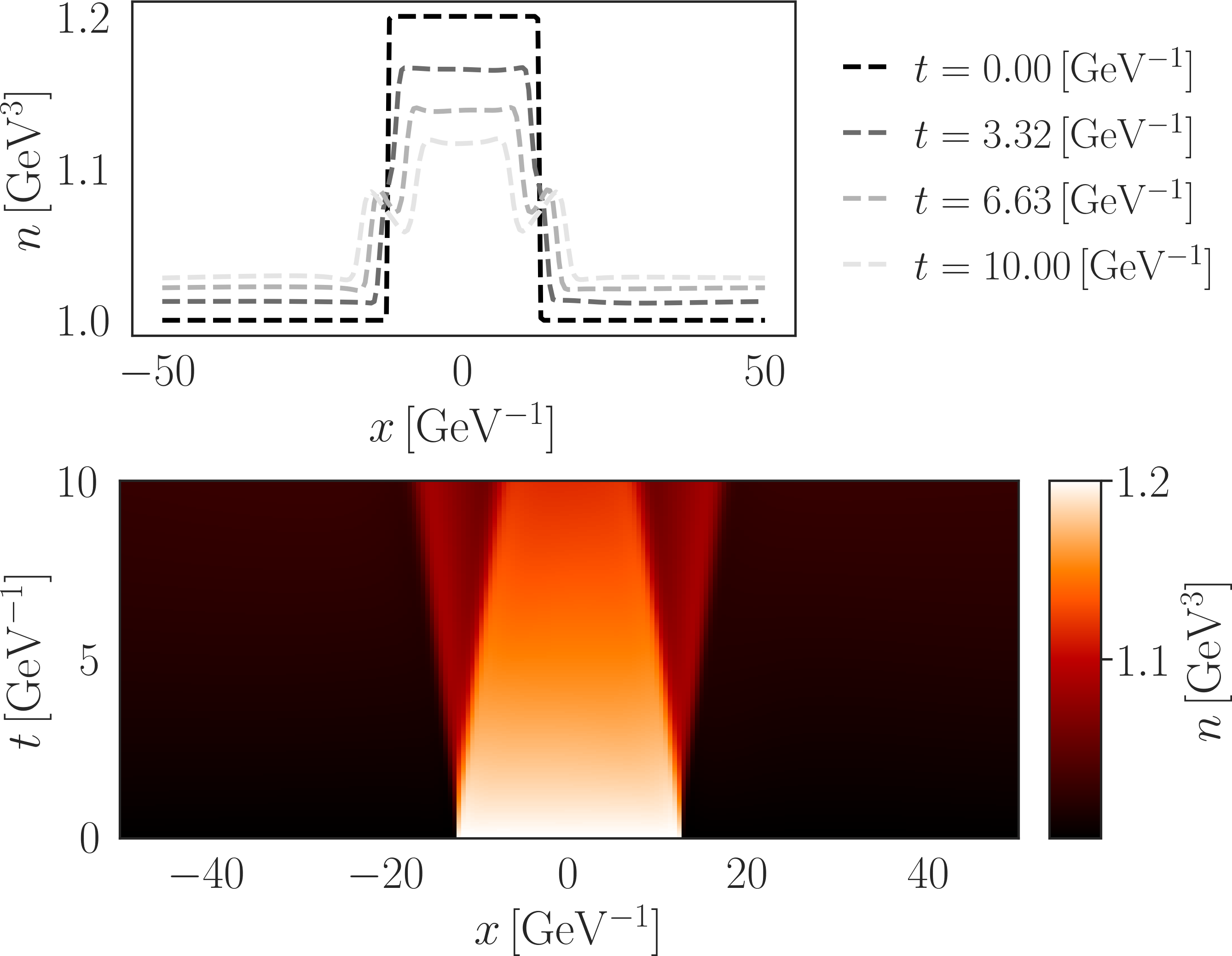}
  \makebox[0pt][l]{\small (a)}
\end{minipage}\hfill
\begin{minipage}[t]{0.49\textwidth}
  \centering
  \includegraphics[width=\linewidth]{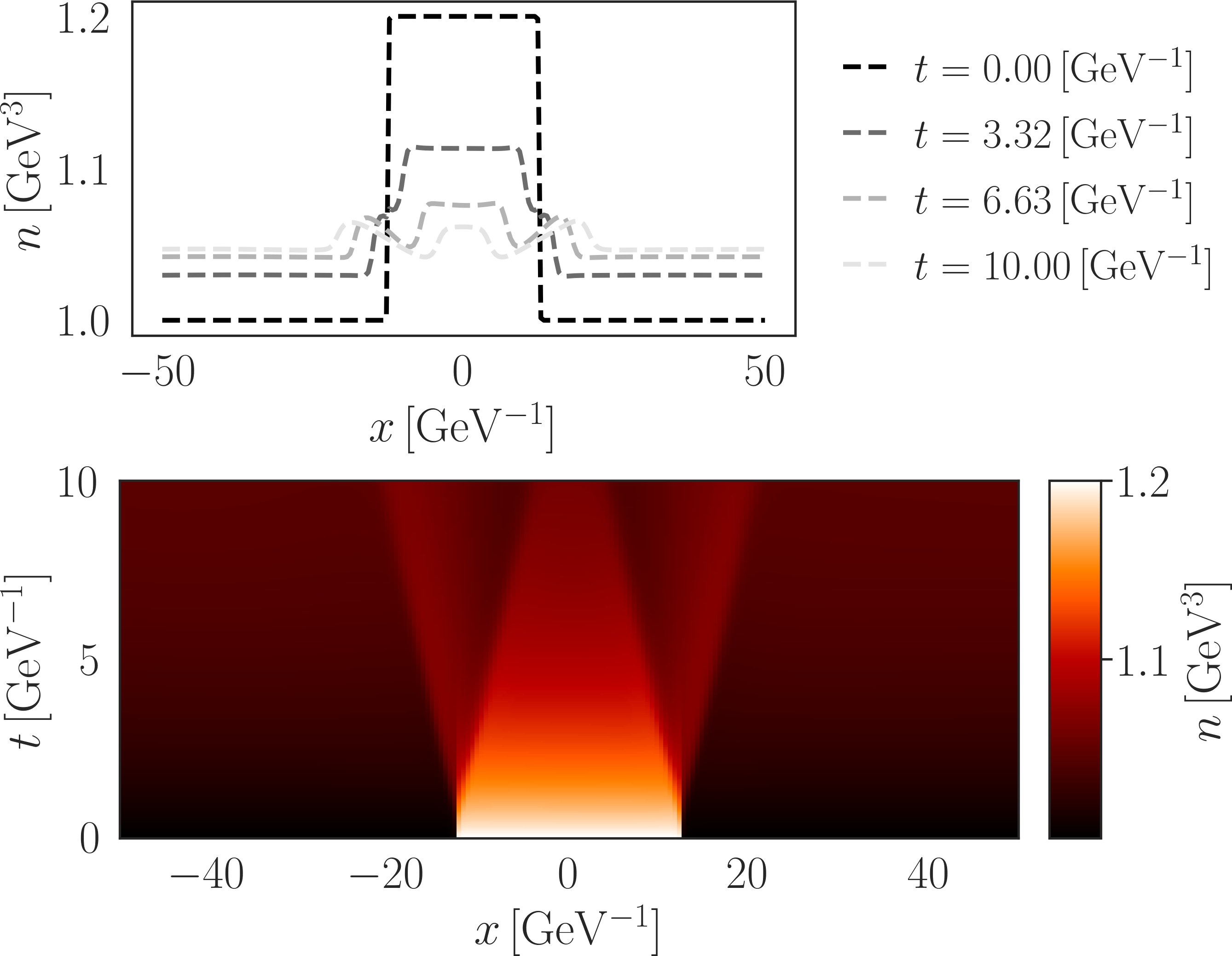}
  \makebox[0pt][l]{\small (b)}
\end{minipage}

\vspace{0.75em}

\begin{minipage}[t]{0.49\textwidth}
  \centering
  \includegraphics[width=\linewidth]{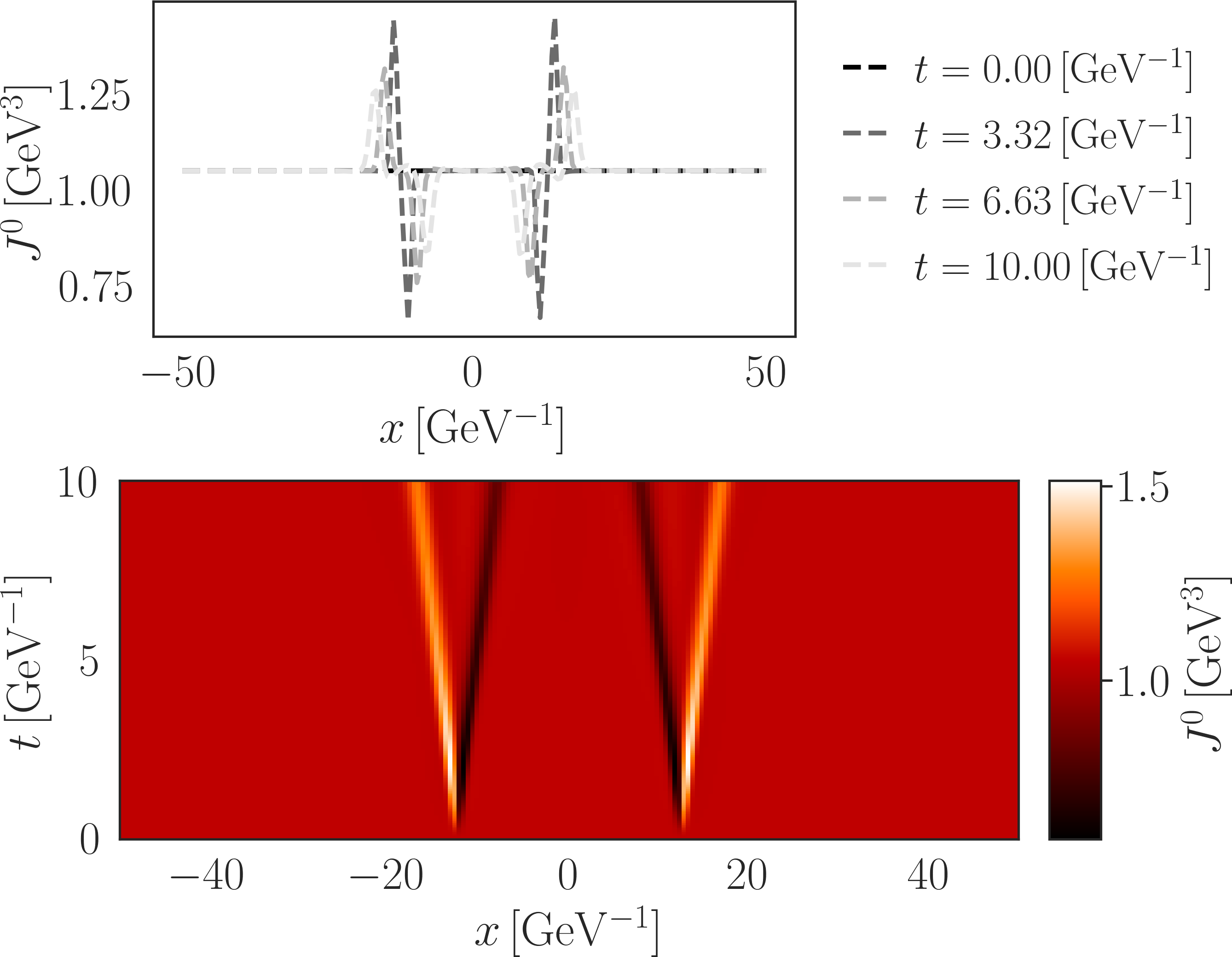}
  \makebox[0pt][l]{\small (c)}
\end{minipage}\hfill
\begin{minipage}[t]{0.49\textwidth}
  \centering
  \includegraphics[width=\linewidth]{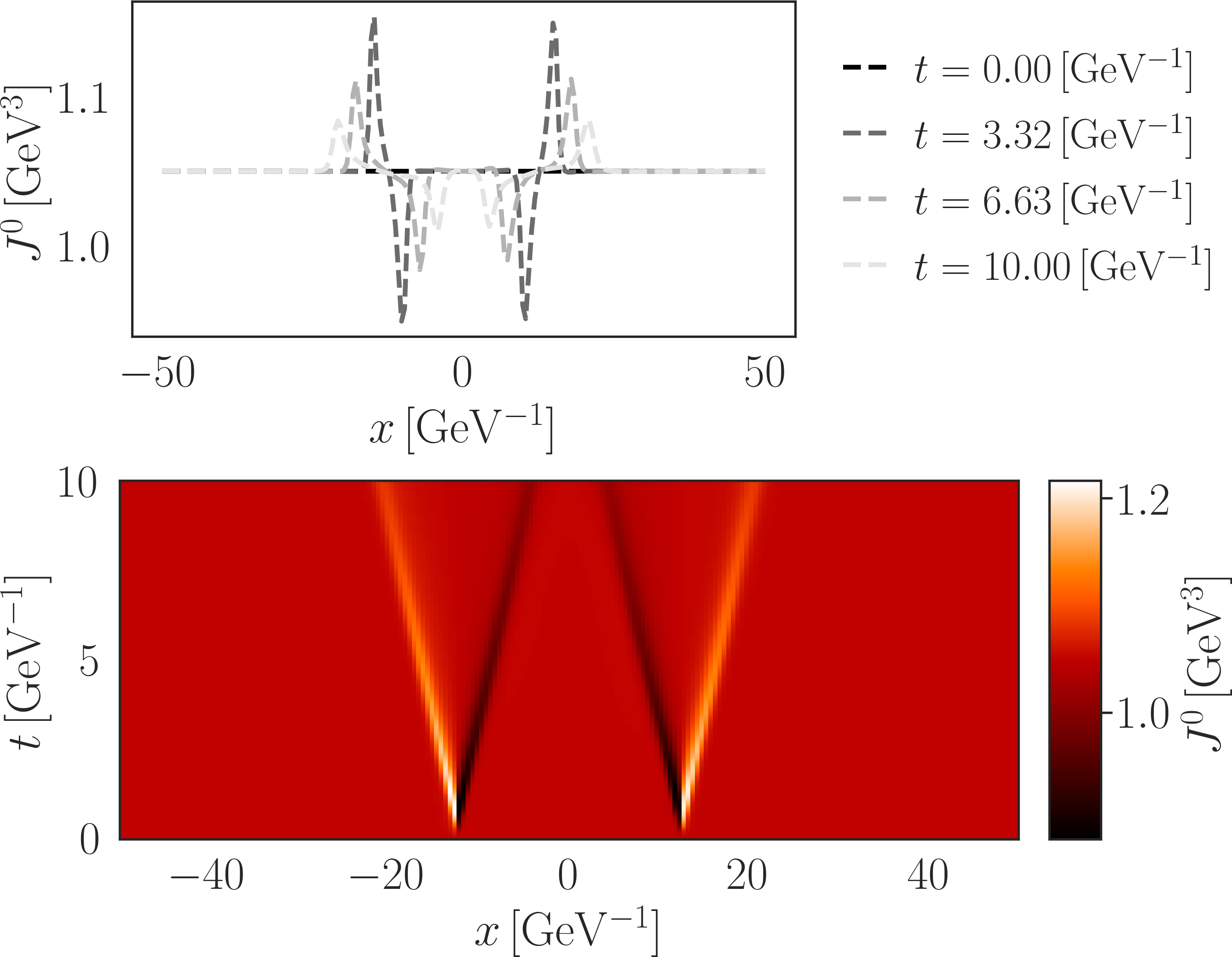}
  \makebox[0pt][l]{\small (d)}
\end{minipage}

\vspace{0.75em}

\begin{minipage}[t]{0.49\textwidth}
  \centering
  \includegraphics[width=\linewidth]{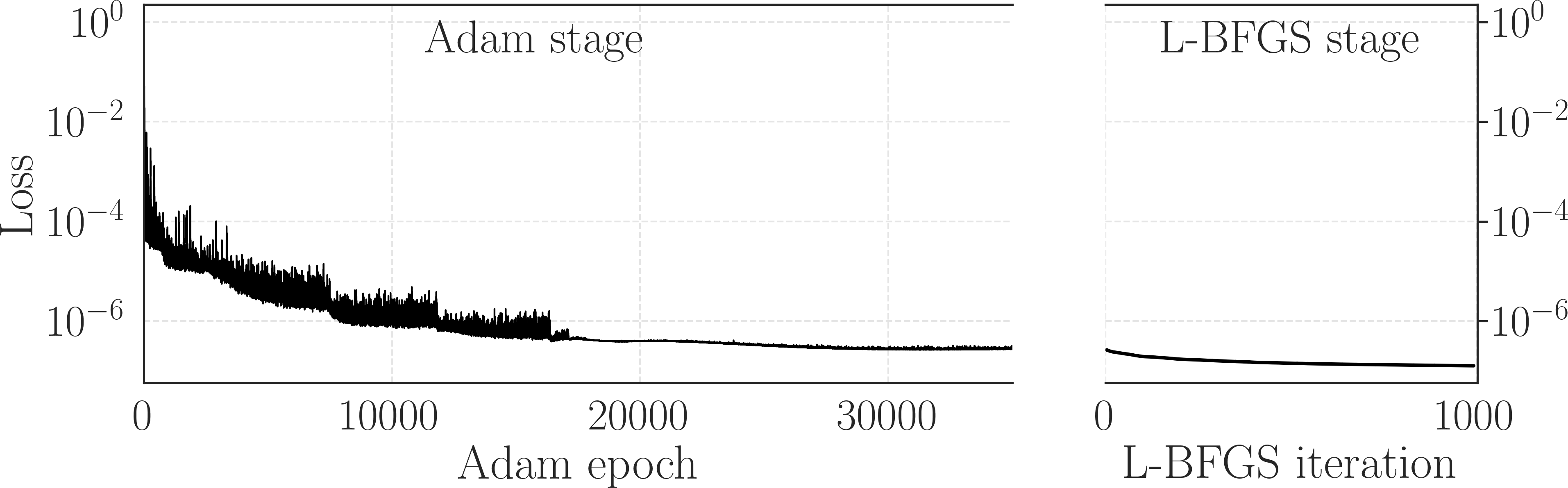}
  \makebox[0pt][l]{\small (e)}
\end{minipage}\hfill
\begin{minipage}[t]{0.49\textwidth}
  \centering
  \includegraphics[width=\linewidth]{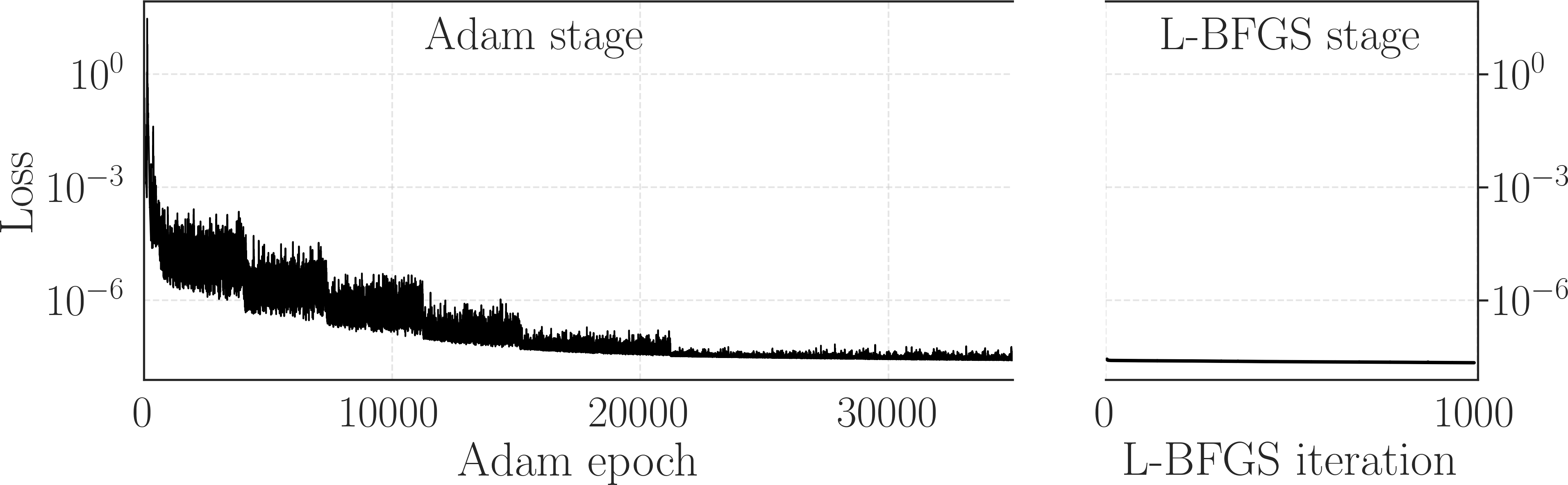}
  \makebox[0pt][l]{\small (f)}
\end{minipage}

\caption{Results from the SA-PINN-ACTO for the second setup.
For these simulations, we use $|N_{\rm PDE}|=50,000+500$, and $35,000$ Adam epochs, followed by L-BFGS fine-tuning. (a) Evolution of $n$ for $c_{\rm ch}=0.5$.
(b) Evolution of $n$ for $c_{\rm ch}=0.9$.
(c) Evolution of $J^0$ for $c_{\rm ch}=0.5$. Total charge $\int_{-L}^{L}{J^0\,dx}$ conserved at all times up to a fraction of $6.5\times 10^{-4}$ of the initial charge.
(d) Evolution of $J^0$ for $c_{\rm ch}=0.9$. Total charge $\int_{-L}^{L}{J^0\,dx}$ conserved at all times up to a fraction of $1.6\times 10^{-4}$ of the initial charge.
(e) Total loss history for $c_{\rm ch}=0.5$. Best loss was $1.272\times 10^{-7}$. Total training time: $1039.00$ seconds (Adam: $1003.95$ seconds; L-BFGS: $35.05$ seconds).
(f) Total loss history for $c_{\rm ch}=0.9$. Best loss was $2.234\times 10^{-8}$. Total training time: $1030.27$ seconds (Adam: $995.50$ seconds; L-BFGS: $34.77$ seconds).}
\label{fig:setup2-PINN}
\end{figure*}

Our results obtained using KT can be found in Fig.~\ref{fig:setup2-KT}. In the case where $c_{\text{ch}} = 0.5$ (subplot (a)), the evolution propagates more slowly in $x$. This is reflected in both the snapshots and the heatmaps, where the broadening of the discontinuous features persists and remains closer to $x=0$ for longer than in the larger characteristic speed counterpart, in which the wave fronts travel further in $x$ in the same amount of time. The discontinuities in this initial condition are preserved throughout the evolution. This is also evidenced in $J^0$, where the discontinuous spikes come from the relationship to $n$, and they shrink but do not smoothen out in time. Once again, we see smaller overall gradients for the case of $c_{\text{ch}} = 0.9$, subplots (b) and (d). While there are no conclusive data for shock formation in any of the investigated cases, the prerequisite of wavefronts accumulating near some discontinuity is precluded more effectively by larger characteristic speeds, as evidenced by the faster dissipation of the shock-like structure in the $c_{\text{ch}} = 0.9$ case.

Unlike the KT results, the SA-PINN-ACTO simulation does not preserve the discontinuous features throughout the evolution. Instead, Fig.~\ref{fig:setup2-PINN} shows a clear damping and smoothing of structure for both $n$ and $J^0$. This result is most notable in the $c_{\text{ch}} = 0.9$ simulation (subplots (b) and (d)). Consistent with this smoothing, Table~\ref{tab:relL2} shows larger errors for $J^0$ in this setup, of order $10^{-2}$--$10^{-1}$. The loss function steadily decreases throughout the Adam stage and decreases further during the L-BFGS stage. Loss minima are of order $10^{-7}$ and $10^{-8}$ for $c_{\text{ch}} = 0.5$ and $c_{\text{ch}} = 0.9$, respectively.

\subsection{Third setup: Gaussian initial condition on BDNK background}\label{ssec:third_setup}
Lastly, we simulate a Gaussian initial condition for $n$ and a simple pedestal for $J^0$, described by
\begin{equation}
    \begin{cases}
        n(t=0,x)
        =
        \left[0.2e^{-\left(\frac{7x}{L}\right)^2}+1\right]
        \times 10^{-3}
        \;{\rm GeV}^{3}
        \\
        J^0(t=0,x)
        =
        1.05
        \times 10^{-3}
        \;{\rm GeV}^{3}.
    \end{cases}
\end{equation}
That is, $n(t=0,x)$ is a Gaussian of amplitude $0.2\times 10^{-3}\;{\rm GeV}^{3}$ and standard deviation $L/(7\sqrt{2})$ standing on a pedestal of $10^{-3}\;{\rm GeV}^{3}$, and $J^0(t=0,x)$ is a pedestal of $1.05\times 10^{-3}\;{\rm GeV}^{3}$.
We perform this simulation first for $c_{\rm ch}=0.5$, and then for $c_{\rm ch}=0.9$. In this setup, temperature and velocity are not constant; rather, we use $T$ and $\vec{v}$ profiles that result from conformal BDNK background simulations performed in~\cite{Clarisse:2025lli}; these profiles are shown in Fig.~\ref{fig:setup3-profiles}. The simulations run up to $t=t_{\rm end}=20\,{\rm GeV}^{-1}$. Here, $C_B=1/4\pi$.

\begin{figure*}[!t]
\centering

\begin{minipage}[t]{0.49\textwidth}
  \centering
  \includegraphics[width=\linewidth]{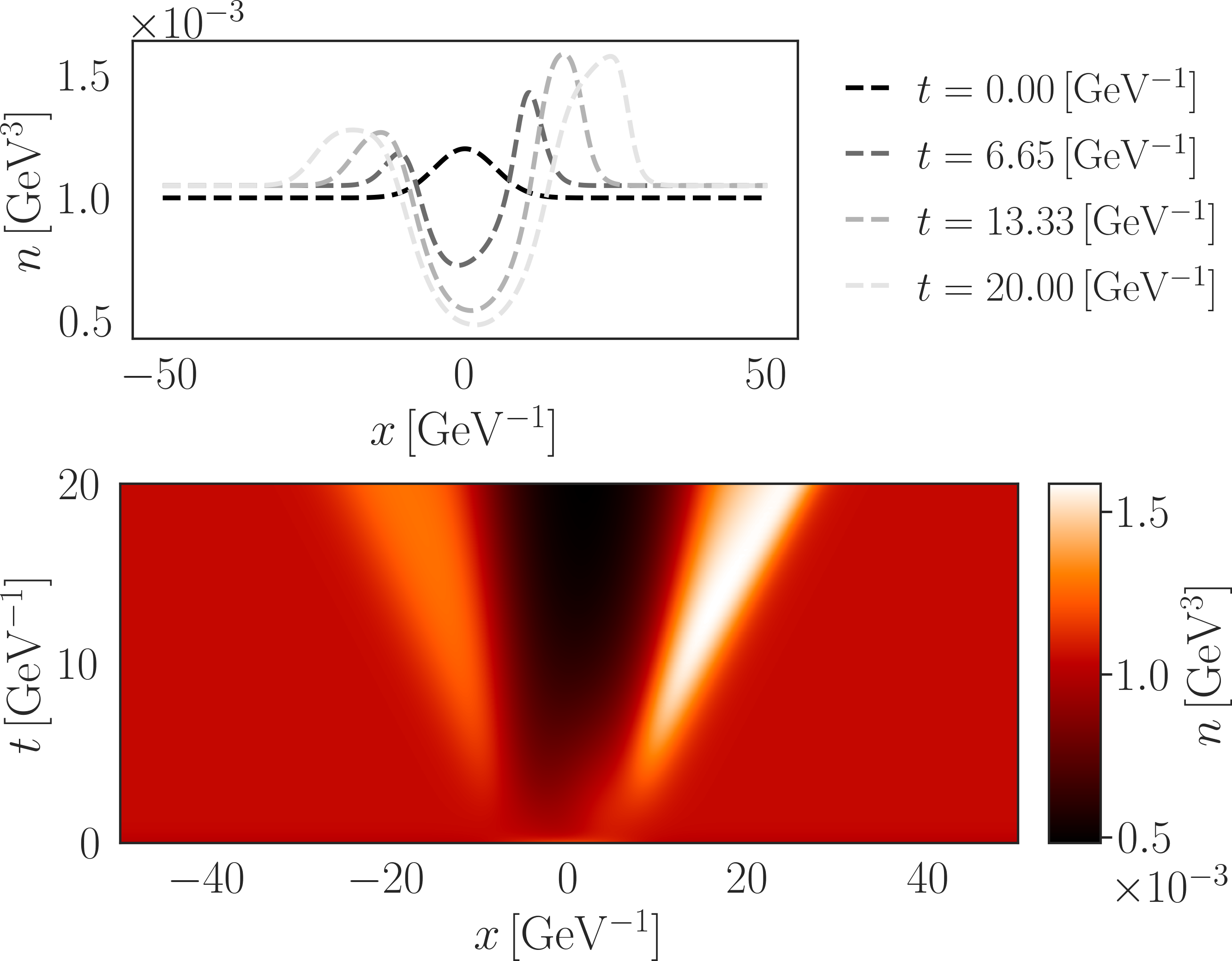}
  \makebox[0pt][l]{\small (a)}
\end{minipage}\hfill
\begin{minipage}[t]{0.49\textwidth}
  \centering
  \includegraphics[width=\linewidth]{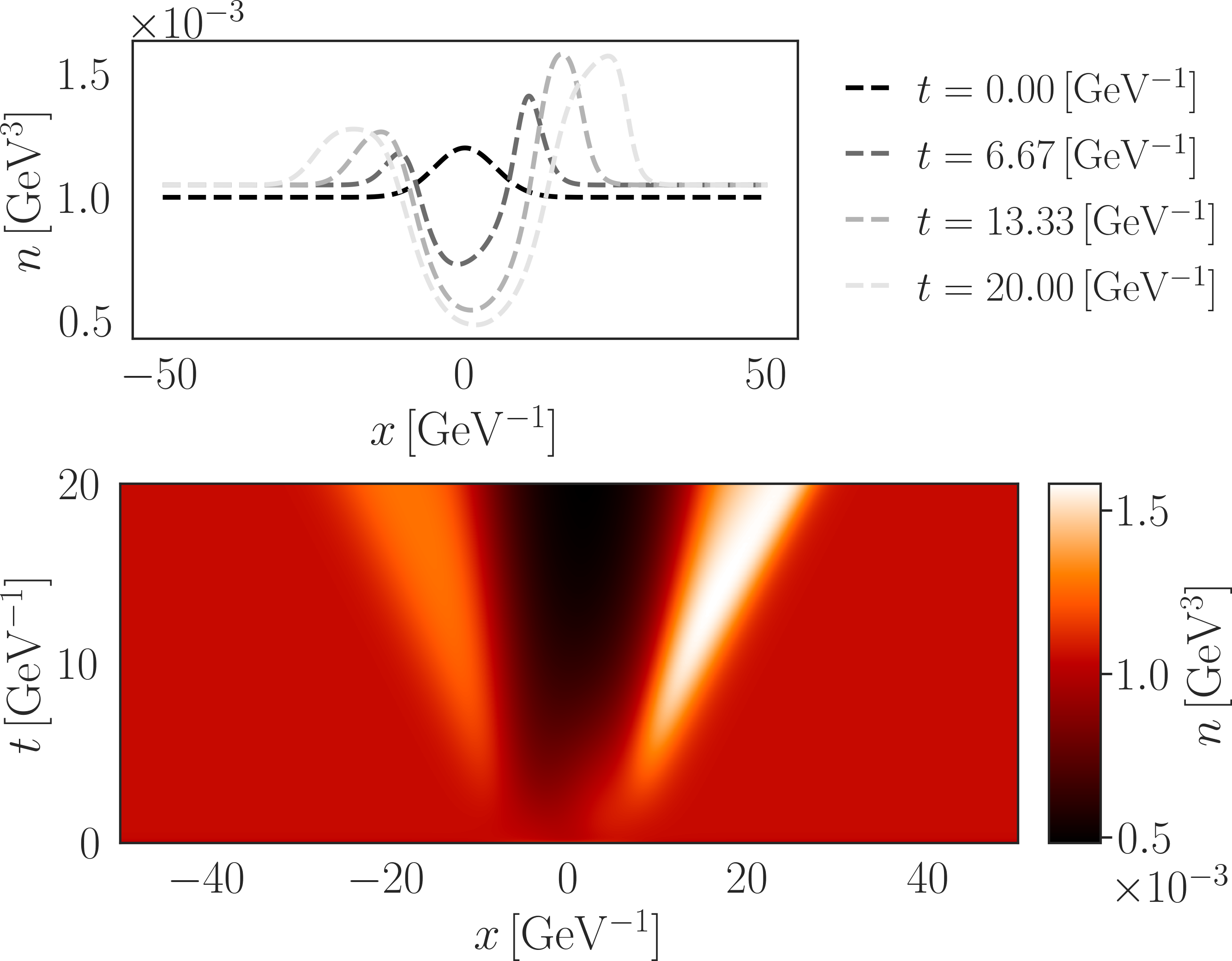}
  \makebox[0pt][l]{\small (b)}
\end{minipage}

\vspace{0.75em}

\begin{minipage}[t]{0.49\textwidth}
  \centering
  \includegraphics[width=\linewidth]{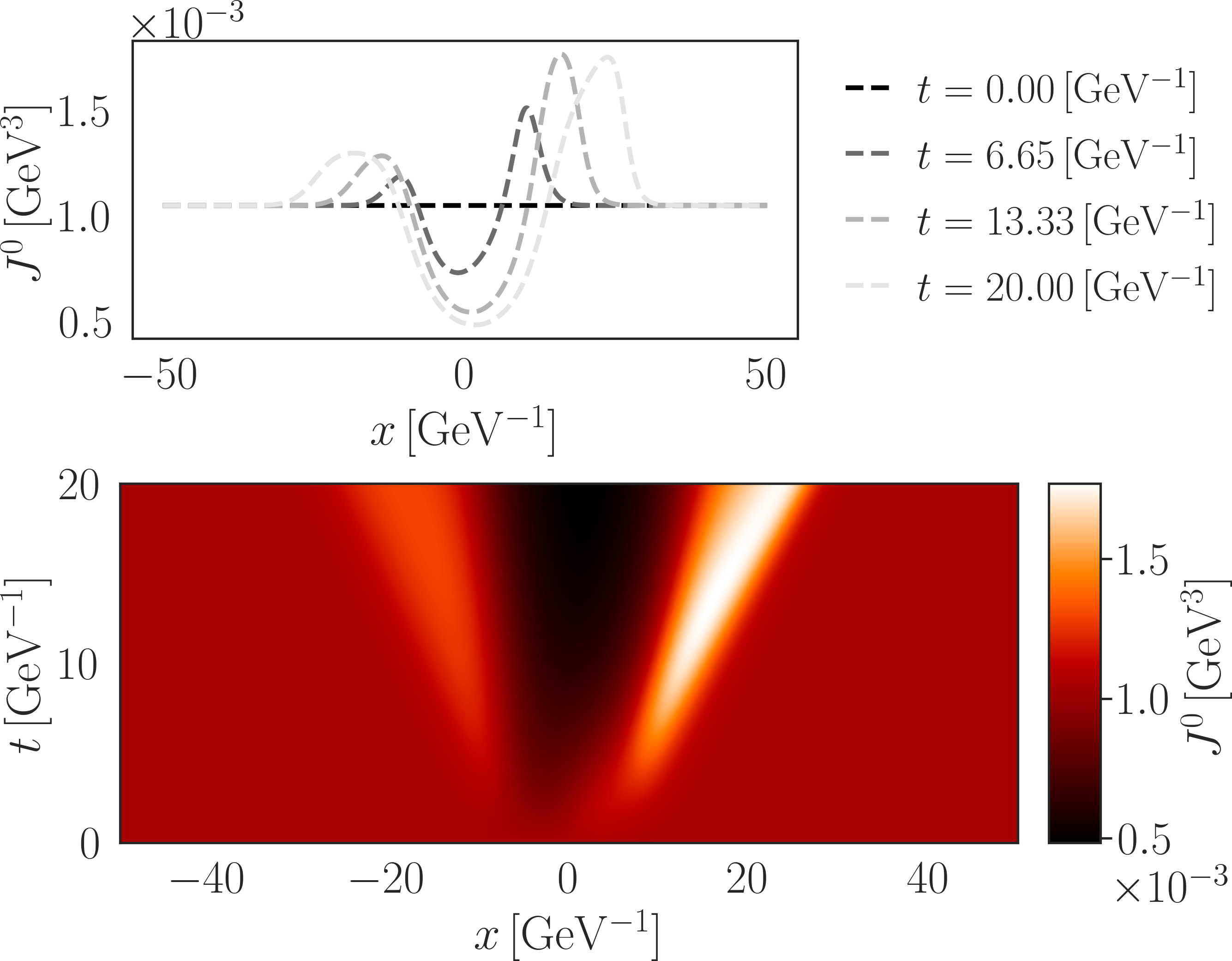}
  \makebox[0pt][l]{\small (c)}
\end{minipage}\hfill
\begin{minipage}[t]{0.49\textwidth}
  \centering
  \includegraphics[width=\linewidth]{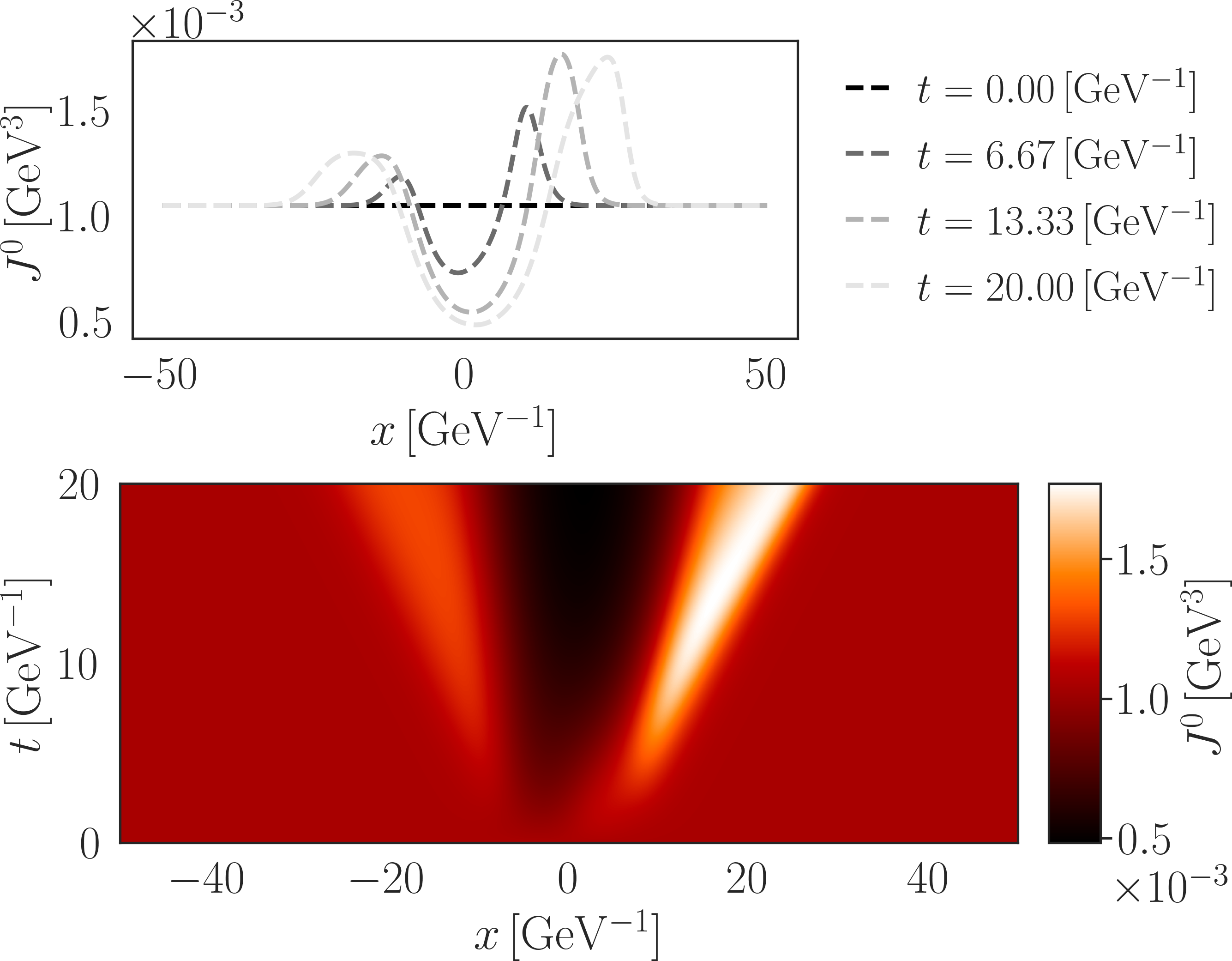}
  \makebox[0pt][l]{\small (d)}
\end{minipage}

\caption{Results from KT for the third setup.
(a) Evolution of $n$ for $c_{\rm ch}=0.5$.
(b) Evolution of $n$ for $c_{\rm ch}=0.9$.
(c) Evolution of $J^0$ for $c_{\rm ch}=0.5$. Total charge $\int_{-L}^{L}{J^0\,dx}$ conserved at all times up to a fraction of $9.3\times 10^{-16}$ of the initial charge.
(d) Evolution of $J^0$ for $c_{\rm ch}=0.9$. Total charge $\int_{-L}^{L}{J^0\,dx}$ conserved at all times up to a fraction of $5.6\times 10^{-15}$ of the initial charge.
}
\label{fig:setup3-KT}
\end{figure*}

\begin{figure*}[!t]
\centering

\begin{minipage}[t]{0.49\textwidth}
  \centering
  \includegraphics[width=\linewidth]{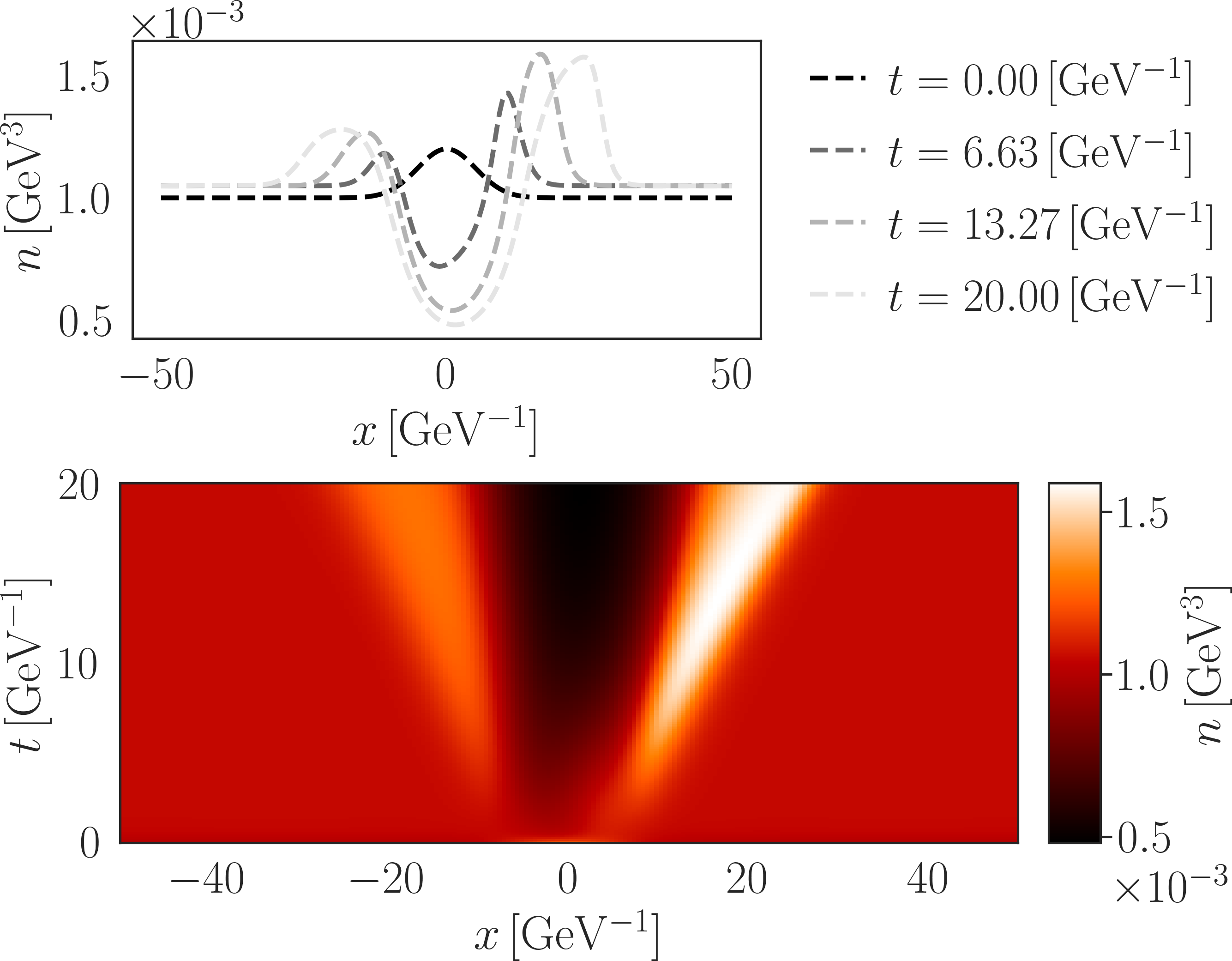}
  \makebox[0pt][l]{\small (a)}
\end{minipage}\hfill
\begin{minipage}[t]{0.49\textwidth}
  \centering
  \includegraphics[width=\linewidth]{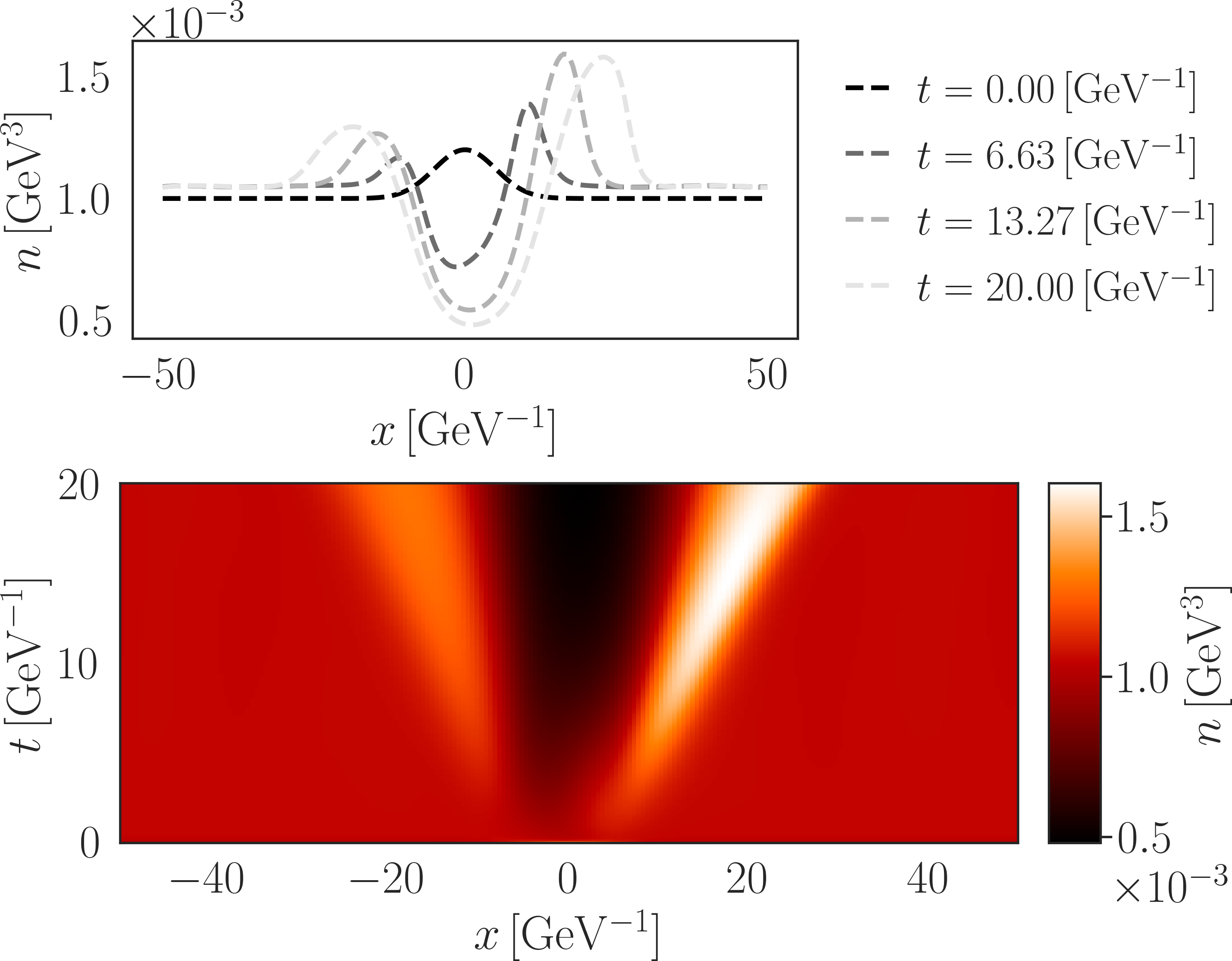}
  \makebox[0pt][l]{\small (b)}
\end{minipage}

\vspace{0.75em}

\begin{minipage}[t]{0.49\textwidth}
  \centering
  \includegraphics[width=\linewidth]{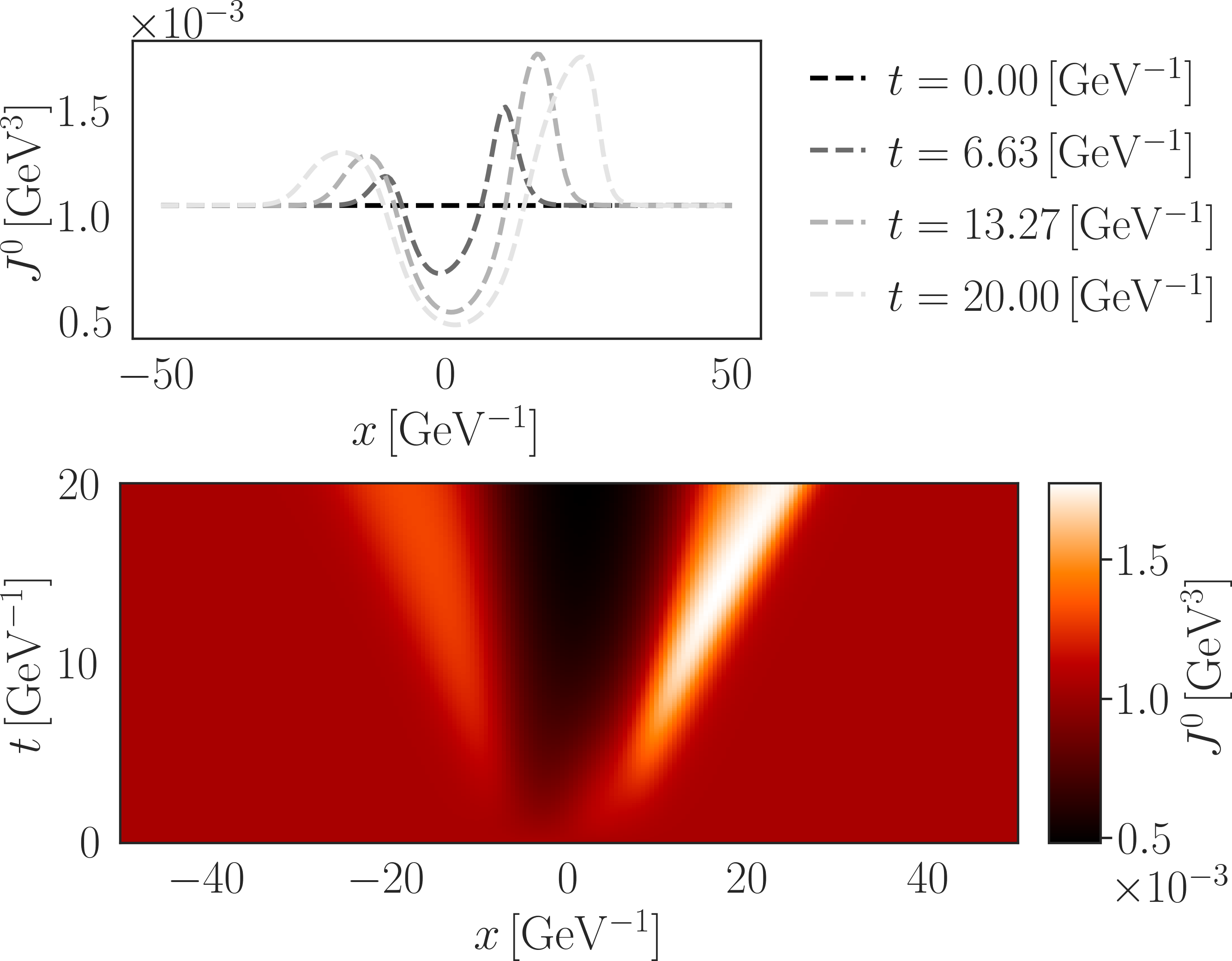}
  \makebox[0pt][l]{\small (c)}
\end{minipage}\hfill
\begin{minipage}[t]{0.49\textwidth}
  \centering
  \includegraphics[width=\linewidth]{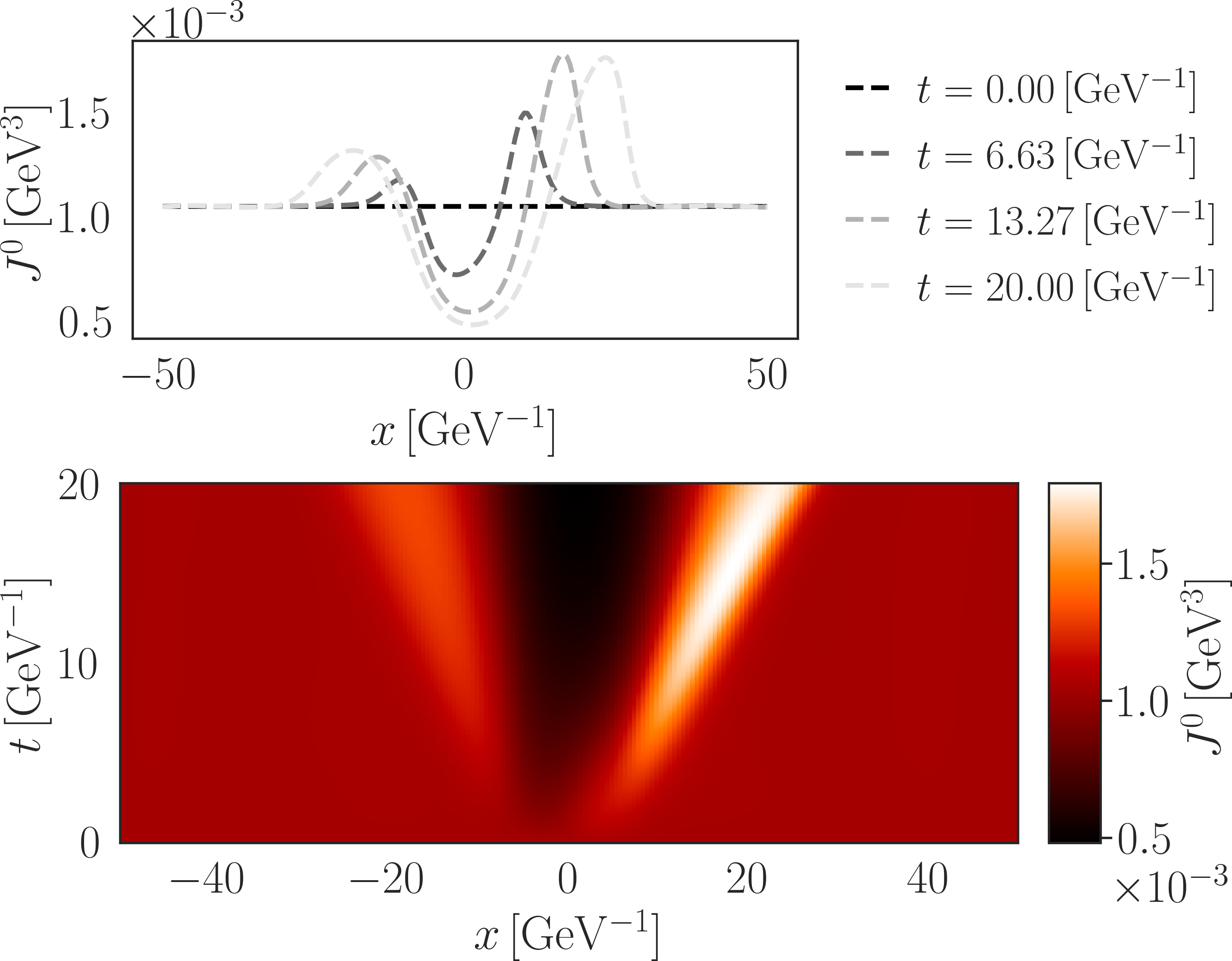}
  \makebox[0pt][l]{\small (d)}
\end{minipage}

\vspace{0.75em}

\begin{minipage}[t]{0.49\textwidth}
  \centering
  \includegraphics[width=\linewidth]{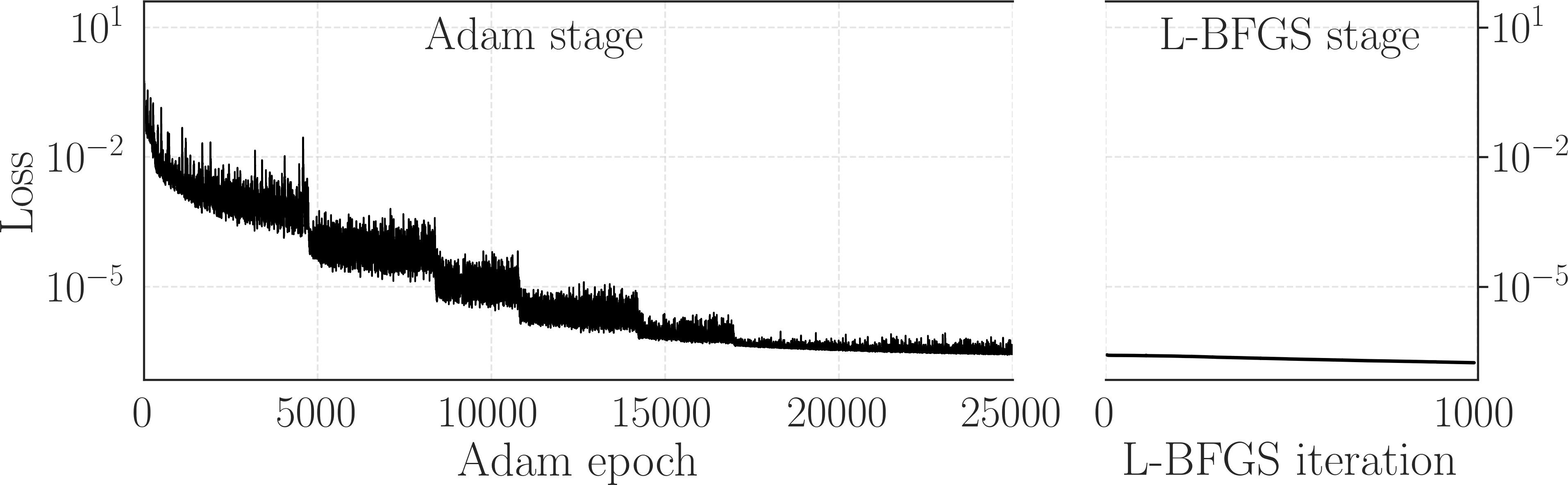}
  \makebox[0pt][l]{\small (e)}
\end{minipage}\hfill
\begin{minipage}[t]{0.49\textwidth}
  \centering
  \includegraphics[width=\linewidth]{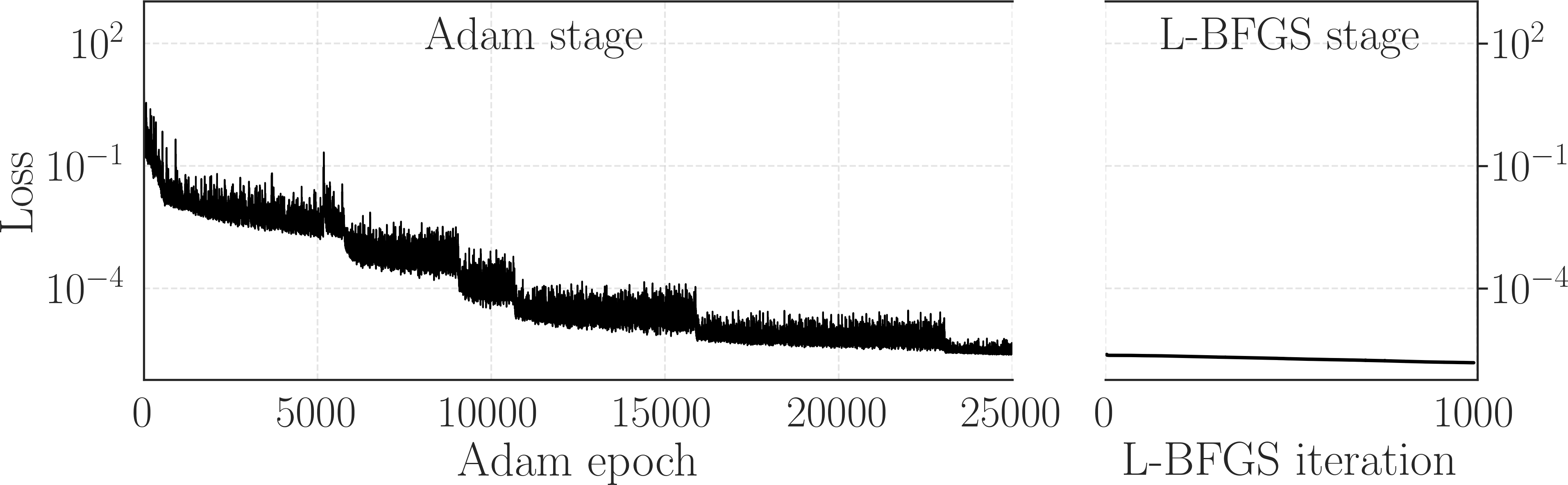}
  \makebox[0pt][l]{\small (f)}
\end{minipage}

\caption{Results from the SA-PINN-ACTO for the third setup. For these simulations, we use $|N_{\rm PDE}|=20,000+500$, and $25,000$ Adam epochs, followed by L-BFGS fine-tuning. (a) Evolution of $n$ for $c_{\rm ch}=0.5$.
(b) Evolution of $n$ for $c_{\rm ch}=0.9$.
(c) Evolution of $J^0$ for $c_{\rm ch}=0.5$. Total charge $\int_{-L}^{L}{J^0\,dx}$ conserved at all times up to a fraction of $4.9\times 10^{-4}$ of the initial charge.
(d) Evolution of $J^0$ for $c_{\rm ch}=0.9$. Total charge $\int_{-L}^{L}{J^0\,dx}$ conserved at all times up to a fraction of $1.6\times 10^{-3}$ of the initial charge.
(e) Total loss history for $c_{\rm ch}=0.5$. Best loss was $1.728\times 10^{-7}$. Total training time: $844.62$ seconds (Adam: $806.14$ seconds; L-BFGS: $38.48$ seconds).
(f) Total loss history for $c_{\rm ch}=0.9$. Best loss was $1.517 \times 10^{-6}$. Total training time: $844.27$ seconds (Adam: $805.74$ seconds; L-BFGS: $38.53$ seconds).}
\label{fig:setup3-PINN}
\end{figure*}

The initial conditions provide a difference in the evolution that is shared across both the KT solutions, Fig.~\ref{fig:setup3-KT} and the SA-PINN-ACTO solutions, Fig.~\ref{fig:setup3-PINN}. Both simulations report a left-moving wave in both $n$ and $J^0$ that is smaller in magnitude than the right-moving wave counterpart which propagates with a larger maximum height. These two waves have maxima slightly above $1.5\times10^{-3}\;{\rm GeV}^{3}$, and are separated by a shared minimum centered near $x=0$ whose magnitude drops slightly below $0.5\times10^{-3}\;{\rm GeV}^{3}$. In contrast to the first setup (\ref{ssec:first_setup}), we no longer see a stark difference in the extrema of the evolution between $c_{\mathrm{ch}} = 0.5, 0.9$. There are still subtle differences in the evolution depending on the characteristic speed, with sharper features appearing in the right-moving waves for the smaller $c_{\mathrm{ch}} = 0.5$ case, but this effect is less pronounced than in the first setup. Additionally, we would like to note that the minimum loss achieved in the $c_{\mathrm{ch}} = 0.5$ case is smaller than the $c_{\mathrm{ch}} = 0.9$ case, as seen in Fig.~\ref{fig:setup3-PINN} caption.

The agreement between the two numerical methods, KT and SA-PINN-ACTO, gives credence to the reliability of the predicted PDE solutions. That two fundamentally different numerical approaches yield such closely matching evolutions, especially considering that the PINN is informed solely through the PDE residual in the loss function and has no additional structural constraints on the dynamics, is a significant result. This is supported by Table~\ref{tab:relL2}, which shows that the relative $L^2$ errors between these methods are of order $10^{-3}$ for the first (\ref{ssec:first_setup}) and third (\ref{ssec:third_setup}) setups.

\begin{table*}[t]
\centering
\setlength{\tabcolsep}{8pt}
\renewcommand{\arraystretch}{1.2}

\begin{tabular}{c c c @{\;\;\;}||@{\;\;\;} c c}
\hline\hline
\rule{0pt}{2.7ex}\rule[-1.2ex]{0pt}{0pt}
Background & Setup & $c_{\rm ch}$ & $E_{\rm rel}[n]$ & $E_{\rm rel}[J^0]$ \\
\hline
\multirow{4}{*}{\parbox{0.25\textwidth}{Trivial\\($T=0.3\,{\rm GeV}$, $\vec{v}=0$)}}
  & \multirow{2}{*}{\parbox{0.25\textwidth}{$n$ Gaussian, $J^0$ Gaussian\\(\ref{ssec:first_setup})}}
      & $0.5$ & $1.794\times10^{-3}$ & $3.449\times10^{-3}$ \rule{0pt}{2.5ex} \\
  &   & $0.9$ & $2.013\times10^{-3}$ & $2.560 \times10^{-3}$ \\[2pt]
  & \multirow{2}{*}{\parbox{0.25\textwidth}{$n$ shock, $J^0$ pedestal\\(\ref{ssec:second_setup})}}
      & $0.5$ & $7.096\times10^{-3}$ & $1.401\times10^{-1}$ \\
  &   & $0.9$ & $7.259\times10^{-3}$ & $5.897\times10^{-2}$ \\[2pt]

\multirow{2}{*}{\parbox{0.25\textwidth}{BDNK\\(see Fig.~\ref{fig:setup3-profiles})}}
  & \multirow{2}{*}{\parbox{0.25\textwidth}{$n$ Gaussian, $J^0$ pedestal\\(\ref{ssec:third_setup})}}
      & $0.5$ & $2.246\times10^{-3}$ & $2.299\times10^{-3}$ \\
  &   & $0.9$ & $8.283\times10^{-3}$ & $8.638\times10^{-3}$ \\
\hline\hline
\end{tabular}
\caption{Relative $L^2$ error $E_{\rm rel}$ between SA-PINN-ACTO and Kurganov-Tadmor results
(see Eq.~\eqref{eq:l2_validation}), over the entire spacetime domain $(t,x)\in[0,t_{\rm end}]\times[-L,L]$, for both $n$ and $J^0$.}
\label{tab:relL2}
\end{table*}

\section{Conclusions}\label{sec:conclusions}

In this work, we have reformulated the relativistic BDNK diffusion equation in flux-conservative form and solved the resulting equations in $(1+1)$D using two independent numerical methods: a second-order KT finite‐volume solver and a physics-informed neural network (PINN). On the finite volume side, we implemented an SSP-RK2 Kurganov-Tadmor scheme and verified its convergence for both smooth and discontinuous initial data. On the machine learning side, we introduced the SA-PINN-ACTO method, which combines the self-adaptive collocation weights technique of the SA-PINN~\cite{MCCLENNY2023111722} with an algebraic (also known as hard) enforcement of initial and periodic boundary conditions via only an output transform, which we call the ACTO transform. This guarantees that those conditions are exactly satisfied without requiring any modification to the network architecture, while allowing the network to focus its attention on the regions of the domain where it is harder to properly approximate the solution to a given PDE, as well as having to minimize only the PDE residual.

Across the set of smooth test problems, the SA-PINN-ACTO produced solutions that closely match the convergent KT solutions, with spacetime relative $L^2$ errors of order $10^{-3}$. The KT solver also displayed the expected second-order behavior in the smooth tests. For the shock initial condition, however, while the KT method correctly preserved the discontinuities, the PINN smoothed them out and reached error levels in the $10^{-2}$--$10^{-1}$ range, an expected behavior due to the intrinsic smoothness of neural-network representations and well-established challenges PINNs face near sharp gradients~\cite{liu2024discontinuity,ABBASI2025131440,doi:10.1137/22M1522504}. We remark that the PINN remained much slower than the KT method in all cases. For the tests involving dynamical BDNK backgrounds, the PINN again achieved relative errors of order $10^{-3}$ and successfully reproduced the asymmetric propagation induced by the background fields.
In the large gradient regime, both the PINN and KT method exhibit hydrodynamic frame dependent behavior, as can be seen in Fig.~\ref{fig:setup1-KT}. Despite this expected frame dependence, the residuals remain small in each frame.

We remark that the PINN solutions preserve, for example, total charge across all simulations and symmetry in simulations where the initial data and backgrounds are symmetric. These properties were not imposed as constraints in the learning process; rather, they are natural consequences of the governing laws of the system, in this case the BDNK diffusion equation. This illustrates a central point of physics-informed machine learning: physical structure can arise from the equations themselves, even without being imposed through predefined constraints on the model.

We found that the KT method remains significantly faster and more accurate, especially for discontinuous solutions, and it continues to be the appropriate tool when high-precision shock capture is required. The PINN approach, on the other hand, offers greater flexibility: it does not require rewriting the equations in flux-conservative form, provides a differentiable continuous solution representation, and effortlessly extends to inverse problems, irregular geometries, a large variety of types of PDEs, and even the inclusion of available empirical data during training. Furthermore, different from KT which is defined for a set of first-order PDEs, our PINN technique can also equally handle sets of second-order PDEs. This is discussed in detail in Appendix~\ref{app:2ndorder} where we compare PINN results for BDNK diffusion using the first-order flux-conservative formulation developed here and also the standard formulation of this theory in terms of second-order PDEs. These complementary strengths highlight the distinct roles of traditional solvers and physics-informed networks.

Future work should focus on accelerating the training process of PINNs, and importantly, improving their performance near discontinuities, potentially through hybrid finite volume-machine learning schemes that can take the best from each framework. We emphasize that the methods developed here can also be extended to higher-dimensional BDNK systems, or to inference of transport coefficients from partial data. In this context, it would be very interesting to implement the prescription for BDNK initial data recently discussed in \cite{Gavassino:2026xjw}, to better gauge the domain of applicability of BDNK hydrodynamics and its hydrodynamic-frame robustness in numerical simulations. Overall, the close agreement between the two fundamentally different approaches indicates that the SA-PINN-ACTO framework introduced in this paper provides a viable and flexible tool for solving PDEs and motivates further exploration of physics-informed machine learning in relativistic viscous hydrodynamics, with potential applications in heavy-ion collisions and astrophysics.

\begin{acknowledgments}
We thank Thomas Woehrle for helpful discussions on neural networks. V. C.-C. acknowledges the use of the Illinois Campus Cluster, a computing resource that is operated by the Illinois Campus Cluster Program (ICCP) in conjunction with the National Center for Supercomputing Applications (NCSA) and which is supported by funds from the University of Illinois Urbana-Champaign. V. C.-C. also acknowledges the Lorella M. Jones Undergraduate Research Scholarship and the Department of Physics, University of Illinois Urbana-Champaign. N.M. was partially supported by U.S. Department of Energy, Office of Nuclear Physics, Contract DE-FG02-03ER41260. J.N. and N.C. are partly supported by the U.S. Department of Energy, Office of Science, Office of Nuclear Physics under Award No. DE-SC0023861.

\end{acknowledgments}

\appendix
\section{Multilayer perceptrons (MLPs)}\label{app:mlp}

An MLP is a type of feedforward neural network that, mathematically, behaves as a parametric map $f_\theta:\mathbb{R}^{d_{\rm in}}\to\mathbb{R}^{d_{\rm out}}$ given by
\begin{equation}
\label{eq:mlp_def}
f_\theta(z) = \sigma_{L+1}\left(W_{L+1}\,\sigma_{L}\!\left(\cdots \sigma_1(W_1 z + b_1)\cdots\right) + b_{L+1}\right),
\end{equation}
obtained by composing affine transforms with fixed elementwise nonlinearities, with a set of learnable parameters $\theta=\{(W_\ell,b_\ell)\}_{\ell=1}^{L+1}$, where $W_\ell\in\mathbb{R}^{d_\ell\times d_{\ell-1}}$ are weight matrices,
$b_\ell\in\mathbb{R}^{d_\ell}\;(\ell=1,\dots,L{+}1)$ are bias vectors, and
$\{d_\ell\}_{\ell=0}^{L+1}$ are the layer widths (i.e., the numbers of neurons in each layer), with
$d_0=d_{\rm in}$ and $d_{L+1}=d_{\rm out}$. Each $\sigma_\ell$ acts elementwise for $1\le \ell\le L+1$.

Information flows strictly layer by layer in a forward, fully connected manner; there are no recurrences or convolutions. In our work, the input is $z=(t,x)$ (the $(1+1)$D spacetime coordinate $x^\mu$), and the output is the prediction vector
\begin{equation}
\mathbf{u}_\theta(t,x) \;:=\; f_\theta(z) \;=\; \left(J^0_\theta(t,x),\,\alpha_\theta(t,x)\right).
\end{equation}

Hidden layers (that is, the layers between the input and output layers) use a smooth nonlinearity. Here we use $\sigma(\cdot)=\tanh(\cdot)$. Therefore, each neuron in the hidden layers takes the full vector of outputs from the previous layer, takes the inner product of that vector with the neuron's vector of learned weights, adds its learned bias, and finally applies the $\tanh$ transformation. This process results in the neuron's output. This output, along with all the outputs of all the neurons in the same layer, is sent to the next layer, and the process is repeated until the last layer is reached. This is reflected by Eq.~\eqref{eq:mlp_def}. Without such nonlinear transformations, MLPs would not work, as the final output of the network would simply be a first-degree polynomial in the inputs $t$ and $x$. For the output layer, we use $\sigma(\cdot)={\rm id}(\cdot)$; in a vanilla PINN, it is not in our interest to apply a transformation to the output of the neural network.

Training does not result in a modification of the architecture of the MLP; it simply updates the elements of $\theta$ to minimize the physics-informed loss defined in Eq.~\eqref{eq:loss}. Training a PINN is therefore an optimization problem on the set of trainable parameters $\theta$. Because Adam optimization is gradient-based, it requires $\partial\mathcal{L}/\partial\theta$. We obtain these gradients via automatic differentiation (AD).

\section{First- vs. second-order formulation of the BDNK diffusion problem for the SA-PINN-ACTO}
\label{app:2ndorder}
Instead of solving the flux-conservative formulation~\eqref{eq:flux_conservative_formulation}, the SA-PINN-ACTO (or, generally speaking, any PINN) can directly solve the BDNK diffusion problem in its second-order formulation, given by the second-order partial differential equation of motion~\eqref{eq:2nd-order-PDE}, by choosing a PDE residual
\begin{equation}\label{eq:loss_pde_2nd_order}
    \mathcal{L}_{\rm PDE}
    \equiv \frac{1}{|N_{\rm PDE}|}
    \sum_{i\in N_{\rm PDE}}
    \lambda_i^2 R_{{\rm 2nd},i}^2,
\end{equation}
where
\begin{equation}
    R_{{\rm 2nd},i}
    \equiv
    [\partial_{\mu} (n u^{\mu}) + \partial_{\mu} ( \lambda T u^{\mu} u^{\nu} \partial_{\nu} \alpha ) - \partial_{\mu} ( \sigma T \Delta^{\mu\nu} \partial_{\nu} \alpha )]_i \,/s_{\alpha},
\end{equation}
and building the network with only the raw output $\hat{\alpha}/s_\alpha$, which would then get denormalized and ACTO-transformed into the final output $\alpha$. For the $(1+1)$D BDNK diffusion problem in the rest frame, for example, this residual would be simply
\begin{equation}
    R_{{\rm 2nd},i}
    \equiv
    [
    \partial_t n
    +
    \partial_t(\lambda T\partial_t\alpha)
    -
    \partial_x(\sigma T\partial_x \alpha)
    ]_i \,/s_{\alpha}.
\end{equation}
Once the PINN learns an $\alpha$ that approximately satisfies this equation, $J^0$ can be simply reconstructed from the constitutive relation in Eq.~\eqref{eq:Jmu}.

However, note that the ACTO transform enforces only $\alpha(t=0)$, but not $\partial_t \alpha(t=0)$ (see Sec.~\ref{sssec:heic}), which means that the network would have the task of learning the second-order initial condition for $\alpha$. Therefore, we would have to add the loss
\begin{equation}
    \mathcal{L}_{\rm IC}
    =
    \frac{1}{|N_{\rm IC}|}
    \sum_{i\in N_{\rm IC}}
    |
    \partial_t\alpha_\theta(0,x_i)
    -
    \partial_t \alpha_{IC}(x_i)
    |^2,
\end{equation}
and train the network on minimizing
\begin{equation}
    \mathcal{L} = \mathcal{L}_{\rm PDE} + \mathcal{L}_{\rm IC}.
\end{equation}

\begin{figure*}[!t]
\centering

\begin{minipage}[t]{0.49\textwidth}
  \centering
  \includegraphics[width=\linewidth]{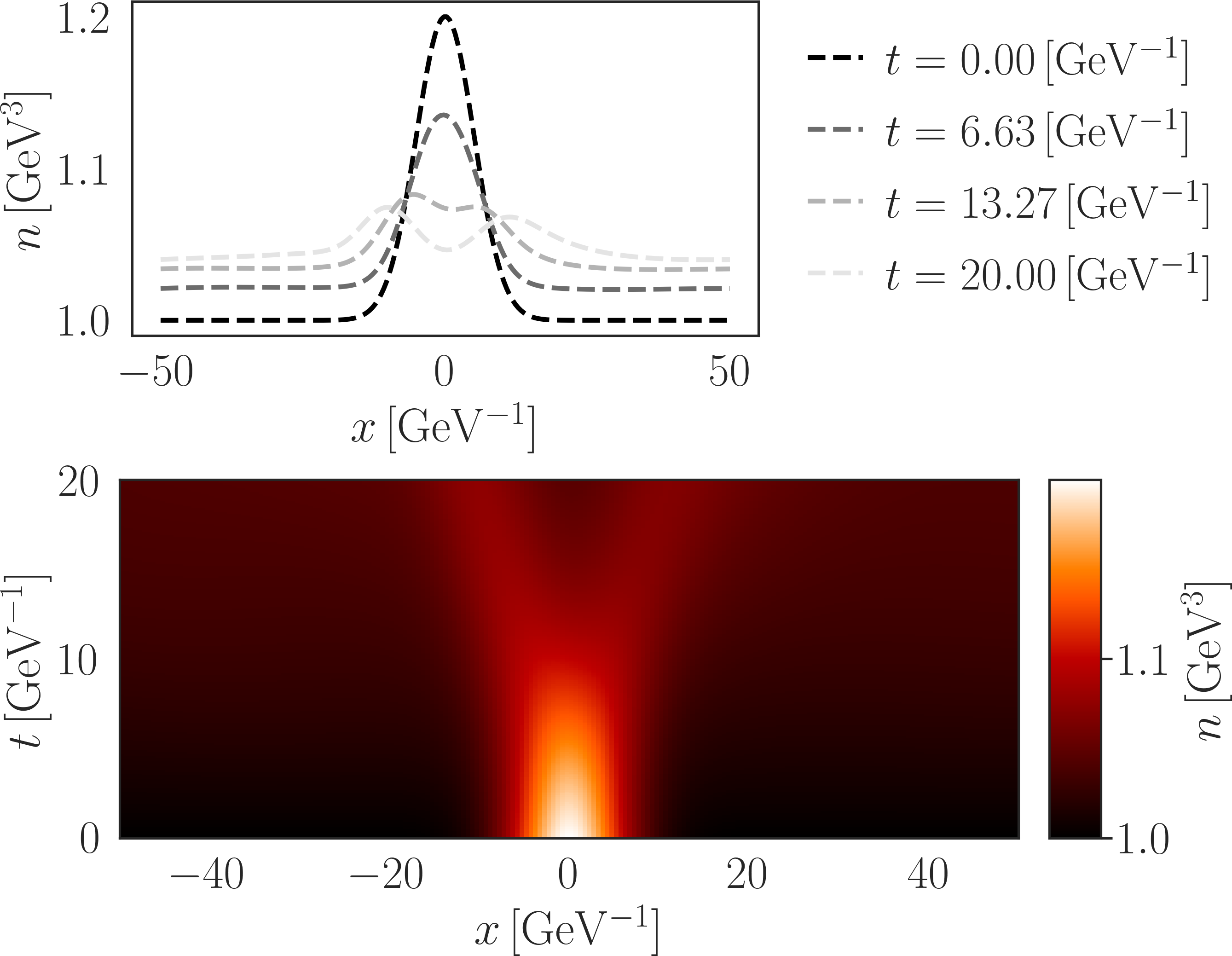}
  \makebox[0pt][l]{\small (a)}
\end{minipage}\hfill
\begin{minipage}[t]{0.49\textwidth}
  \centering
  \includegraphics[width=\linewidth]{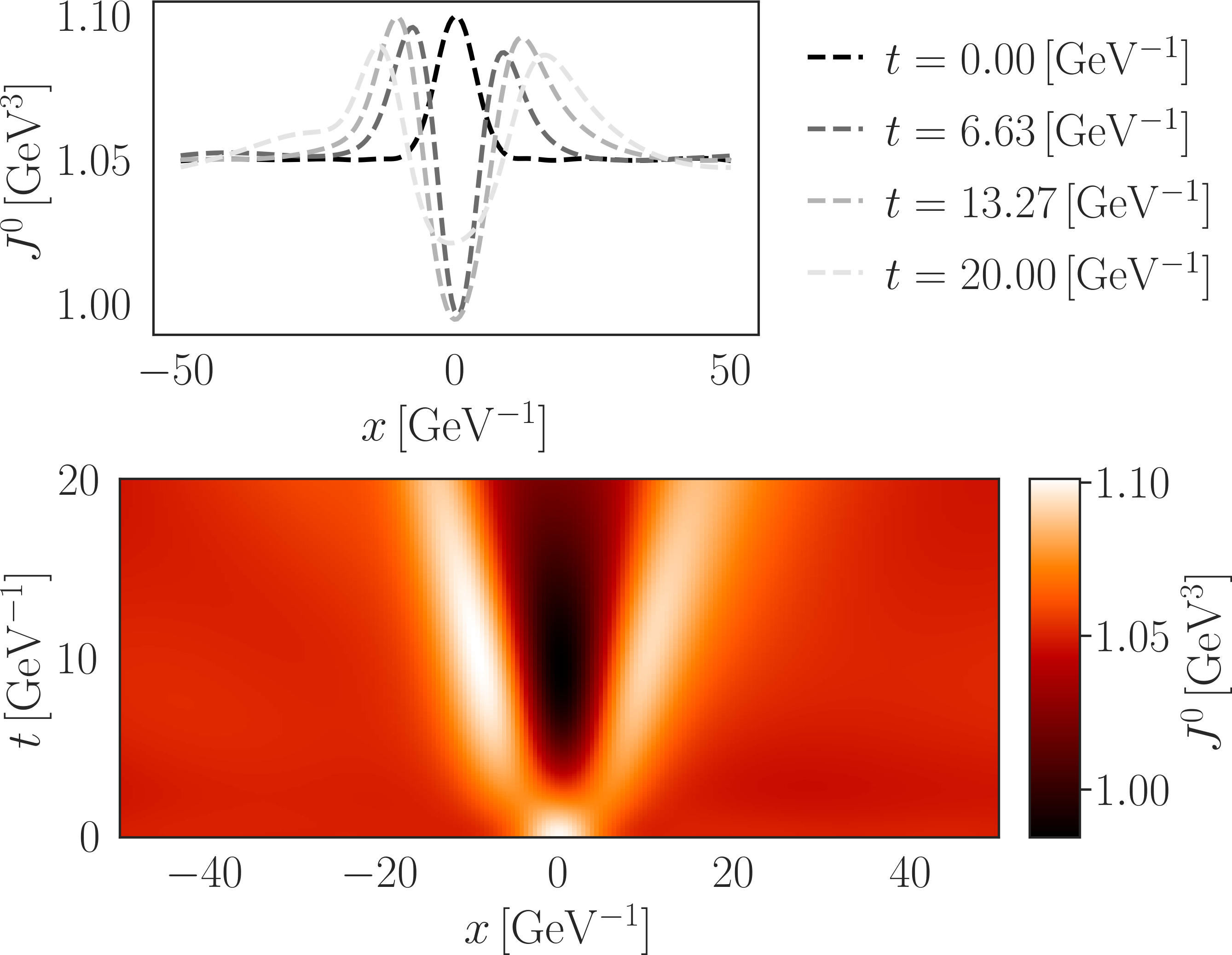}
  \makebox[0pt][l]{\small (b)}
\end{minipage}

\caption{Results from the SA-PINN-ACTO for the first setup, with $c_{\rm ch}=0.5$. For these simulations, we use $|N_{\rm PDE}|=20,000+500$, $|N_{\rm IC}|=1,000$, and $50,000$ Adam epochs, followed by L-BFGS fine-tuning. For this simulation, we obtain a spacetime relative $L^2$ error of $2.712\times 10^{-3}$ for $n$ and $5.485\times 10^{-3}$ for $J^0$. (a) Evolution of $n$ for $c_{\rm ch}=0.5$.
(b) Evolution of $J^0$ for $c_{\rm ch}=0.5$. Total charge $\int_{-L}^{L}{J^0\,dx}$ conserved at all times up to a fraction of $1.6\times 10^{-3}$ of the initial charge.
Best loss was $3.055\times 10^{-9}$ (PDE loss: $2.345\times 10^{-9}$; IC loss: $7.109\times 10^{-10}$). Total training time: $1328.42$ seconds (Adam: $1216.03$ seconds; L-BFGS: $112.39$ seconds).}
\label{fig:setup1-PINN-2nd-order}
\end{figure*}

Solving the problem this way has the strong advantage that it becomes unnecessary to derive a flux-conservative formulation of the original second-order PDE, which is typically needed for finite-volume schemes such as Kurganov-Tadmor. However, the second-order formulation requires, obviously, computing second-order derivatives via automatic differentiation (AD). This increases compute time and typically slows down training, which, in the end, makes the PINN slower. Moreover, the PINN would now have to learn a solution that satisfies certain relations between the second-order derivatives of the output (i.e., a solution that satisfies a second-order PDE), which naturally makes the learning task more complex. For these reasons, in this work we opted to solve the BDNK problem in its first-order (flux-conservative) form, but results of comparable accuracy can be achieved with a second-order PINN formulation, although with meaningfully longer training: for example, using $|N_{\rm PDE}|=20,000+500$ and $|N_{\rm IC}|=1,000$ collocation points, $50,000$ Adam pre-training epochs, up to $2,000$ L-BFGS iterations, and an initial learning rate of $1\times10^{-3}$, scheduled to decrease to $40\%$ of its previous value every time the loss plateaus for $2,000$ Adam epochs, we obtain, for the first setup, with $c_{\rm ch}=0.5$, the results shown in Fig.~\ref{fig:setup1-PINN-2nd-order}. 

However, note that even when only a second- or higher-order PDE formulation is available and no flux-conservative form is known, first-order physics-informed neural networks (FO-PINNs)~\cite{GLADSTONE2025106161} can achieve higher accuracy, more stable training, and faster convergence; an FO-PINN would most likely outperform the ``naive'' second-order SA-PINN-ACTO implemented in this appendix.

\section{Summary of background BDNK solution}

\begin{figure*}
\centering

\begin{minipage}[t]{0.49\linewidth}
  \centering
  \includegraphics[width=\linewidth]{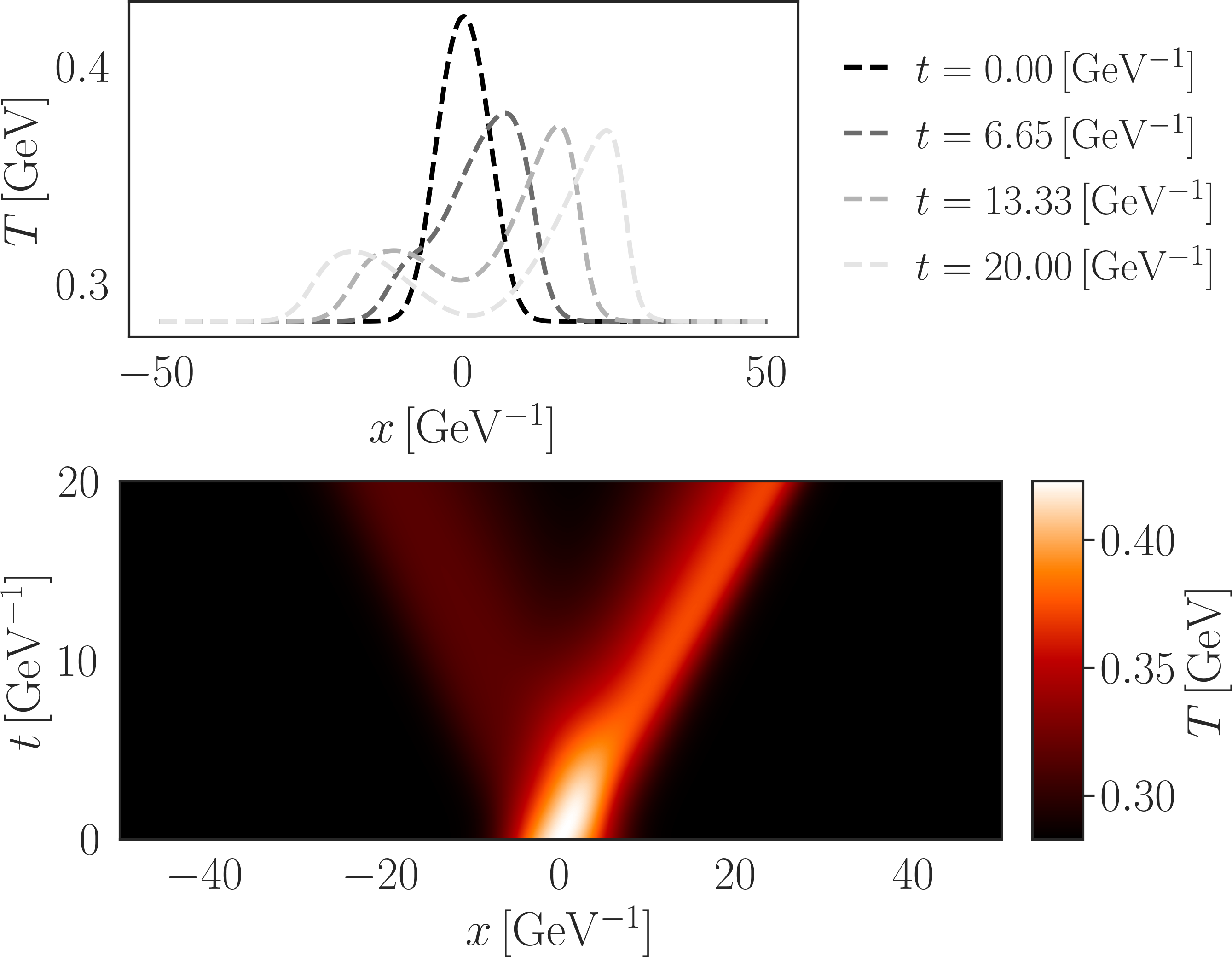}
  \makebox[0pt][l]{\small (a)}
\end{minipage}\hfill
\begin{minipage}[t]{0.49\linewidth}
  \centering
  \includegraphics[width=\linewidth]{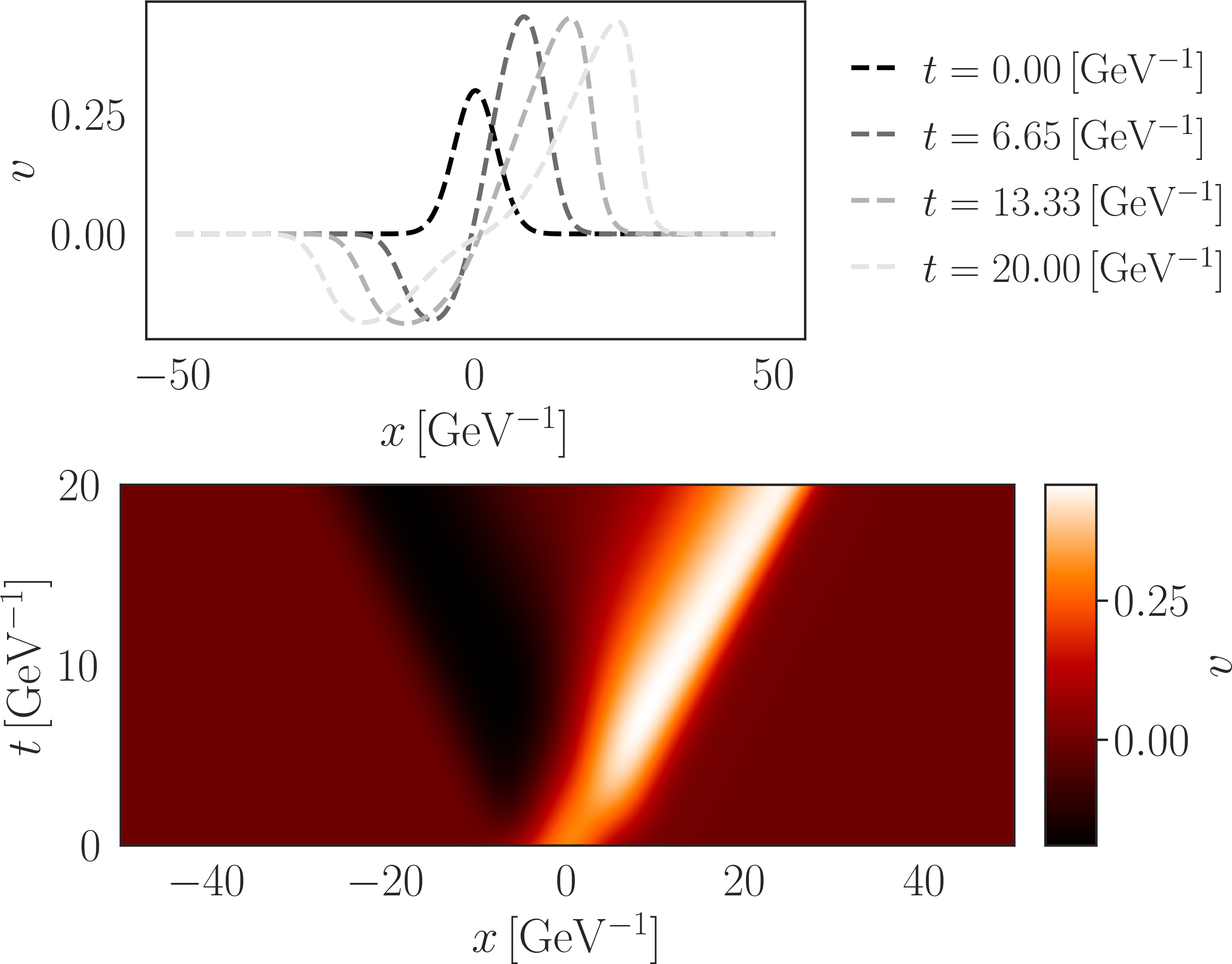}
  \makebox[0pt][l]{\small (b)}
\end{minipage}

\caption{(a) Temperature profile for the third setup.
(b) Velocity profile for the third setup.}
\label{fig:setup3-profiles}
\end{figure*}

The temperature and linear velocity profiles $T$, $\vec{v}$ used in the third setup, shown in Fig.~\ref{fig:setup3-profiles}, were generated from the numerical simulation of conformal $(1+1)$D BDNK hydrodynamics in Cartesian coordinates $(t,x)$, at zero chemical potential, developed in~\cite{Clarisse:2025lli}. This section serves as a high-level overview of how the corresponding equations of motion were solved.

In~\cite{Clarisse:2025lli}, the BDNK equations for energy-momentum conservation $\partial_\mu T^{\mu\nu}=0$ were recast in first-order flux-conservative form by defining some key new variables. In fact, in addition to studying the evolution of the stress tensor elements, we define a new field parallel to the flow $C_\mu = T u_\mu$ and the corresponding derivative $X_{\mu\nu} = \partial_\mu C_\nu$ used in the first-order form of the PDEs. The result is the energy-momentum tensor being recast into the ideal part plus viscous corrections in the following way,
\begin{equation}\label{defineTnew}
    T^{\mu\nu} = T^{\mu\nu}_0 + \frac{1}{T^2}H^{\mu\nu \alpha\rho}\left (\delta_\rho^\lambda+2u_\rho u^\lambda\right )X_{\alpha\lambda}- \frac{1}{T^2}H^{\mu\nu\alpha\rho} \Gamma_{\alpha\rho}^\lambda C_\lambda.
\end{equation}
Further details about the validity of this approach are covered in~\cite{Clarisse:2025lli}.

The resulting equations of motion are
\be \label{EOM_vec}
\partial_0 \m{ T^{0\nu} \\ C_0 \\ C_i \\ X_{ij} \\ X_{i0} }
+ \partial_k \m{  T^{k\nu} \\ 0 \\ 0_i \\ -\delta^k_i X_{0j} \\ - \delta^k_i X_{00} }
= \m{ - \Gamma ^\mu_{\mu\lambda} T^{\lambda \nu} - \Gamma^\nu_{\mu\lambda} T^{\mu\lambda} \\ X_{00} \\ X_{0i} \\ 0_{ij} \\ 0_{i}},
\ee
which allows for a clear conservation law form in $(1+1)$D that is favorable for finite volume methods,
\begin{equation}
	\partial_t \mathbf q + \partial_x \mathbf F^x = \mathbf S.
\end{equation}
Alongside the aforementioned KT method discussed previously in Sec.~\ref{ssec:KT}, we employ the standard second-order Total Variation Diminishing Runge--Kutta scheme (TVDRK2)~\cite{Shu:1988iw}. 
The time integration proceeds from step $l$ to $l+1$ with intermediate values denoted by the superscript~$(1)$:
\begin{align}
\mathbf{q}^{(1)} &= \mathbf{q}_l - h_t\, \mathbf{Z}(\mathbf{q}_l), \\
\mathbf{q}_{l+1} = \mathbf{q}^{(2)} &= \tfrac{1}{2}\mathbf{q}_l + \tfrac{1}{2}\mathbf{q}^{(1)} - \tfrac{1}{2}h_t\, \mathbf{Z}(\mathbf{q}^{(1)}).
\end{align}
The spatial residual is defined as
\begin{equation}
\mathbf{Z}(\mathbf{q}) = \partial_x \mathbf{F}^x(\mathbf{q}) - \mathbf{S}(\mathbf{q}),
\end{equation}
and the Runge--Kutta coefficients are
\begin{equation}
\alpha_{ij} = \{\alpha_{10}=1,\, \alpha_{20}=1/2,\, \alpha_{21}=1/2\}, \quad  
\beta_{ij} = \{\beta_{10}=1,\, \beta_{20}=0,\, \beta_{21}=1\}.
\end{equation}
The CFL condition~\cite{1967_courant_friedrichs_lewy} imposes the constraint 
\begin{equation}
0 < c_{\text{CFL}} \leq 0.5,
\end{equation}
which relates the temporal and spatial resolutions through
\begin{equation}
h_t \leq c_{\text{CFL}} h_x.
\end{equation}

The specific details of the initial conditions for all of these fields, to then predict the evolution of $T, v$ that was used in this work correspond to a system with vanishing $\mathcal{A}$ (the BDNK out-of-equilibrium correction to the energy density) and $\mathcal{Q}^\mu$ (the BDNK energy diffusion vector), but nonzero shear stress tensor $\pi^{\mu\nu} \neq 0$. Hence, the only dissipative contribution to the initial energy--momentum tensor $T^{\mu\nu}$ arises from the shear tensor $\sigma^{\mu\nu}$. This setup ensures that the BDNK evolution begins close to an equilibrium state, consistent with the validity regime of the \emph{first-order} formulation. Following the procedure detailed in~\cite{Bemfica:2017wps}, the general $(3+1)$ expressions for the initial derivatives of flow velocity and energy density to properly initialize all the BDNK fields are as follows: 
\begin{align}
    \partial_0 \varepsilon \big|_{t=0} &= 
    \frac{4 \varepsilon \gamma}{3 + 2 (\overline{u})^2} 
    \left[
    \frac{u^l u^m \partial_l u_m}{1 + (\overline{u})^2} 
    - \frac{\partial_l u^l - u^l \partial_l \varepsilon}{2\varepsilon}
    \right], \\
    \gamma\, \partial_0 u_j \big|_{t=0} &= 
    \left[
    \gamma^2 \partial_l u^l 
    - u^l u^m \partial_l u_m 
    - \frac{u^l \partial_l \varepsilon}{4\varepsilon}
    \right]
    \frac{u_j}{3 + 2 (\overline{u})^2}
    - u^l \partial_l u_j 
    - \frac{\partial_j \varepsilon}{4\varepsilon},
    \label{eq:d0uj}
\end{align}
where $(\overline{u})^2 = u_i u^i$. The shear viscosity used in this simulation is such that $\eta/s = 1/4\pi$, and the temperature and velocity profiles used as background fields are shown in Sec.~\ref{sec:test_cases_and_results}. The hydrodynamic frame choice used is Frame 1 outlined in~\cite{Clarisse:2025lli} Sec. II F, where the frame parameters are $a_1=25/4$, $a_2 = 25/7$ and maximum propagation speed is $c_+ = 1.0$.

\section{Results of convergence tests}\label{app:convergence_tests_results}

\begin{figure*}[!t]
\centering

\begin{minipage}[t]{0.49\textwidth}
  \centering
  \includegraphics[width=\linewidth]{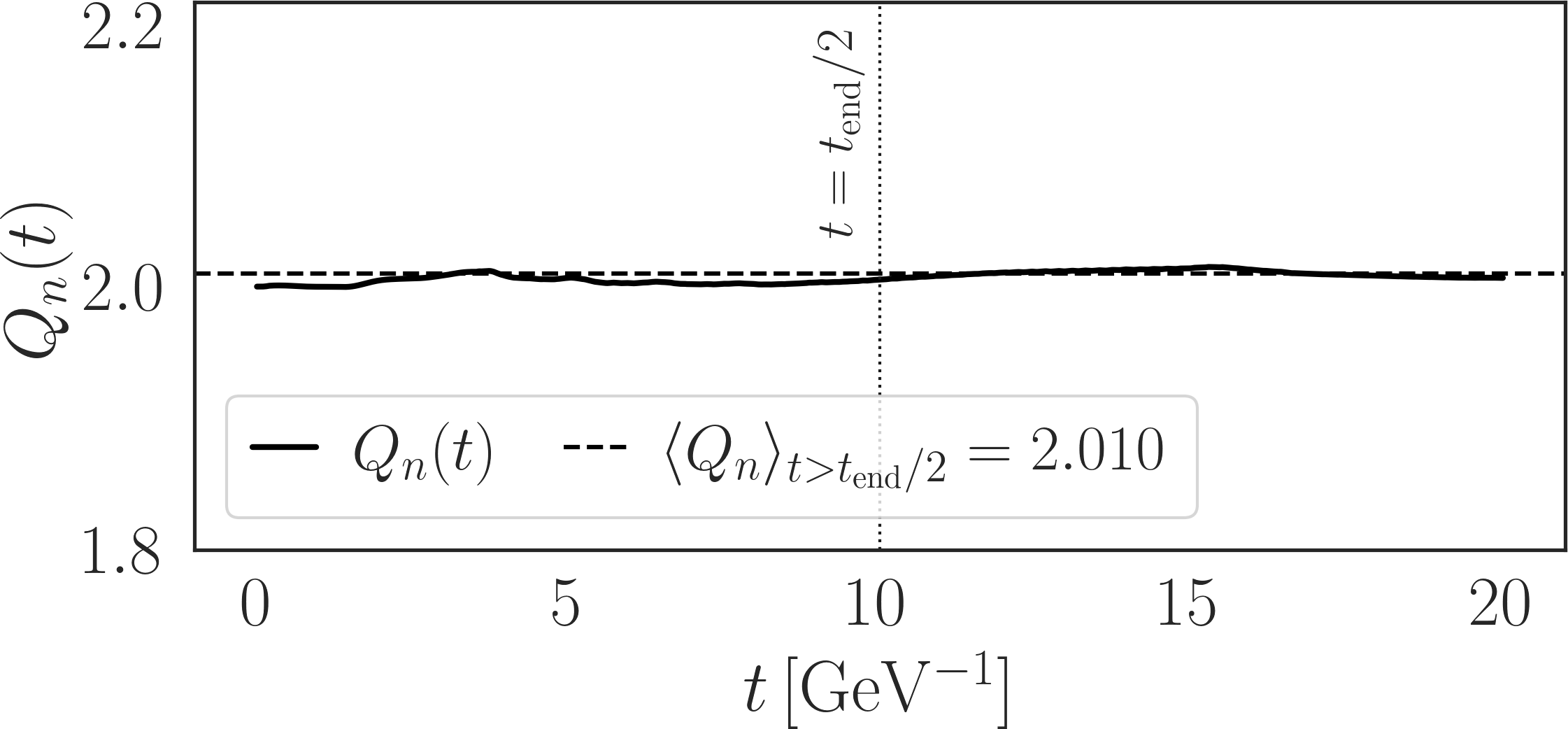}
  \makebox[0pt][l]{\small (a)}
\end{minipage}\hfill
\begin{minipage}[t]{0.49\textwidth}
  \centering
  \includegraphics[width=\linewidth]{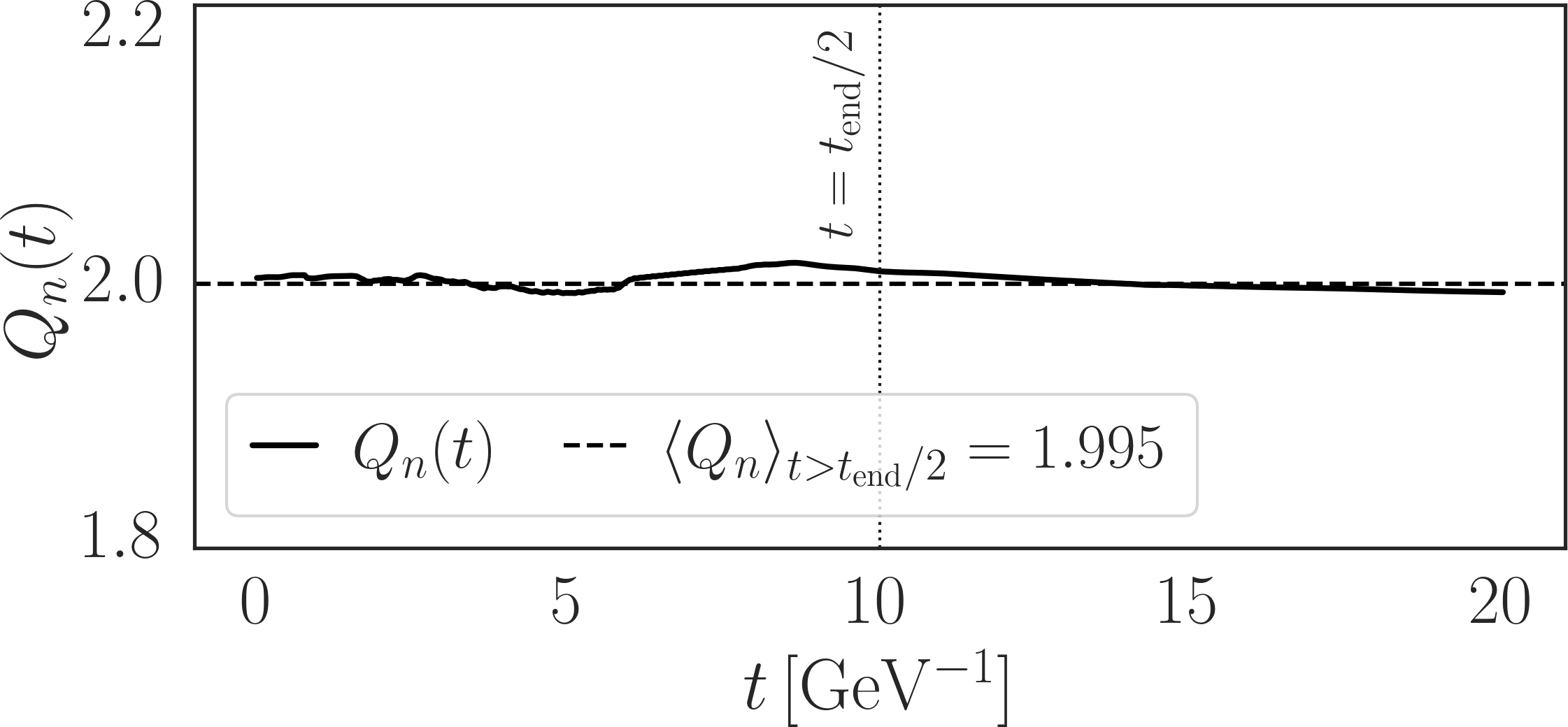}
  \makebox[0pt][l]{\small (b)}
\end{minipage}

\vspace{0.75em}

\begin{minipage}[t]{0.49\textwidth}
  \centering
  \includegraphics[width=\linewidth]{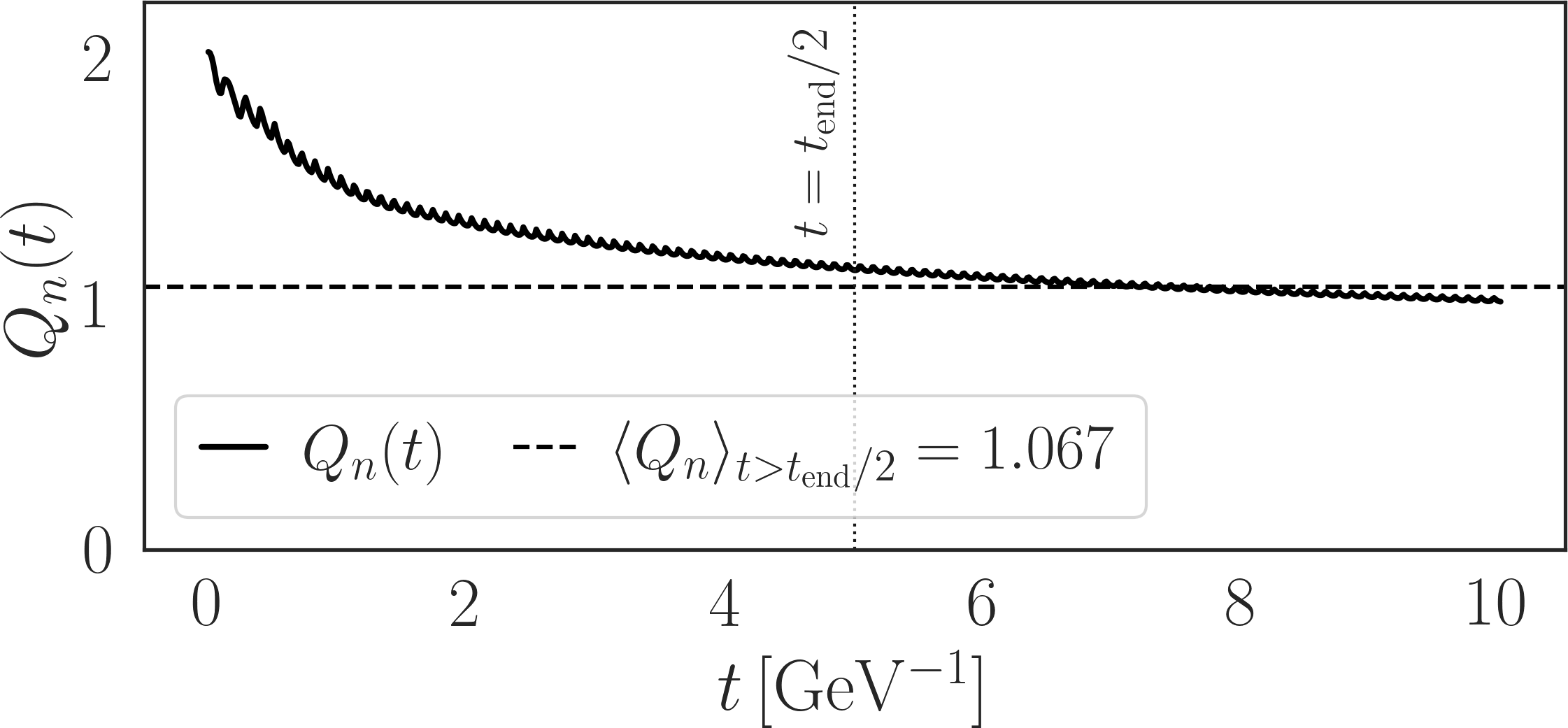}
  \makebox[0pt][l]{\small (c)}
\end{minipage}\hfill
\begin{minipage}[t]{0.49\textwidth}
  \centering
  \includegraphics[width=\linewidth]{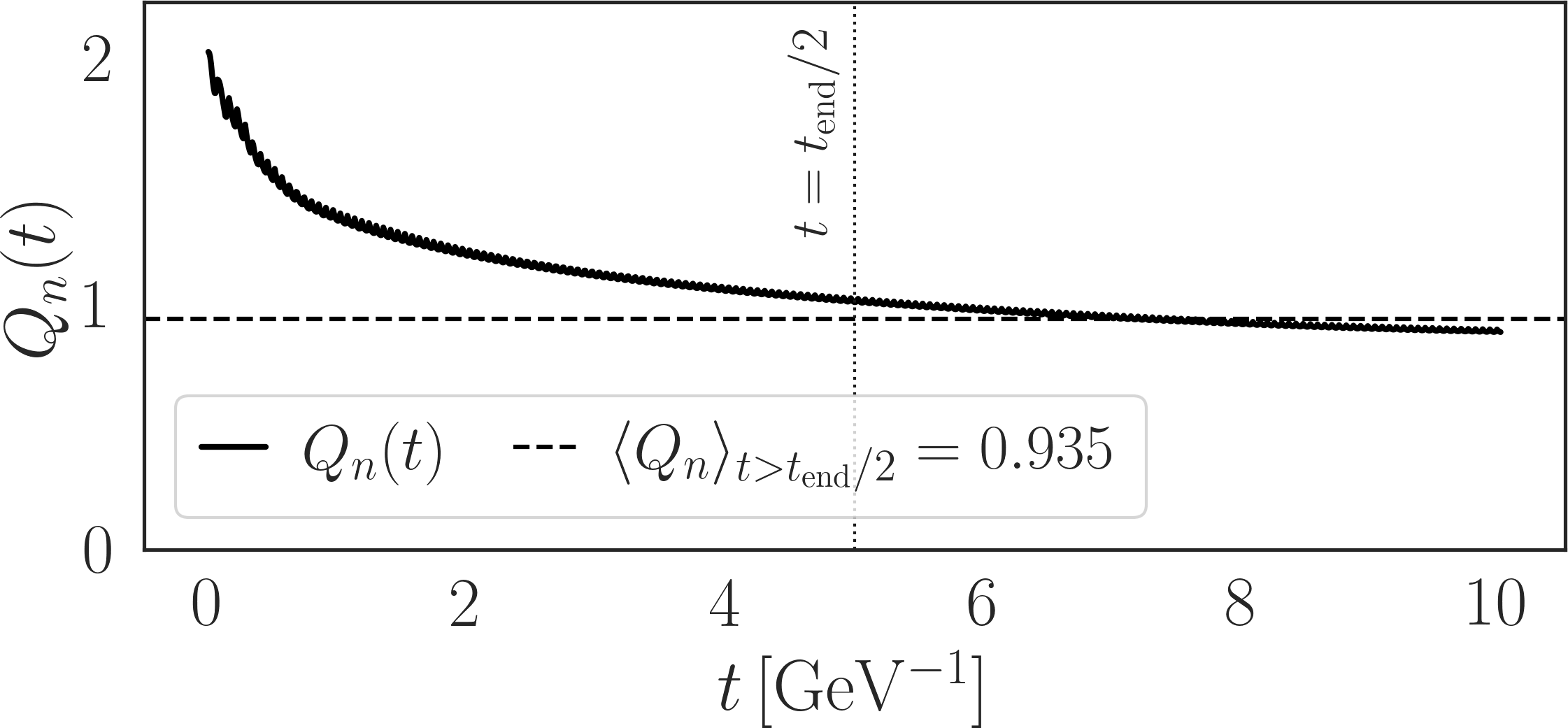}
  \makebox[0pt][l]{\small (d)}
\end{minipage}

\vspace{0.75em}

\begin{minipage}[t]{0.49\textwidth}
  \centering
  \includegraphics[width=\linewidth]{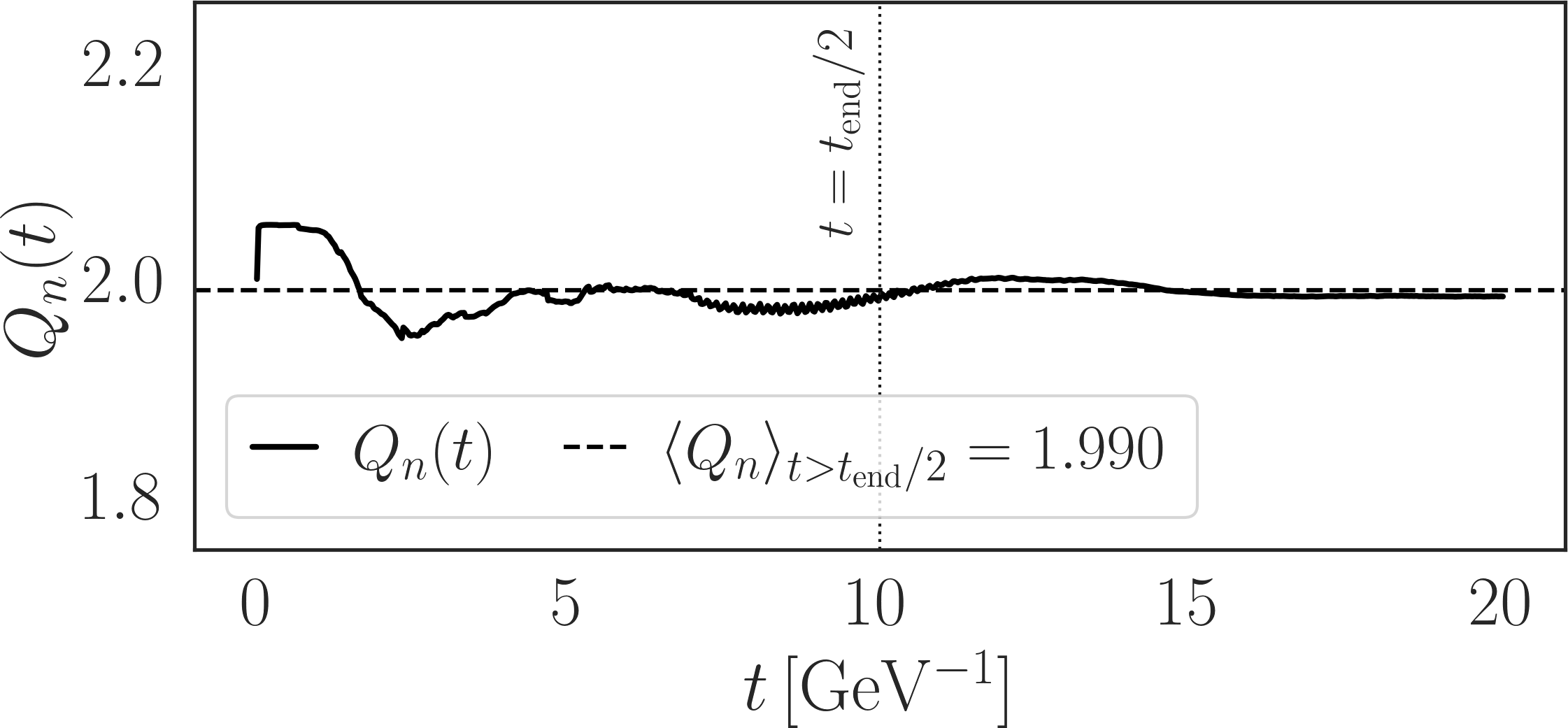}
  \makebox[0pt][l]{\small (e)}
\end{minipage}\hfill
\begin{minipage}[t]{0.49\textwidth}
  \centering
  \includegraphics[width=\linewidth]{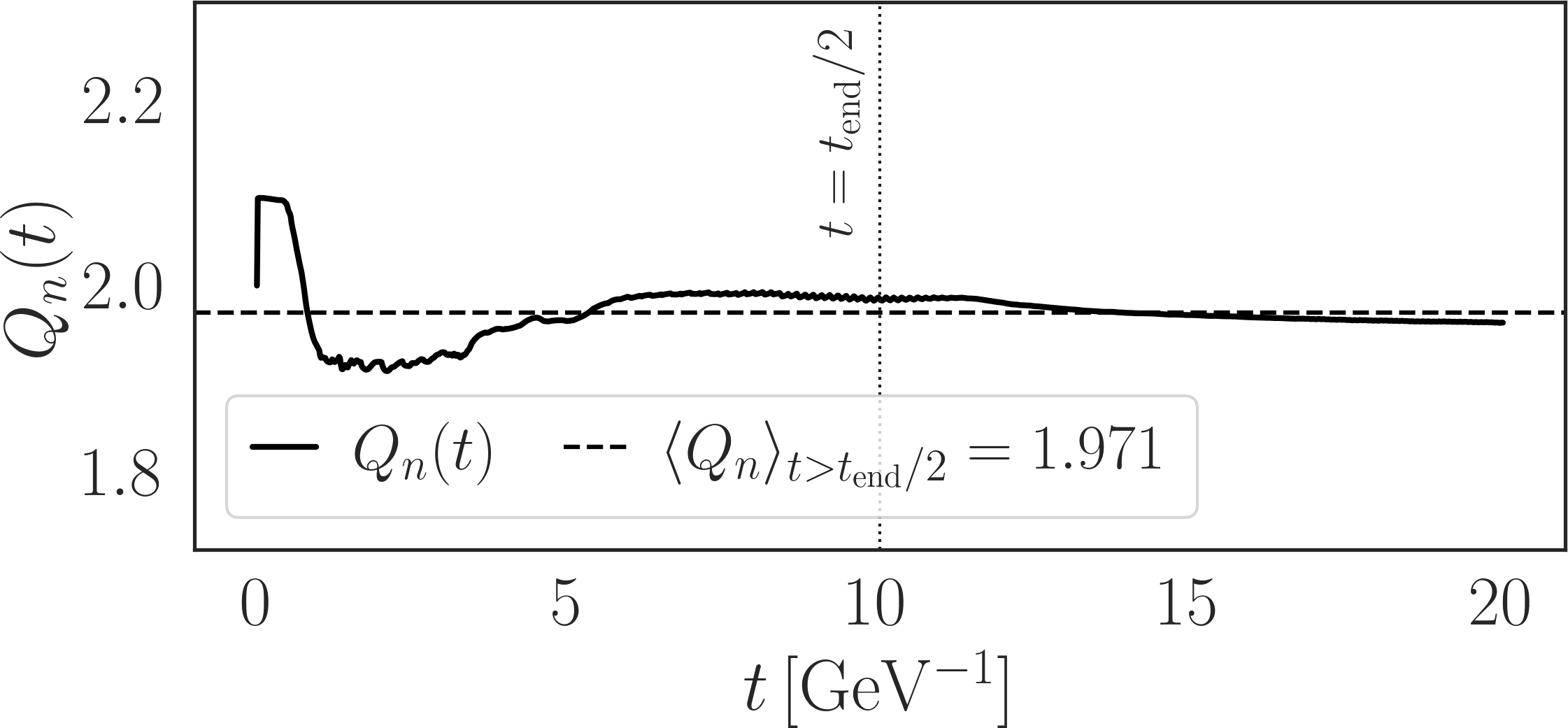}
  \makebox[0pt][l]{\small (f)}
\end{minipage}

\caption{Results of convergence tests of the KT simulations for:
(a) first setup, with $c_{\rm ch}=0.5$;
(b) first setup, with $c_{\rm ch}=0.9$;
(c) second setup, with $c_{\rm ch}=0.5$;
(d) second setup, with $c_{\rm ch}=0.9$;
(e) third setup, with $c_{\rm ch}=0.5$;
(f) third setup, with $c_{\rm ch}=0.9$.
Execution times for all three runs for each of (a)-(f) are $9.19$ seconds, $11.83$ seconds, $12.90$ seconds, $17.58$ seconds, $9.87$ seconds, and $13.18$ seconds, respectively.}
\label{fig:convergence_tests_results}
\end{figure*}

The results from the convergence tests of the KT simulations are shown in Fig.~\ref{fig:convergence_tests_results}. For the simulations in Secs.~\ref{ssec:first_setup} and~\ref{ssec:third_setup}, which have smooth initial conditions, we use $1000$, $2000$, and $4000$ spatial cells; for the smooth shock case in Sec.~\ref{ssec:second_setup}, we use $2000$, $4000$, and $8000$ spatial cells.
As seen in Fig.~\ref{fig:convergence_tests_results}, $Q_n$ converges to $\sim 2$ for the first and third setups, as expected for continuous solutions, and to $\sim 1$ for the second setup, as expected for solutions with discontinuities.

\bibliography{references}

\end{document}